\documentclass[aps]{revtex4}
\usepackage{makeidx}
\usepackage{latexsym, amscd}
\usepackage{amsmath}
\usepackage{graphicx}

\begin{document}

\title{Spatial control of the competition between self-focusing and self-defocusing
nonlinearities in one- and two-dimensional systems}

\author{Nguyen Viet Hung$^{1}$, Marek Trippenbach$^{2}$, Eryk Infeld$^{3}$,
and Boris A. Malomed$^{4}$}

\address{$^{1}$Advanced Institute for Science and Technology (AIST),
Hanoi University of Science and Technology (HUST), Hanoi, Vietnam}

\address{$^{2}$Institute of Theoretical Physics, Physics Department, Warsaw
University, Ho\.{z}a 69, PL-00-681 Warsaw, Poland}

\address{$^{3}$National Centre for Nuclear Research, Ho\.{z}a 69, PL-00-681
Warsaw, Poland}

\address{$^{4}$Department of Physical Electronics, School of Electrical
Engineering, Faculty of Engineering, Tel Aviv University, Tel Aviv
69978, Israel}

\begin{abstract}

We introduce a system with competing self-focusing (SF)\ and
self-defocusing (SDF) terms, which have the same scaling dimension.
In the one-dimensional (1D) system, this setting is provided by a
combination of the SF cubic term multiplied by the delta-function,
$\delta (x)$, and a spatially uniform SDF
quintic term. This system gives rise to the most general family of \textit{%
1D Townes solitons}, the entire family being unstable. However, it is completely
stabilized by a finite-width regularization of the $\delta $-function. The
results are produced by means of numerical and analytical methods. We also
consider the system with a symmetric pair of regularized $\delta $%
-functions, which gives rise to a wealth of symmetric, antisymmetric, and
asymmetric solitons, linked by a bifurcation loop, that accounts for the
breaking and restoration of the symmetry. Soliton families in 2D versions of
both the single- and double-delta-functional systems are also studied. The 1D
and 2D settings may be realized for spatial solitons in optics, and in
Bose-Einstein condensates.
\end{abstract}

\maketitle

\section{Introduction}

The creation of various self-trapped modes (loosely called
\textquotedblleft solitons" in this paper) in many physical systems
is complicated by competition between self-focusing (SF) and
self-defocusing (SDF)\ nonlinearities. The first example, which was
studied in detail, is the competition between the
second-harmonic-generating, i.e., quadratic, and cubic
nonlinearities in optics. In such systems, the adjustment of the
mismatch between the fundamental and second harmonics makes it
possible to render the quadratic nonlinearity effectively
self-focusing, while the cubic term may be self-defocusing. The
interplay of these competing interactions gives rise to diverse
soliton states, both one- and two-dimensional (1D and 2D)
\cite{chi2,chi2-review}.

Furthermore, a subject of many works was the competition between cubic
SF and quintic SDF\ terms. The cubic-quintic (CQ) nonlinearity --
usually, with opposite signs of the two terms -- occurs in the various
photonic media. Known examples include the light propagation in
diverse fluids \cite{liquids}, specialty glasses \cite{glasses},
ferroelectric films \cite{ferroelectric}, and colloidal suspensions
of metallic nanoparticles \cite{colloid}. The colloids offer
remarkable flexibility, making it possible to adjust parameters of
the CQ nonlinearity (the signs and magnitudes of both the cubic and
quintic terms) through the selection of the diameter of the
nanoparticles and the filling factor of the suspension.

The realization of the CQ nonlinearity was also theoretically elaborated in
terms of the Gross-Pitaevskii equation \cite{Pitaevskii} for Bose-Einstein
condensates (BEC), where the quintic term accounts for three-body
collisions, provided that inelastic effects may be neglected \cite{BEC}. In
this context, the adjustment of the nonlinearity may be provided by the
Feshbach resonance which affects the sign and strength of the cubic term
\cite{FR}.

It has been theoretically demonstrated that the use of the
CQ nonlinearity combining the SF and SDF terms opens a way to the
creation of stable multidimensional solitons, including 2D
\cite{Manolo} and 3D
\cite{nine-authors} self-trapped vortices, see a review in Ref. \cite{review}%
. Recently, 2D fundamental solitons of this type were produced
experimentally in colloidal waveguides \cite{liquids2}. The use of the CQ
nonlinearity is necessary for this purpose, because the SF\ cubic
interaction can create only the family of \textit{Townes' solitons} in the
2D setting, which is fully destabilized by the occurrence of the critical
collapse in the same geometry \cite{Berge}. The Townes' family is
degenerate, in the sense that the norm (total power) of the solitons takes
the single value, which does not depend on the propagation constant.
However, the 2D system including the SDF quintic term suppresses the
collapse and thus lends stability to the self-trapped states, with both the
fundamental and vortical internal structure \cite{Manolo,review}.

A degenerate family the Townes-like solitons, subject to the instability
driven by the critical collapse, is known in the 1D setting as well. It is
described by the nonlinear Schr\"{o}dinger (NLS) equation with the quintic
SF term, in the absence of cubic ones \cite{Salerno}. The addition of the SF
cubic nonlinearity does not eliminate the collapse, but it lifts the
degeneracy and stabilizes all the solitons against small perturbations \cite%
{Pelin}. It is well known too that, while the 1D NLS equation with the CQ
combination of nonlinear terms is not integrable, it admits an exact
analytical solutions for the full soliton family \cite{Pushkarov}.

A remarkable peculiarity of the 1D NLS equation is that it admits another
type of the nonlinearity, which gives rise to another degenerate soliton
family, that may also be considered as a variety of the Townes' solitons.
This equation contains the SF\ cubic nonlinear term concentrated, in the
ideal form, at a single spatial point, $x=0$, i.e., multiplied by
Dirac's delta-function, $\delta (x)$. This model was introduced in Ref. \cite%
{Azbel}, and it was later shown that the entire family of the solitons,
whose norm does not depend on the propagation constant either (the
characteristic feature of solitons of the Townes' type), is completely
unstable \cite{Dror}. Lifting of the degeneracy, and partial or complete
stabilization of the latter family is possible in the system with two $%
\delta $-functions \cite{Dong,Yasha}, or if the ideal $\delta $-function is
replaced by its finite-width counterpart \cite{Dror}.

The fact that two different nonlinear terms in the 1D\ NLS equation, $\delta
(x)|U|^{2}U$ and $|U|^{4}U$, where $U(x)$ is the complex wave field, give
rise to Townes-like soliton families is a unique feature of the 1D setting,
which is explained by the coinciding \textit{scaling dimension} of both
these terms. This circumstance suggests to consider a general family of the
Townes' solitons, produced by the 1D NLS equation including both terms. In
this work, we demonstrate that such a family can be found in an exact form,
and it remains degenerate (with the norm independent of the propagation
constant), hence it is unstable too. In fact, another
situation is more interesting, which is the main subject of the present work, \textit{viz}., the
competition between such cubic SF and quintic SDF\ terms, which still share
the identical scaling dimension. This situation is relevant to both physical
settings mentioned above: in optical waveguides, the local cubic
nonlinearity may be induced by means of locally implanted resonant dopants
\cite{Kip}, while in BEC it can be implemented by means of the Feshbach
resonance imposed by a tightly focused laser beam \cite{laser}.

The most interesting issue to be considered in the framework of these
systems is the stability of solitons. We conclude that, in the case of the
ideal Dirac's $\delta $-function multiplying the SF cubic term, the soliton
family can be found in an exact analytical form, remaining degenerate and
unstable. However, the regularization, which replaces the ideal $\delta $%
-function by an approximation with a finite intrinsic scale, immediately
stabilizes the entire family. This result, which is meaningful in terms of
the above-mentioned physical realizations, as any locally induced
nonlinearity has a finite spatial size, is obtained below by means of a
combination of numerical and analytical methods. Interestingly, parts of the
soliton families for which the SF or SDF term is the dominant one, turn out
to be stable, severally, in accordance with the \textit{%
Vakhitov-Kolokolov} (VK) criterion \cite{VK,Berge} or the
\textit{anti-VK} one \cite{anti,Merhasin}. As concerns analytical
methods, we use the perturbation theory, the Thomas-Fermi
approximation (TFA), and also produce particular exact solutions in
the model with the regularized $\delta $-function. Using numerical
methods, we subsequently extend the analysis to the system with two
regularized $\delta $-functions, and, finally, to the 2D version of
the model, with both the single and double regularized $\delta
$-functions.

The rest of the paper is organized as follows. The 1D system is formulated
in Section II, where we present the basic equations and analytical
solutions. Numerical and additional analytical results for the 1D systems,
with the single and double regularized $\delta $-functions, are reported in
Section III. The 2D systems are introduced in Section IV, where numerical
results are reported for them too. The paper is concluded by Section IV.

\section{One-dimensional models}

\subsection{Basic equations and solutions}

We start by considering the1D NLS equation which includes competing
SF cubic and SDF quintic terms:
\begin{equation}
iU_{z}=-\frac{1}{2}U_{xx}-\delta (x)|U|^{2}U+\sigma |U|^{4}U,  \label{1D1}
\end{equation}%
where constant $\sigma $ is positive in the case of the competing SF-SDF
nonlinearities, and $\delta (x)$ is Dirac's $\delta $-function. It
imposes a condition for the jump of the first derivative at $x=0$:%
\begin{equation}
U_{x}\left( x=+0\right) -U_{x}\left( x=-0\right) =-2\left\vert
U(x=0)\right\vert ^{2}U(x=0),  \label{jump}
\end{equation}%
while function $U(x)$ itself is continuous at this point. Note that,
although Eq. (\ref{1D1}) admits the invariance with respect to the scaling
transformation,
\begin{equation}
z\equiv x_{0}^{2}\tilde{z},~x\equiv x_{0}\tilde{x},~U\equiv x_{0}^{-1/2}%
\tilde{U},  \label{scaling}
\end{equation}%
with arbitrary factor $x_{0}$, it cannot alter coefficient $\sigma $ in
front of the quintic term, due to the above-mentioned fact that both
nonlinear terms in Eq. (\ref{1D1}) have the same scaling dimension. The
presence of the \emph{irreducible} parameter $\sigma $ is an essential
peculiarity of the system.

In the absence of the SDF quintic term ($\sigma =0$), Eq.~~(\ref{1D1}) has
an exact stationary solution with arbitrary propagation constant $\mu >0$,
in the form of a soliton pinned to the $\delta $-function \cite{Azbel,Dror}:
\begin{equation}
U(x,z)\equiv U(x)e^{i\mu z}=(2\mu )^{1/4}e^{-\sqrt{2\mu }|x|+i\mu
z}. \label{exact0}
\end{equation}%
The total power (norm) of this pinned state is%
\begin{equation}
N=\int_{-\infty }^{+\infty }\left\vert U(x)\right\vert ^{2}dx=1,
\label{N0}
\end{equation}%
which is independent of the propagation constant, $\mu $ [note that the norm
is invariant with respect to transformation (\ref{scaling})]. According to
the VK criterion, which states that inequality
\begin{equation}
dN/d\mu >0  \label{VK>0}
\end{equation}%
is a necessary stability condition for solitons supported by a SF
nonlinearity \cite{VK,Berge}, all solutions (\ref{exact0}) might be
neutrally stable, but in fact they all are unstable \cite{Dror}.

In the presence of the SDF quintic term ($\sigma >0$), a family of exact
solutions to Eq. (\ref{1D1}) can be found in the analytical form as well:%
\begin{eqnarray}
U(x,z) &=&e^{i\mu z}U(x)=e^{i\mu z}\frac{\left( 3\mu /\sigma \right) ^{1/4}}{%
\sqrt{\sinh \left( \sqrt{8\mu }|x|+\xi \right) }},  \label{exact} \\
\xi &=&\ln \left( \sqrt{\frac{3}{2\sigma }}+\sqrt{\frac{3}{2\sigma }-1}%
\right) .  \label{xi}
\end{eqnarray}%
As follows from Eq. (\ref{xi}), this solution exists provided that the
coefficient in front of the SDF quintic term is not too large,%
\begin{equation}
0<\sigma <\sigma _{\max }\equiv 3/2.  \label{max}
\end{equation}

The total norm of the solution family (\ref{exact}) is again degenerate (it
does not depend on $\mu $, depends on $\sigma $):%
\begin{equation}
N\left( \sigma \right) =\sqrt{\frac{3}{2\sigma }}\ln \left( \frac{\sqrt{3}+%
\sqrt{3-2\sigma }+\sqrt{2\sigma }}{\sqrt{3}+\sqrt{3-2\sigma }-\sqrt{2\sigma }%
}\right) .  \label{N}
\end{equation}%
In particular, in the limit of $\sigma \rightarrow 0$ expression (\ref{N})
continuously goes over into norm (\ref{N0}), $N(\sigma =0)=1$, while in the
opposite limit of $\sigma \rightarrow \sigma _{\max }$, see Eq. (\ref{max}),
the norm diverges, $N\approx (1/2)\ln \left( 1/\left( \sigma _{\max }-\sigma
\right) \right) $. In the latter limit, the peak power (squared amplitude)
of the solution given by Eqs. (\ref{exact}) and (\ref{xi}) diverges too:%
\begin{equation}
U^{2}(x=0)=\frac{\sqrt{3\mu /\sigma }}{\sinh \xi }\approx \sqrt{\frac{3\mu }{%
\sigma _{\max }-\sigma }}.  \label{u^2}
\end{equation}%
Once again, the fact that $dN/d\mu =0$ formally suggests that the soliton
family (\ref{exact}) might be neutrally stable according to the VK
criterion, but in reality such a family of solitons with degenerate total
power is \emph{completely unstable} \cite{Dror}. Thus, the combination of
the cubic and quintic terms in Eq. (\ref{1D1}) gives rise to the family of
the solitons of the Townes' type (with the $\mu $-independent norm) even in
the case when the two terms are set to compete.

In the case of $\sigma <0$ (the \textit{cooperating}, rather than competing,
nonlinearities), when both nonlinear terms in Eq. (\ref{1D1}) have the SF
sign, a family of exact soliton solutions can be readily found too:%
\begin{eqnarray}
U(x,z) &=&e^{i\mu z}\frac{\left( -3\mu /\sigma \right) ^{1/4}}{\sqrt{\cosh
\left( \sqrt{8\mu }|x|+\tilde{\xi}\right) }},  \label{cosh} \\
\tilde{\xi} &=&\ln \left( \sqrt{-\frac{3}{2\sigma }+1}+\sqrt{-\frac{3}{%
2\sigma }}\right) ,  \label{xi-tilde}
\end{eqnarray}%
cf. the exact solutions given by Eqs. (\ref{exact}) and (\ref{xi}) for the
competing nonlinearities, with $\sigma >0$. The total power of this family is%
\begin{equation}
N\left( \sigma <0\right) =\sqrt{-\frac{6}{\sigma }}\arctan \left( \sqrt{-%
\frac{3}{2\sigma }+1}-\sqrt{-\frac{3}{2\sigma }}\right) .  \label{N-}
\end{equation}%
Like the respective expression (\ref{N}) for $\sigma >0$, this norm does not
depend on the propagation constant, $\mu $, hence this family too is the
degenerate one, of the Townes' type, and is completely unstable. However,
unlike its counterpart (\ref{N}), Eq. (\ref{N-}) demonstrates that, with the
increase $-\sigma $ from zero to infinity, the total power drops from $%
N(\sigma \rightarrow -0)=1$ to $N\left( \sigma \rightarrow -\infty
\right) =0 $. Thus, the interplay of the cubic and quintic terms in
Eq. (\ref{1D1}) gives rise to the most general Townes-soliton
family, which comprises both cases of the competing and
collaborating types.

The solitons can be stabilized by a regularization of Dirac's $\delta $%
-function, for which we adopt the commonly known Gaussian form:
\begin{equation}
\delta _{\mathrm{Dirac}}(x)\rightarrow \delta
_{\mathrm{Gauss}}(x)\equiv \left( 1/\sqrt{\pi }a\right)
e^{-x^{2}/a^{2}}  \label{delta}
\end{equation}%
with scale constant $a>0$. Indeed, rewriting, accordingly, the corrected
jump condition (\ref{jump}) in the form of%
\begin{equation}
U_{x}\left( x=+0\right) -U_{x}\left( x=-0\right) =-\frac{2}{\sqrt{\pi }a}%
\int_{-\infty }^{+\infty }\left\vert U(x)\right\vert
^{2}U(x)e^{-x^{2}/a^{2}}dx,  \label{jump2}
\end{equation}%
and looking for a stationary solution to Eq. (\ref{1D1}) with $\sigma =0$ as
$A\exp \left( -\sqrt{2\mu }|x|+i\mu z\right) $, cf. expression (\ref{exact0}%
), the calculation of the integral on the right-hand side of Eq. (\ref{jump2}%
) yields the first correction to the squared amplitude for small $a$ (or for
small $\mu $, if $a=1$ is fixed, see below),%
\begin{equation}
A^{2}\approx \sqrt{2\mu }\left( 1+3\sqrt{2\mu /\pi }a\right) ,
\end{equation}%
and the respective correction to norm (\ref{N0}),%
\begin{equation}
N\approx \left( 1+3\sqrt{2\mu /\pi }a\right) ,  \label{stabilization}
\end{equation}%
which satisfies the VK criterion (\ref{VK>0}). The consistent stability
analysis for the model with the $\delta $-function regularized as per Eq. (%
\ref{delta}) is developed below in a numerical form.

To further illustrate effects of the regularization of the $\delta $%
-function, it is worthy to briefly consider another functional form of the
smoothing, with scale constant $b$:
\begin{equation}
\delta _{\mathrm{Dirac}}(x)\rightarrow \delta _{\mathrm{sech}}(x)\equiv
\left( \pi b\right) ^{-1}\mathrm{sech}\left( x/b\right) ,  \label{sech}
\end{equation}%
which satisfies the standard normalization condition, $\int_{-\infty
}^{+\infty }\delta _{\mathrm{sech}}(x)dx\equiv 1$. It is easy to see that
Eq. (\ref{1D1}) with the $\delta $-function replaced by expression (\ref%
{sech}) admits two exact solutions for pinned solitons,%
\begin{gather}
U(x,z)=\exp \left( i\mu _{b}z\right) U_{b}\sqrt{\mathrm{sech}\left( \frac{x}{%
b}\right) },  \notag \\
\mu _{b}=\frac{1}{8b^{2}},~U_{b}^{2}=\frac{1}{2\pi \sigma b}\left(
1\pm \sqrt{1-\frac{3}{2}\pi ^{2}\sigma }\right)  \label{b}
\end{gather}%
which exists for $0<\sigma <\tilde{\sigma}_{\max }\equiv 2/\left(
3\pi ^{2}\right) $, cf. Eq. (\ref{max}) [exact solution (\ref{b})
with upper sign is relevant for $\sigma <0$ too]. The total power of
these solutions does not depend on scale constant $b$:%
\begin{equation}
N_{b}^{\left( \pm \right) }=\frac{1}{2\sigma }\left( 1\pm \sqrt{1-\frac{3}{2}%
\pi ^{2}\sigma }\right) .  \label{+-}
\end{equation}

\subsection{The linear-stability analysis}

To study the stability of solitons pinned to the effective delta-functional
nonlinear potential well, we first concentrate on solutions of Eq. (\ref{1D1}%
) with the $\delta $-functions regularized as per Eq. (\ref{delta}).
The stationary solution is taken as $U(x,z)=e^{i\mu z}U(x)$, where
$U(x)$ is assumed real, with propagation constant $\mu >0$. A small
perturbation is added to the solution in the form of
\begin{equation}
\widetilde{U}(x,z)=[U(x)+\delta u(x,z)]e^{i\mu z}  \label{perturb1}
\end{equation}%
with $|\delta u(x,z)|<<|U(x)|$. Substituting this into Eq.
(\ref{1D1}) and linearizing, we derive the evolution equation for
the perturbation,
\begin{equation}
i\left( \delta u\right) _{z}=-\frac{1}{2}\left( \delta u\right)
_{xx}+[3\sigma U^{4}(x)-2\delta(x)U^{2}(x)+\mu ]\delta
u+U^{2}(x)[2\sigma U^{2}(x)-\delta(x)]\delta u^{\ast }.
\label{linearize1}
\end{equation}%
We rewrite this equation in matrix form,
\begin{equation}
i\left(
\begin{array}{l}
\left( \delta u\right) _{z} \\
\left( \delta u^{\ast }\right) _{z}%
\end{array}%
\right) =\left(
\begin{array}{cc}
\widehat{A} & \widehat{B} \\
-\widehat{B} & -\widehat{A}%
\end{array}%
\right) \left(
\begin{array}{l}
\delta u \\
\delta u^{\ast }%
\end{array}%
\right)  \label{linearize2}
\end{equation}%
\begin{equation}
\widehat{A}\equiv -(1/2)\partial _{xx}+3\sigma
U(x)^{4}-2\delta(x)U(x)^{2}+\mu ,
\end{equation}%
\begin{equation}
\widehat{B}\equiv 2\sigma U(x)^{4}-\delta(x)U(x)^{2}.
\end{equation}

Eigenfunctions of perturbation $\delta u(x,z)$ are looked for in the form of%
\begin{equation}
\delta u(x,z)=f(x)e^{\lambda z}+g^{\ast }(x)e^{\lambda ^{\ast }z}
\label{perturb2}
\end{equation}%
where $\lambda $ is a complex stability eigenvalue, and $\{f(x),g(x)\}$ are
respective complex eigenfunctions. The substitution of perturbation (\ref%
{perturb2}) into Eq. (\ref{linearize2}) leads to the following eigenvalue
problem,%
\begin{equation}
i\left(
\begin{array}{cc}
{0} & \widehat{C} \\
\widehat{D} & {0}%
\end{array}%
\right) \left(
\begin{array}{l}
\eta \\
\chi%
\end{array}%
\right) =\lambda \left(
\begin{array}{l}
\chi \\
\eta%
\end{array}%
\right) .  \label{eigen1}
\end{equation}%
Here $\eta \equiv f(x)+g(x),\,\chi \equiv f(x)-g(x)$, and $\widehat{C}\equiv
\widehat{A}+\widehat{B},\,\widehat{D}=\widehat{A}-\widehat{B}$. Equation (%
\ref{eigen1}) was solved numerically. Obviously, the soliton is
unstable if there is at least one eigenvalue with $\lambda _{r}>0$.
Alternatively, the stability of the soliton can be tested in direct
simulations, in which the
perturbations are applied to initial conditions for Eqs. (\ref{1D1}) and (%
\ref{1D2}). In the next section, we use both methods.

\subsection{The double-delta structure}

Next, we consider the model with two identical delta-functions,
\begin{equation}
iU_{z}=-\frac{1}{2}U_{xx}-\left[ \delta (x-1)+\delta \left( x+1\right) %
\right] |U|^{2}U+\sigma |U|^{4}U,  \label{1D2}
\end{equation}%
where coefficient $\sigma \geq 0$ remains irreducible, while the
distance between the $\delta $-functions may be set equal to $2$ by
means of rescaling (\ref{scaling}). Exact analytical solutions to
Eq. (\ref{1D2}) with $\sigma =0$ were found for symmetric,
antisymmetric, and asymmetric solitons pinned to the two $\delta
$-functions \cite{Dong}, the asymmetry parameter being defined as
\begin{equation}
\nu =N^{-1}\left[ \int_{0}^{+\infty }\left\vert U(x)\right\vert
^{2}dx-\int_{-\infty }^{0}\left\vert U(x)\right\vert ^{2}dx\right] .
\label{niu}
\end{equation}

In the case of $\sigma =0$, the symmetric solitons are completely stable,
while the antisymmetric and asymmetric ones are completely unstable.
Stabilization of the asymmetric and antisymmetric solitons is possible, as
shown in Ref. \cite{Dong}, by replacing each ideal $\delta $-functions with
its regularized version, as per Eq. (\ref{delta}).

\subsection{Rescaling of the parameters}

The rescaling transformation (\ref{scaling}) applies as well to Eqs. (\ref%
{1D1}) and (\ref{1D2}) in which the $\delta $-function is replaced by its
regularized version (\ref{delta}). We use this degree of freedom to fix the
regularization spatial scale as $a\equiv 1$, thus replacing Eq. (\ref{1D1})
by
\begin{equation}
iU_{z}=-\frac{1}{2}U_{xx}-\frac{e^{-x^{2}}}{\sqrt{\pi }}|U|^{2}U+\sigma
|U|^{4}U.  \label{rescale1w}
\end{equation}

In the model with the double $\delta $-function, we again fix $a=1$, hence
the distance between the two attraction centers is no longer $2$ [see Eq. (%
\ref{1D2})], but becomes an independent parameter, $2x_{0}$. Thus, the
regularized version of Eq. (\ref{1D2}) is
\begin{equation}
iU_{z}=-\frac{1}{2}U_{xx}-\frac{1}{\sqrt{\pi }}\left[
e^{-(x-x_{0})^{2}}+e^{-(x+x_{0})^{2}}\right] |U|^{2}U+\sigma |U|^{4}U.
\label{rescale2w}
\end{equation}%
It is easy to check that the function multiplying the SF cubic term in Eq. (%
\ref{rescale2w}) keeps the double-well structure at%
\begin{equation}
x_{0}>\left( x_{0}\right) _{\min }=1/\sqrt{2}.  \label{min}
\end{equation}

\section{Numerical and analytical results for the one-dimensional systems}

In our numerical calculations, we chiefly used the
imaginary-time-propagation method for finding stationary solutions, and the
split-step Fourier algorithm for simulations of their perturbed evolution.
To obtain antisymmetric structures, we have additionally used the Newton
conjugate gradient method \cite{Yang1,Yang2} . The linear eigenvalue problem
represented by Eq. (\ref{eigen1}) was solved by means of the Fourier
collocation method.

\subsection{The single-well setting: numerical results}

\begin{figure}[tbp]
\includegraphics[width=10cm]{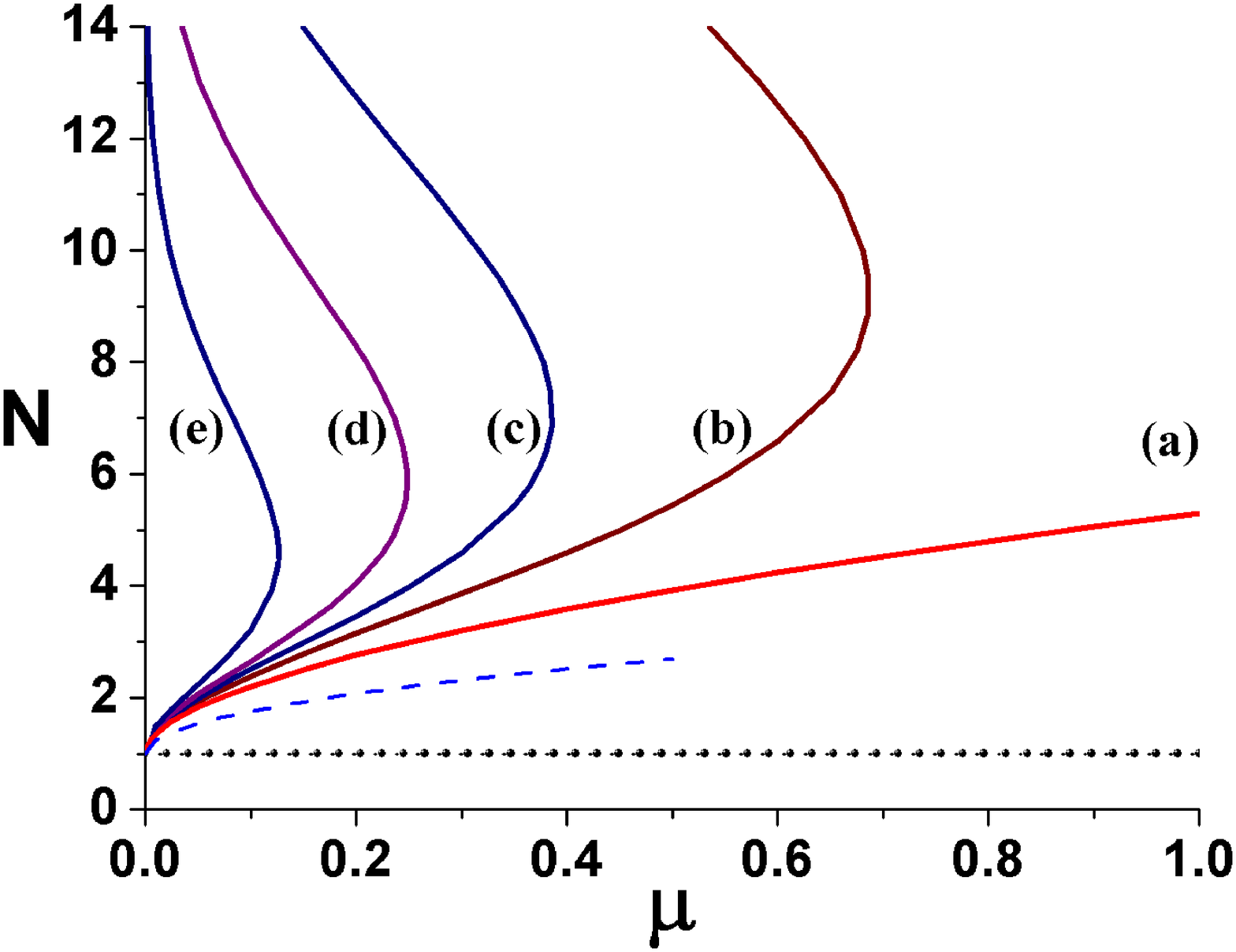}
\caption{(Color online) Families of numerically found stationary
soliton solutions to Eq. (\protect\ref{rescale1w}) for the 1D
single-well configuration. The dependence of the total power on the
propagation constant, $N(\protect\mu )$, is shown for several values
of the quintic
coefficient: (a) $\protect\sigma =0$, (b) $\protect\sigma =0.05$, (c) $%
\protect\sigma =0.075$, (d) $\protect\sigma =0.1$ and (e) $\protect\sigma %
=0.15$. All these curves represent \textit{stable solitons}. The dashed
curve shows analytical approximation (\protect\ref{stabilization}) for $%
\protect\sigma =0.$ The horizontal dotted line corresponds to $N(\protect\mu %
)\equiv 1$, $\protect\sigma =0$. The latter family, given by Eqs. (\protect
\ref{exact}) and (\protect\ref{xi}), is unstable. Qualitatively, the shape
of the $N(\protect\mu )$ curves in this figure is explained by the
Thomas-Fermi approximation based on Eqs. (\protect\ref{TF})-(\protect\ref%
{1/sigma}).}
\label{dia1}
\end{figure}

First, we address soliton solutions generated by Eq. (\ref{1D1}) with the $%
\delta $-function regularized as per Eq. (\ref{delta}), expecting the
stabilization of these states -- at least, for small $\mu $, according to
the analytical result (\ref{stabilization}), which satisfies the VK
criterion. The setting is fully characterized by two independent parameters,
$\sigma $ and $N$. The summary of the results is presented in Fig. \ref{dia1}%
.

\begin{figure}[tbp]
\includegraphics[width=11cm]{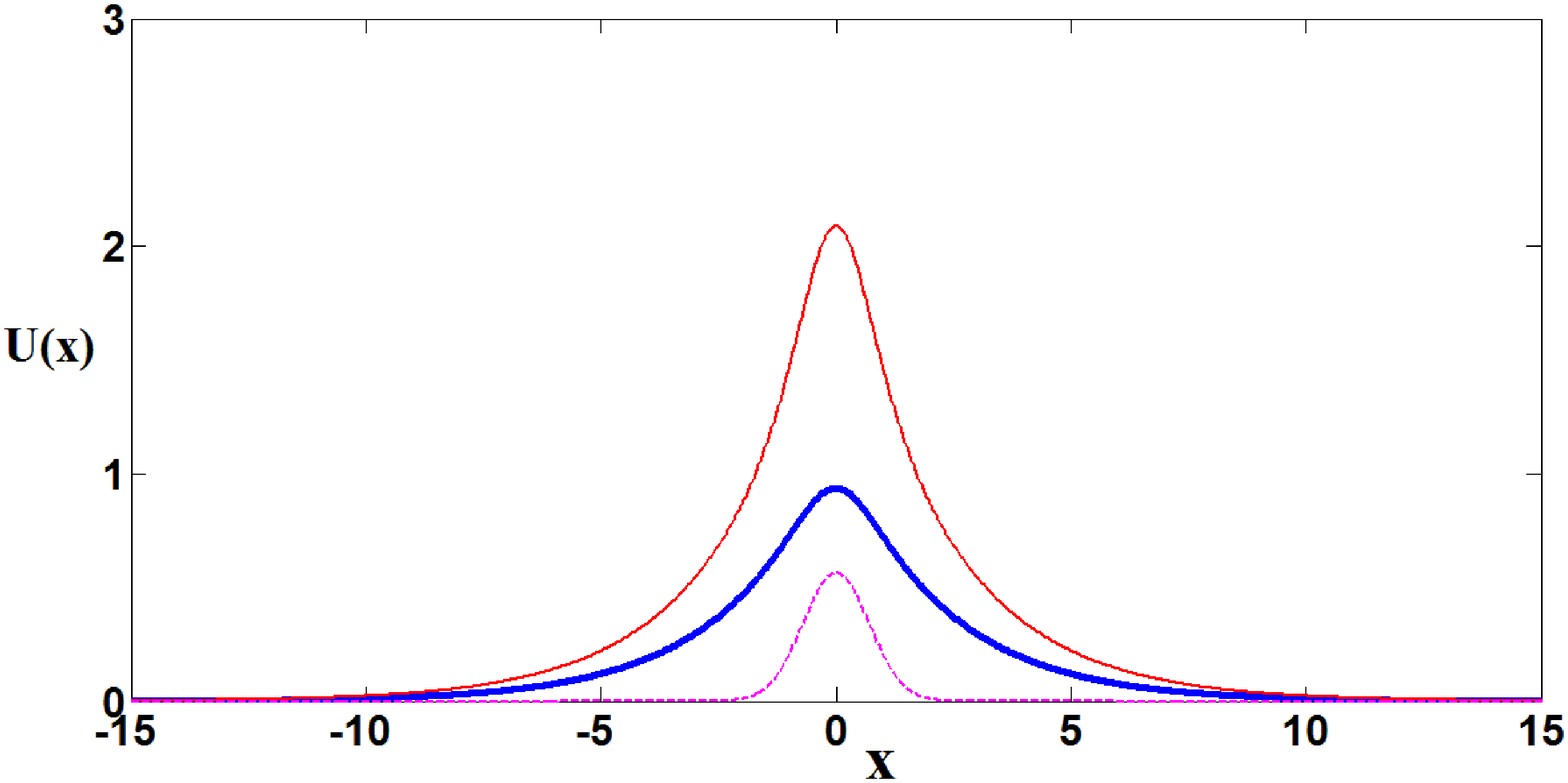}
\caption{(Color online) The thin upper and bold intermediate curves show an
example of a pair of stable 1D solitons pinned to the single regularized $%
\protect\delta $-function, in the model with $\protect\sigma =0.1$.
The dashed bottom curve depicts profile (\protect\ref{delta}) of the
regularized $\protect\delta $-function. These solitons pertain to a
common value of the propagation constant, $\protect\mu =0.1$, while
their total powers (norms) are $N_1=11.16$ and $N_2=2.66$,
respectively.} \label{example}
\end{figure}

A salient feature of Fig. \ref{dia1} is the presence of upper and
lower branches in dependences $N(\mu )$, connected at the rightmost
turning point. To highlight the difference between solitons
pertaining to the same value of $\mu $ but lying on the upper and
lower branches, in Fig. \ref{example} we display profiles of such a
pair of solitons with strongly differing values of the total power.

The numerical analysis, including both the computation of eigenvalues for
small perturbations on the basis of Eq. (\ref{linearize2}) and direct
simulations, demonstrates that \emph{all} the 1D solitons trapped in the
single-well nonlinear potential are indeed \emph{stable} if the Gaussian
regularization (\ref{delta})\ is used instead of the ideal $\delta $%
-function. In particular, the stability of the solution branch with $\sigma
=0$ in Fig. \ref{dia1}, which is produced by the SF-only nonlinearity,
complies with the evident fact that this branch satisfies the VK criterion.
As concerns the branches for $\sigma >0$, their stability for relatively
small values of $N$, beneath the turning points, where the SF term
dominates, may also be explained by the VK criterion. On the other hand,
above the turning points, where, due to large values of $N$, the SDF term is
the dominant one, the stability agrees with the anti-VK criterion, $dN/d\mu
<0$, which applies to solitons supported by the SDF nonlinearity \cite{anti}%
. In fact, such an effective switch between the VK and anti-VK criteria at a
turning point, which secures the stability of the entire soliton branch,
occurs in other systems featuring the competition between SF and SDF terms
\cite{Yang3}.

\begin{figure}[h]
\begin{center}
\includegraphics[width=8cm]{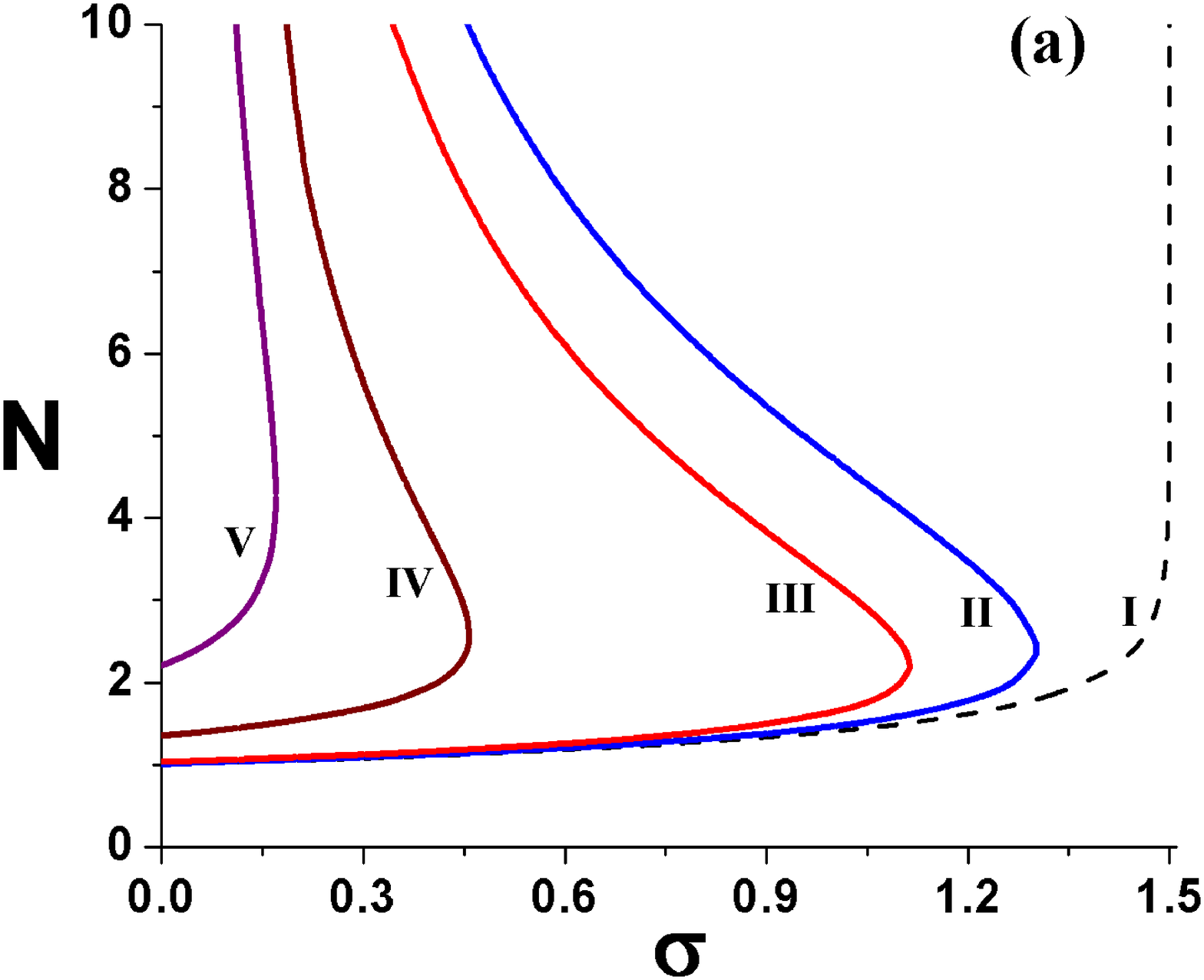} %
\includegraphics[width=8cm]{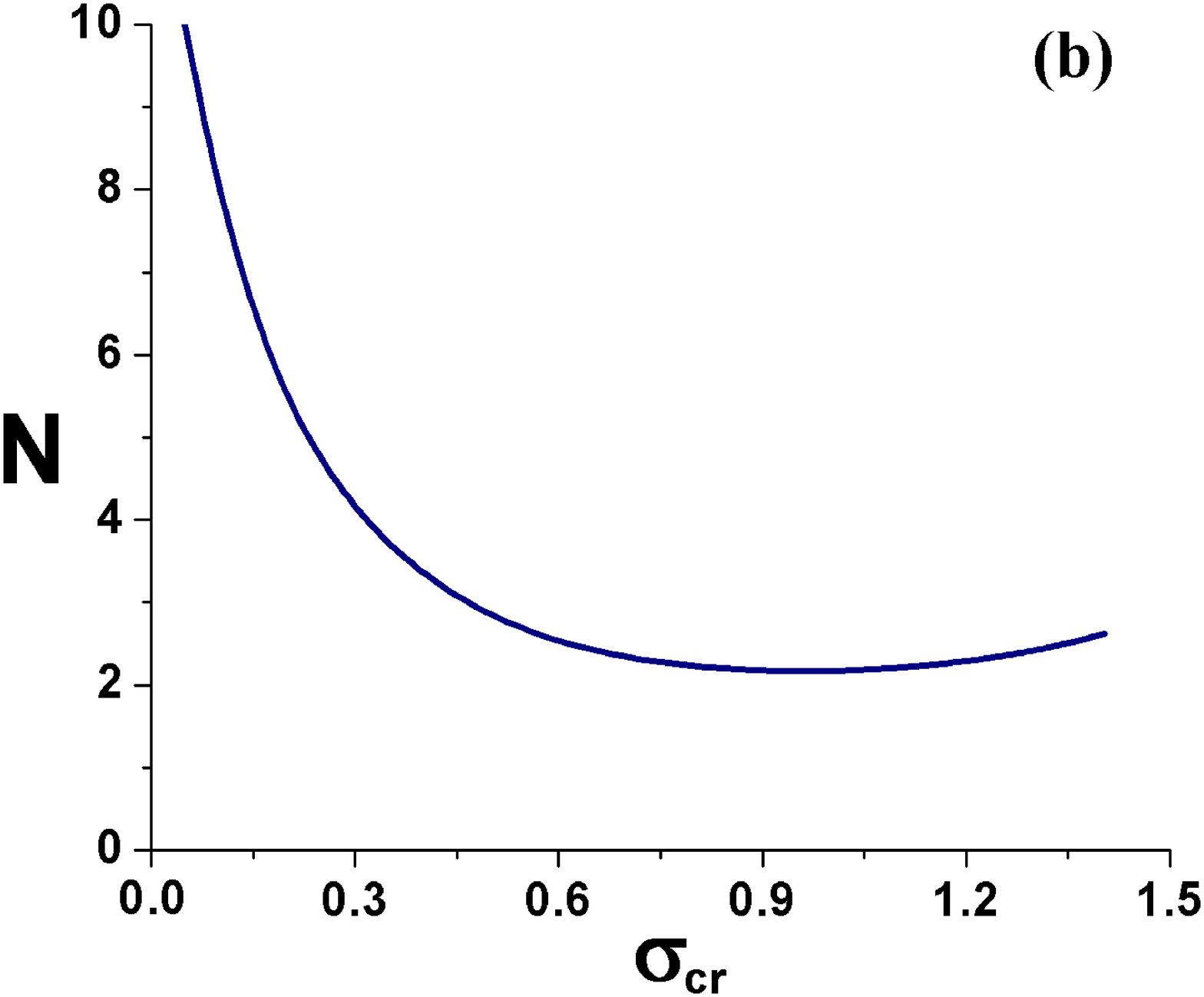}
\end{center}
\caption{(Color online) (a) Dependences $N(\protect\sigma )$ for soliton
families produced by equation (\protect\ref{rescale1w}) with $\protect\mu %
=10^{-1}$ (curve V), $\protect\mu =10^{-2}$ (curve IV), $\protect\mu %
=10^{-4} $ (curve III), and $\protect\mu =10^{-5}$ (curve II). The dashed
curve I represents the exact result (\protect\ref{N}) obtained with the
ideal $\protect\delta $-function (in this case, $N$ does not depend $\protect%
\mu $). (b) Dependence $N(\protect\sigma _{\mathrm{cr}})$, where $\protect%
\sigma _{\mathrm{cr}}$ is the largest value of the quintic coefficient, $%
\protect\sigma $, at which the solitons exist in (a).}
\label{smallmiu}
\end{figure}

Dependences $N(\sigma )$ for different values of $\mu $ are displayed in
Fig. \ref{smallmiu}(a), and compared to the dependence given by Eq. (\ref{N}%
), which was obtained in the exact form for the ideal $\delta $-function,
and which does not depend on $\mu $. As seen in the figure, the largest
value of the quintic SDF coefficient, $\sigma _{\mathrm{cr}}$, up to which
the solitons exist in the system, is close to absolute maximum, $\sigma
_{\max }=3/2$, given by Eq. (\ref{max}), for small $\mu $, at which there
are broad solitons similar to those supported by the ideal $\delta $%
-function (but are stable, on the contrary to that case). With the increase
of $\mu $, the solitons become narrower, and $\sigma _{\mathrm{cr}}$
decreases, so that there appears the dependence between $N$ and $\sigma _{%
\mathrm{cr}}$ displayed in Fig. \ref{smallmiu}(b).

\subsection{The single-well setting: analytical results}

The shape of the $N(\sigma )$ curves in Fig. \ref{smallmiu}(a), i.e., the
fact that, at small $\sigma ,$ one has two solitons corresponding to a
common value of $\mu $, one with the total power $N$ close to $1$, and
another one with apparently larger $N$, can be explained by the particular
exact solutions given by Eqs. (\ref{sech})-(\ref{+-}). Indeed, both
solutions pertain to the common propagation constant, $\mu _{b}=1/\left(
8b^{2}\right) $, and, at small $\sigma $, one has a moderate value of the
power, $N_{b}^{(-)}\approx 3\pi ^{2}/8$, while the other features a large
power,
\begin{equation}
N_{b}^{(+)}\approx 1/\sigma .  \label{N+}
\end{equation}

Another qualitative explanation for the existence of the upper and lower
solution branches in Fig. \ref{smallmiu}(a) is offered by the Thomas-Fermi
approximation (TFA), which is an efficient method for describing the shape
of solitons supported by the SDF nonlinearity \cite{TFA1,TFA2}. In its
simplest form, the TFA neglects the second derivative in Eq. (\ref{rescale1w}%
). For given $\mu >0$, this approximation yields a \emph{pair} of localized
solutions, which may be interpreted as those representing the upper and
lower branches in Fig. \ref{smallmiu}(a):%
\begin{equation}
U\left( x,z\right) =\frac{1}{\sqrt{2\sqrt{\pi }\sigma }}e^{i\mu z}\left\{
\begin{array}{c}
\sqrt{\exp \left( -x^{2}\right) \pm \sqrt{\exp \left( -2x^{2}\right) -4\pi
\sigma \mu }},~\mathrm{at}~\ |x|~\leq x_{0}\equiv \sqrt{-\frac{1}{2}\ln
\left( 4\pi \sigma \mu \right) }, \\
\exp \left[ -x_{0}^{2}/2-\sqrt{2\mu }\left( |x|-x_{0}\right) \right] ,~%
\mathrm{at}~\ |x|~>x_{0}.%
\end{array}%
\right.   \label{TF}
\end{equation}%
The second line in Eq. (\ref{TF}) represents spatially decaying tails,
attached to the TFA-predicted core part of the soliton. The tails are
produced by the linearized version of Eq. (\ref{rescale1w}), where the
second derivative is kept, cf. a similar combined approximation developed
for gap solitons in Ref. \cite{TFA2}. The TFA given by Eq. (\ref{TF}) is
valid under the condition that the expression under the square root is
positive at $x=0,$ i.e., at
\begin{equation}
\mu <\mu _{\max }^{\mathrm{(TFA)}}\equiv \left( 4\pi \sigma \right) ^{-1}.
\label{TF-max}
\end{equation}%
The latter condition offers a qualitative explanation to the existence of
the turning points in Fig. \ref{smallmiu}(a). Further, the TFA makes it
possible to find the asymptotic value of the total power\ corresponding to
the upper branches of $N(\mu )$ in the figure:%
\begin{equation}
\lim_{\mu \rightarrow 0}\left\{ N\left( \mu ;\sigma \right) \right\} _{%
\mathrm{TFA}}=1/\sigma ,  \label{1/sigma}
\end{equation}%
which, incidentally, agrees with Eq. (\ref{N+}). An essential corollary of
Eq. (\ref{1/sigma}) is the fact that $N(\mu )$ remains finite (does not
diverge) at $\mu \rightarrow 0$ along the upper branches in Fig. \ref%
{smallmiu}(a).

\subsection{The double-well setting}

The double-well configuration based on Eq. (\ref{rescale2w}) is controlled
by three parameters, $x_{0}$, $\sigma $ and $N$. In the limit of $\sigma
\rightarrow 0$ (no SDF quintic term), this system turns into the one
considered in Ref. \cite{Dong} (a similar system with a two-component field,
which demonstrates an extremely complex picture of transitions between
symmetric, antisymmetric, and antisymmetric states, was considered in Ref.
\cite{Yasha}). Therefore, we started the numerical analysis of the
double-well setting with small values of $\sigma $, aiming to produce new
results at larger $\sigma $.

In the simulations, symmetric and asymmetric states, with the symmetry
defined with respect to the two identical nonlinear-potential wells, were
generated using the initial guess
\begin{equation}
\left\{ u(x)\right\} _{\mathrm{sym,asym}}^{\mathrm{(in)}}=P~\mathrm{sech}%
(x+x_{0})+Q\text{ }\mathrm{sech}(x-x_{0}),  \label{initial1}
\end{equation}%
with constants $P=Q$ and $P\neq Q$ for the symmetric and asymmetric
ones. Antisymmetric states were created starting from the input
\begin{equation}
\left\{ u(x)\right\} _{\mathrm{antisym}}^{\mathrm{(in)}}=P~\mathrm{sech}%
(qx)\sin \left( kx\right) ,  \label{initial2}
\end{equation}%
with some constants $q$ and $k$.

\begin{figure}[h]
\begin{center}
\includegraphics[width=8cm]{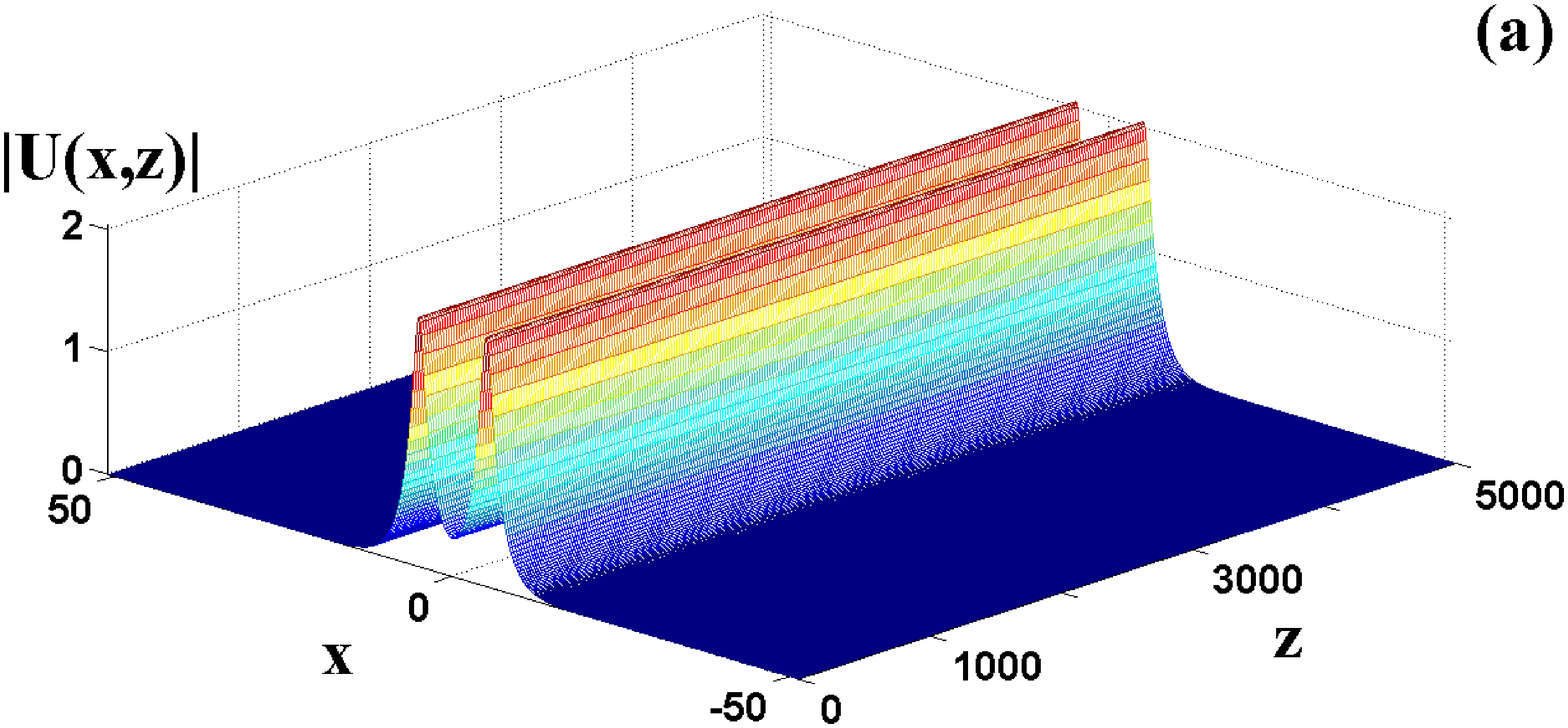}
\includegraphics[width=8cm]{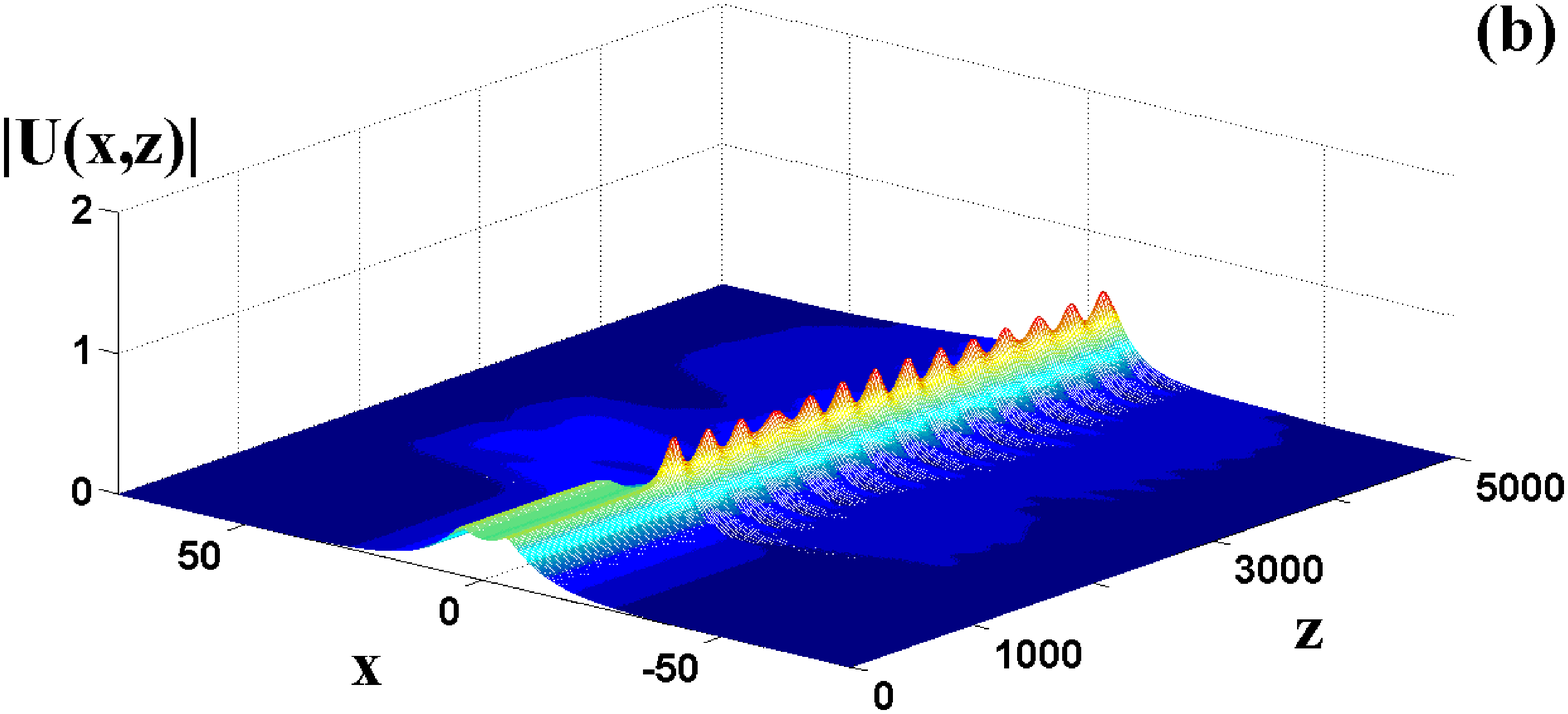}
\end{center}
\caption{(Color online) Examples of the evolution of stable and unstable
symmetric solitons in the 1D double-well model, for $x_{0}=5$, $\protect%
\sigma =0.1$, $\protect\mu =0.14$, and $x_{0}=5$, $\protect\sigma =0.05$, $%
\protect\mu =0.008$, respectively. } \label{sym1}
\end{figure}

\begin{figure}[h]
\begin{center}
\includegraphics[width=8cm]{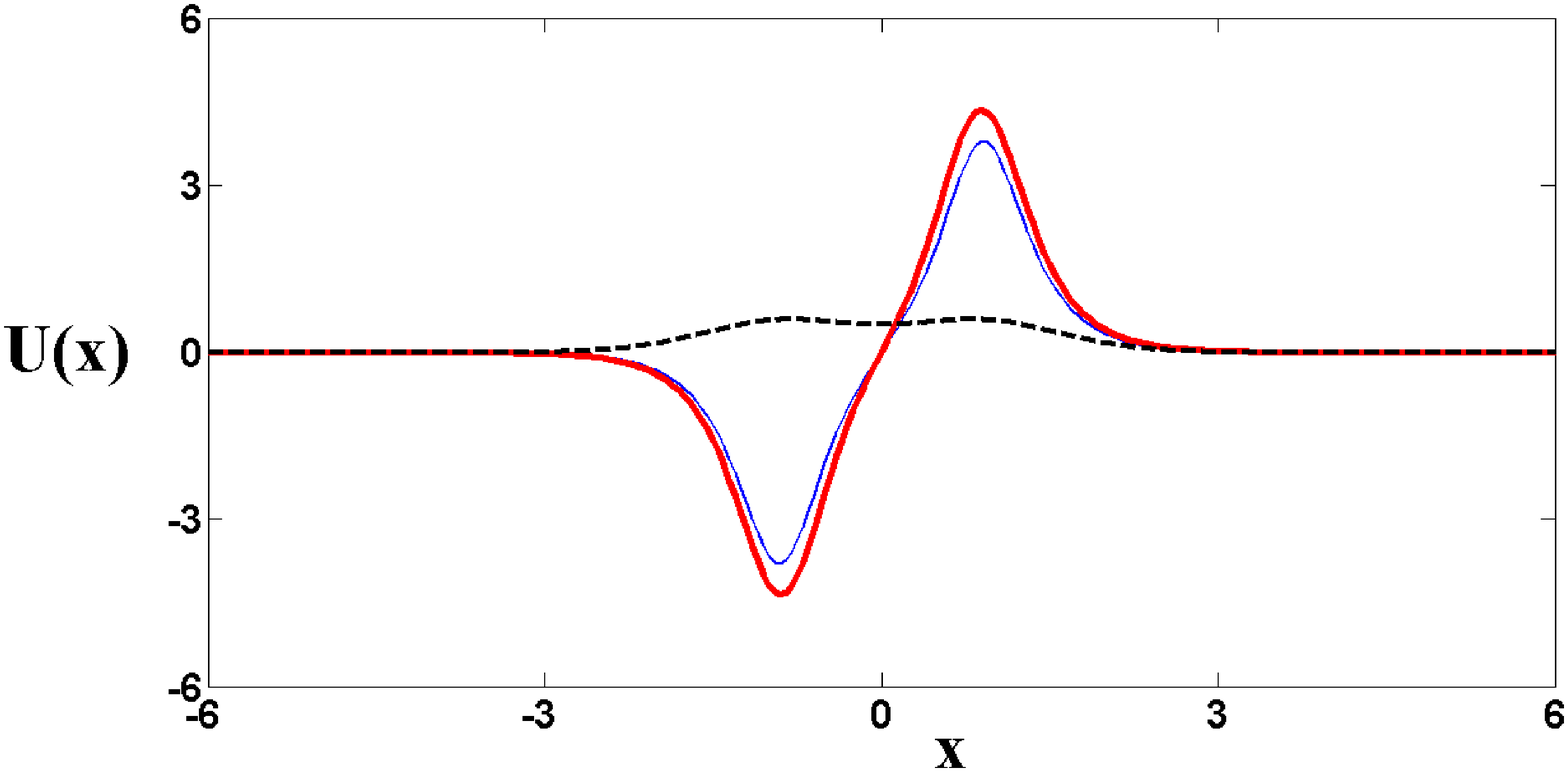}\\[0pt]
\includegraphics[width=8cm]{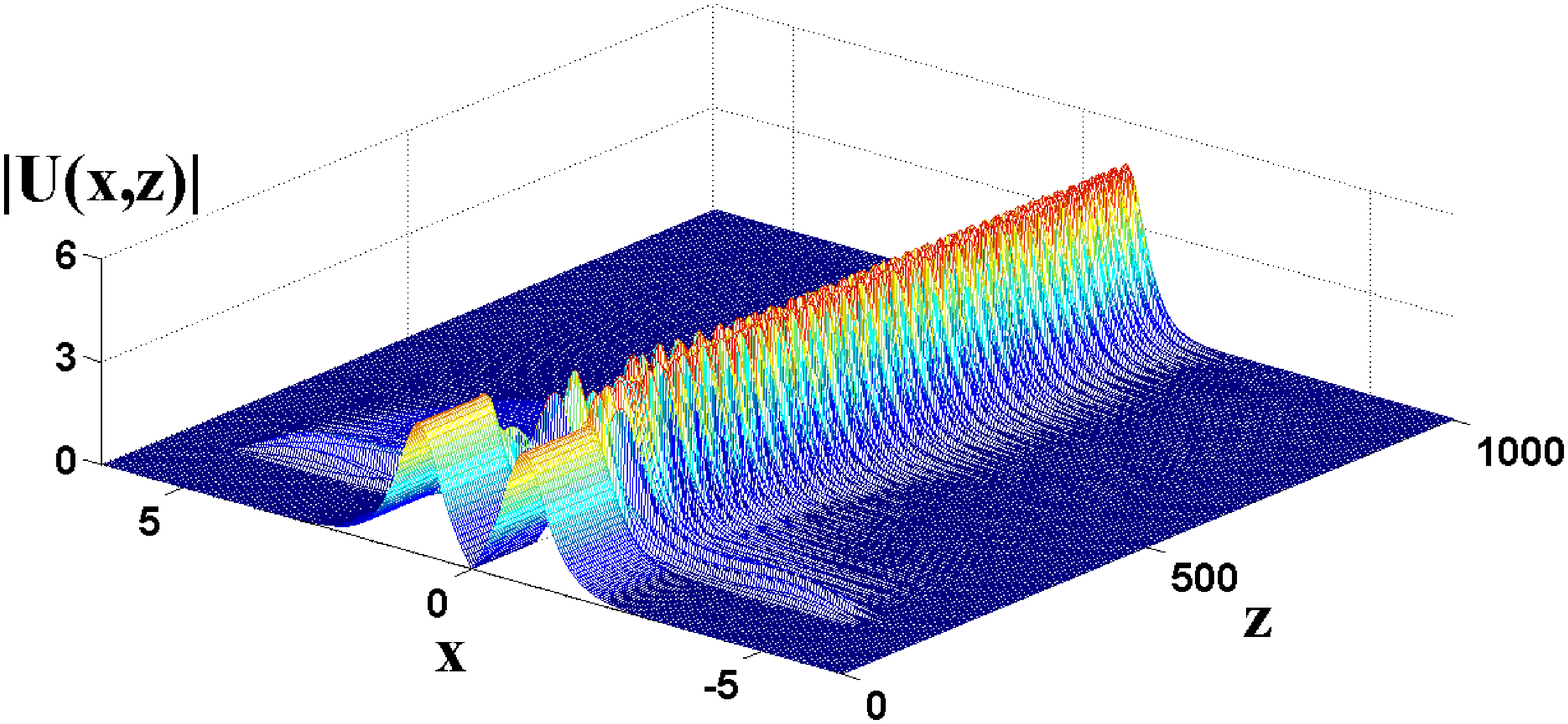} %
\includegraphics[width=8cm]{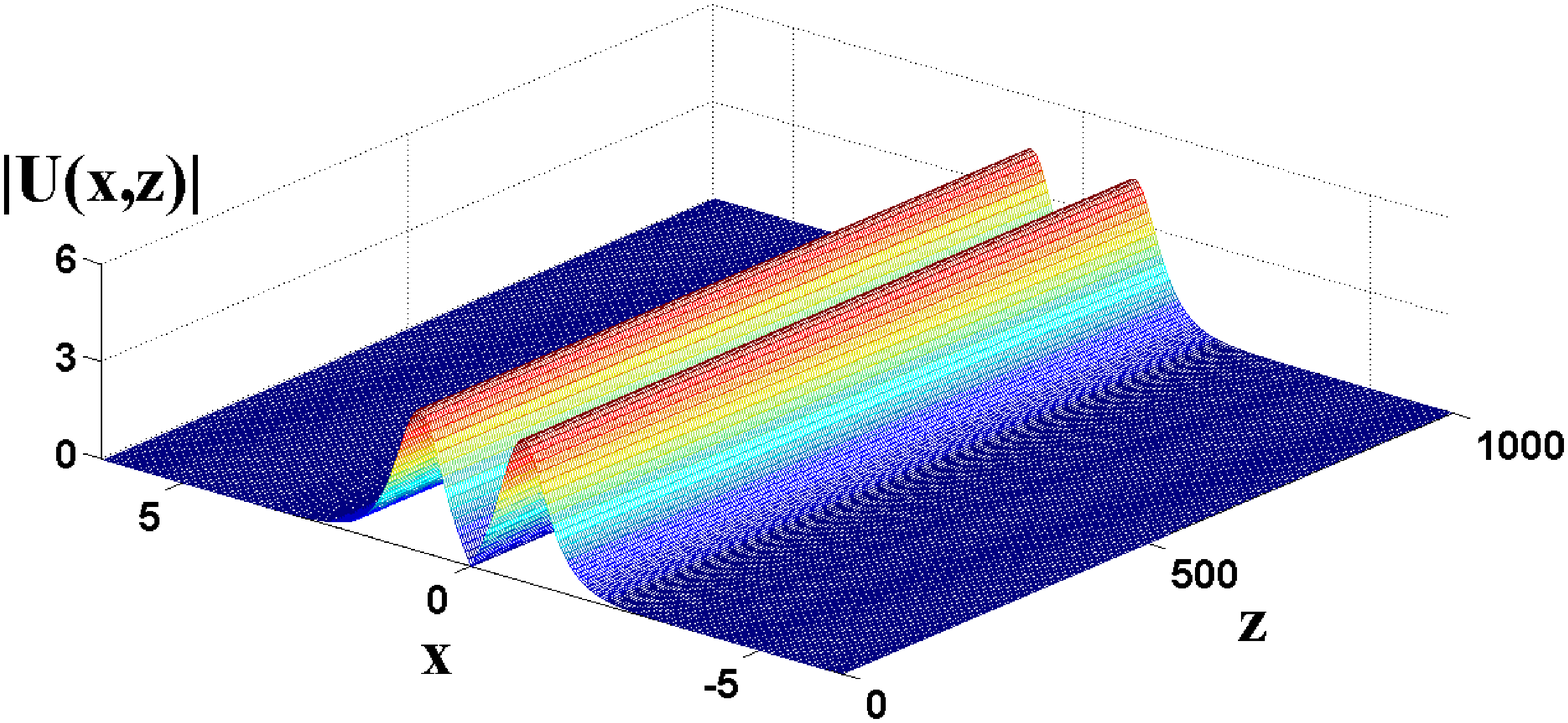}
\end{center}
\caption{(Color online) A stable antisymmetric soliton with norm
$N=26.394$
(the bold curve in the top frame, and the right frame below) produced by Eq. (%
\protect\ref{1D2}) with $\protect\sigma =0.01$, compared to an
unstable antisymmetric one with $N=18.816$ (the thin curve in the
top frame, and the
left frame below) obtained, as in Ref. \protect\cite{Dong}, for $\protect%
\sigma =0$. The other parameters are: $x_{0}=0.9$ (the dashed curve
in the top frame depicts profile of the cubic nonlinear
regularization in the equation (\protect\ref{rescale2w})),
$\protect\mu =3.8688$ [cf. Fig. 8(b) in Ref. \protect\cite{Dong}].}
\label{antisym1}
\end{figure}

\begin{figure}[h]
\includegraphics[width=8cm]{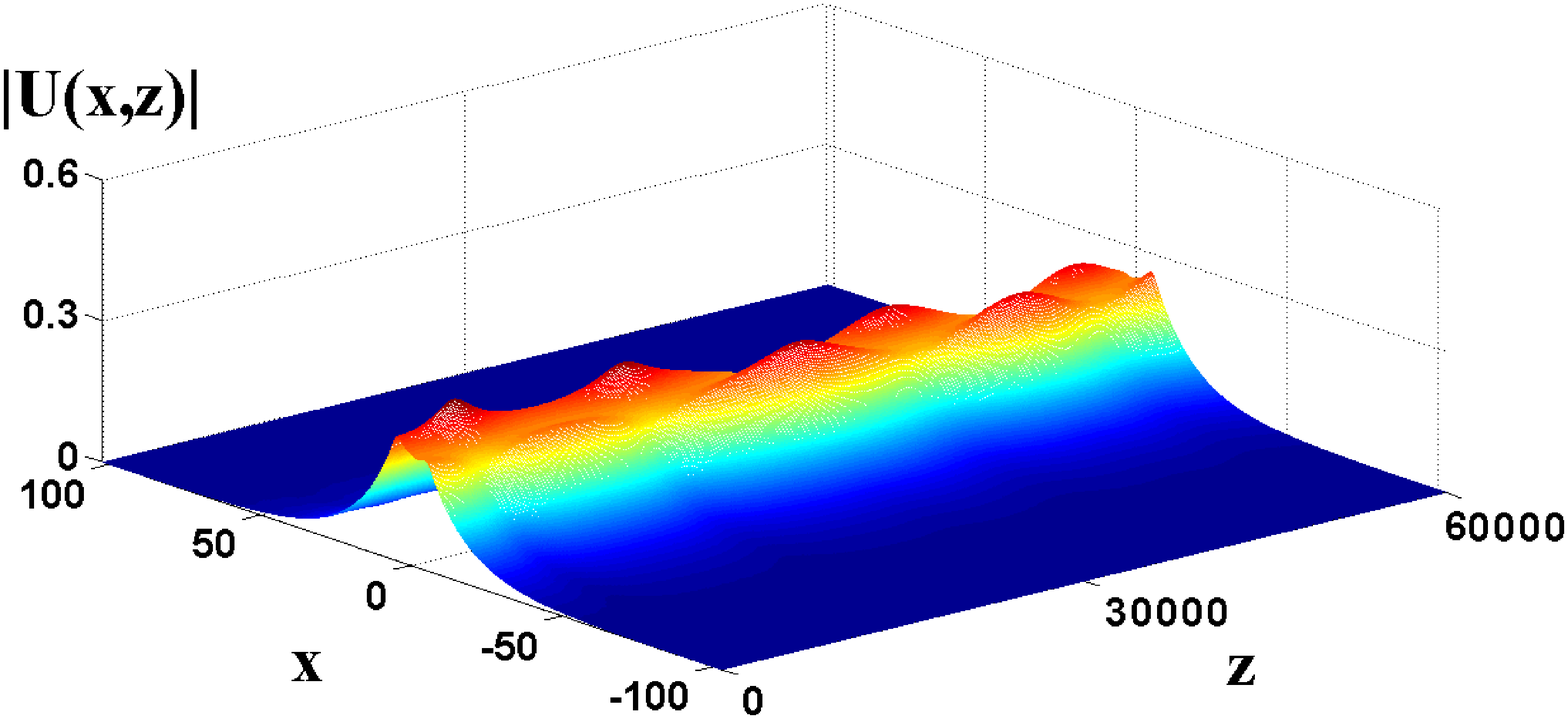} %
\includegraphics[width=8cm]{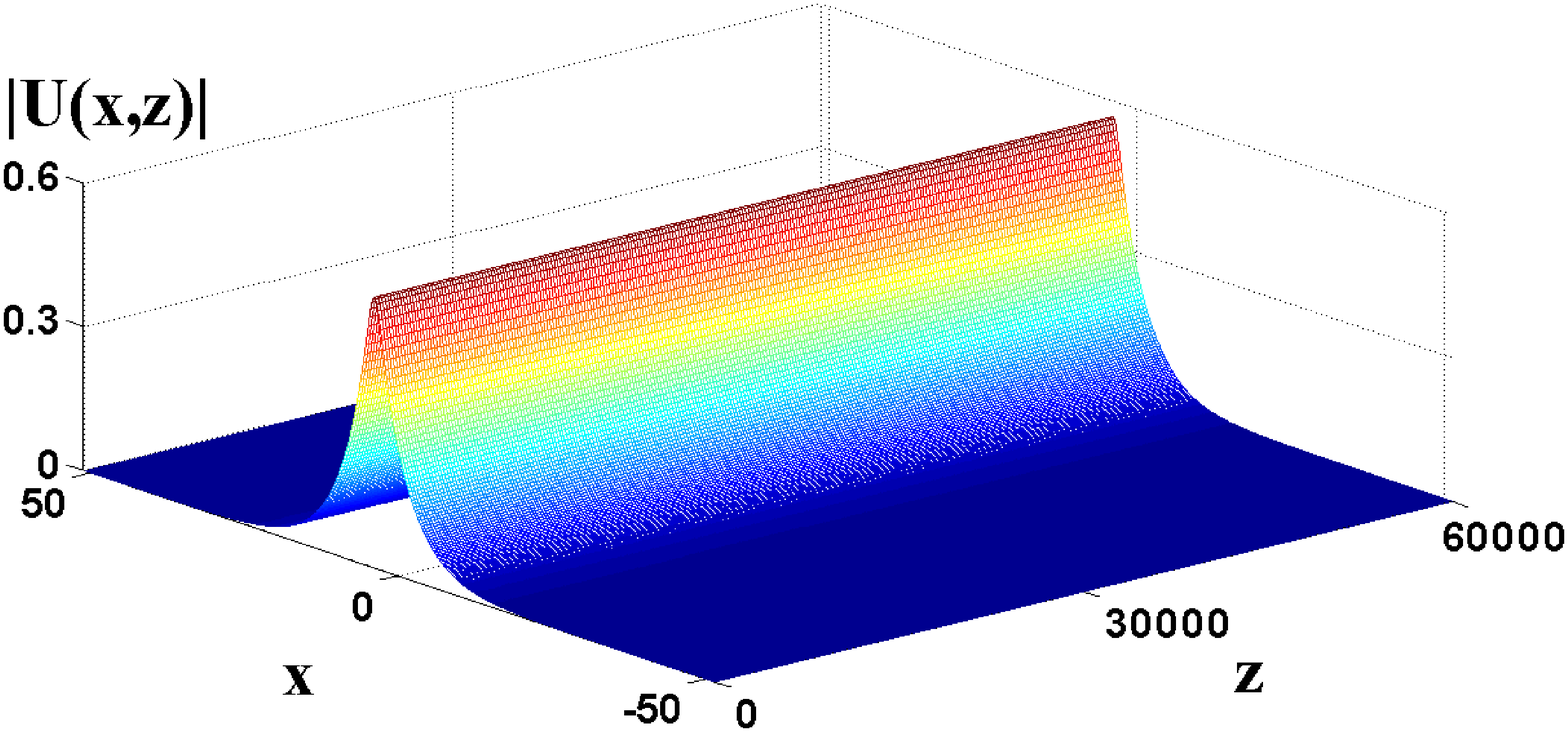}
\caption{(Color online) Evolution of unstable ($x_{0}=5$, $\protect\sigma %
=0.2$, $\protect\mu =0.003$) and stable ($x_{0}=4$, $\protect\sigma =0.1$%
, $\protect\mu =0.025$) asymmetric solitons in the 1D double-well
model.} \label{asym1}
\end{figure}

We start by presenting, in Fig. \ref{sym1}, examples of stable unstable
symmetric solitons, As shown in this figure, unstable symmetric solitons
spontaneously transform themselves into asymmetric breathers trapped in a
single well.
Further, examples of stable and unstable antisymmetric solitons are
displayed in Fig. \ref{antisym1}. It is seen that even a week SDF
term (with $\sigma =0.01$) is able to stabilize the antisymmetric
soliton, while an unstable one spontaneously transforms into a
nearly stationary strongly asymmetric state, trapped in a single
well. Asymmetric solitons may also be stable or unstable, as shown
in Fig. \ref{asym1}, the unstable one transforming into a breather
which stays trapped in the original nonlinear-potential well.
\ \

\begin{figure}[tbp]
\begin{center}
\includegraphics[width=9cm]{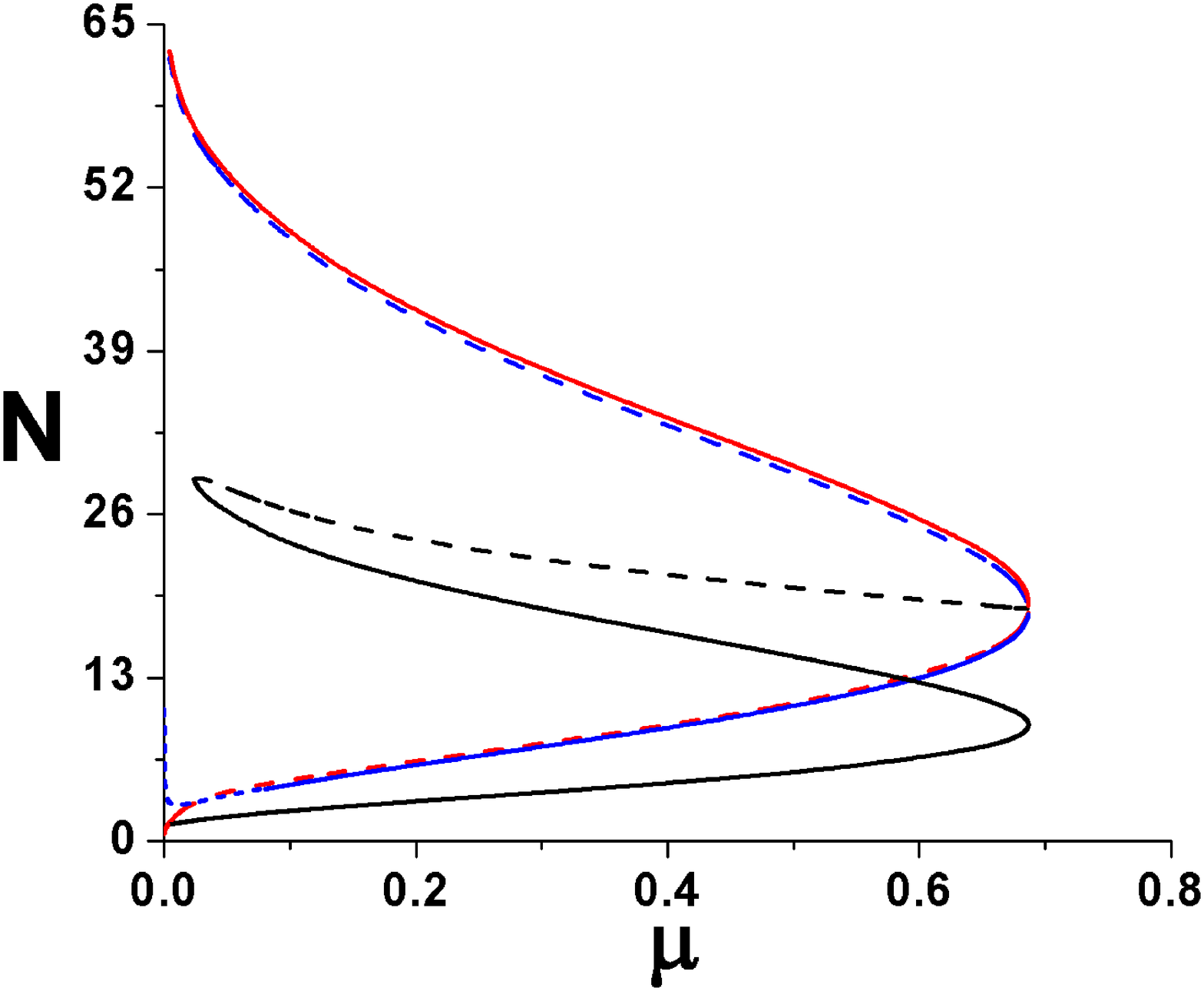}
\includegraphics[width=9cm]{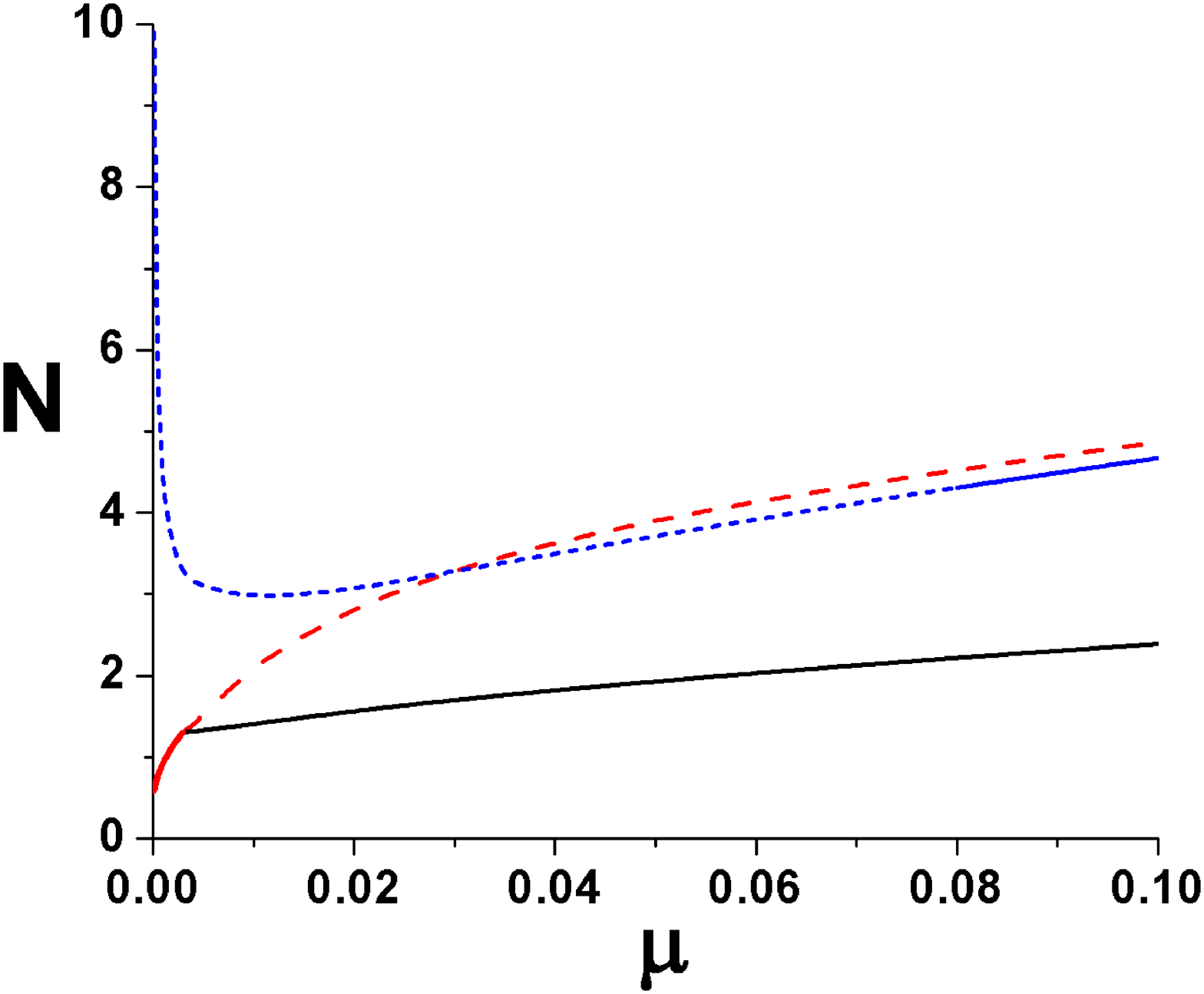}
\end{center}
\caption{(Color online) The bifurcation diagram for the 1D
double-well
system, shown in terms of $N(\protect\mu )$ curves, for $x_{0}=5$ and $%
\protect\sigma =0.05$ (the left panel), and the closeup of the
symmetry-breaking part for smaller values of $\protect\mu $ (the
right panel). Here and in the next figure, red, blue, and black
curves represent symmetric, antisymmetric, and asymmetric modes,
respectively, while solid and dashed segments of the curves refer to
stable and unstable solitons.} \label{dia2b}
\end{figure}

Families of stable and unstable symmetric, antisymmetric, and
asymmetric solitons are linked into a rather complex bifurcation
diagram, which is presented in Fig. \ref{dia2b}. The lower panel of
the figure demonstrates that the asymmetric soliton family branches
out from the symmetric one at a very small value of the total power,
$N\approx 1.3$, and then it disappears at $N\approx 19$. merging
into a common turning point of the symmetric and antisymmetric
branches. Further, it is worthy to note that the stability of the
symmetric and antisymmetric families obeys the anti-VK and VK
criteria, respectively, while the asymmetric branches may feature
either type of the stability.

\begin{figure}[tbp]
\includegraphics[width=12cm]{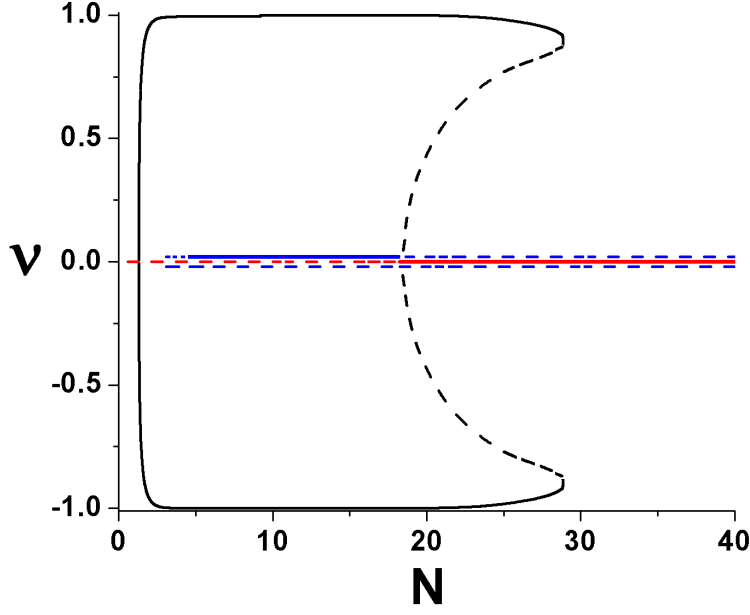}
\caption{(Color online) The asymmetry parameter (\protect\ref{niu})
vs. total power $N$ for the same set of soliton branches in the 1D
double-well system with $x_{0}=5$ and $\protect\sigma =0.05$ which
is shown in Fig. \protect\ref{dia2b}.} \label{dia2c}
\end{figure}

Another adequate picture of the set of soliton families in the double-well
system is provided by the symmetry-breaking bifurcation diagram which is
displayed in Fig. \ref{dia2c}. A characteristic feature of this picture is
the loop connecting the symmetry-breaking and symmetry-restoring
bifurcations. A similar feature was earlier found in dual-core systems
carrying spatially uniform competing SF cubic and SDF quintic nonlinearities
\cite{Albuch}. Note that the loop shown in Fig. \ref{dia2c} has a concave
shape. For a stronger coupling between the two wells, which corresponds to
smaller values of $x_{0}$, it is expected that the loop will shrink and
acquire a convex form, cf. the loops in the 2D model displayed below in
Figs. \ref{dia5a} and \ref{dia4a}. Eventually, the loop will disappear at
still smaller $x_{0}$ \cite{Albuch}.

\begin{figure}[tbp]
\includegraphics[width=8.5cm]{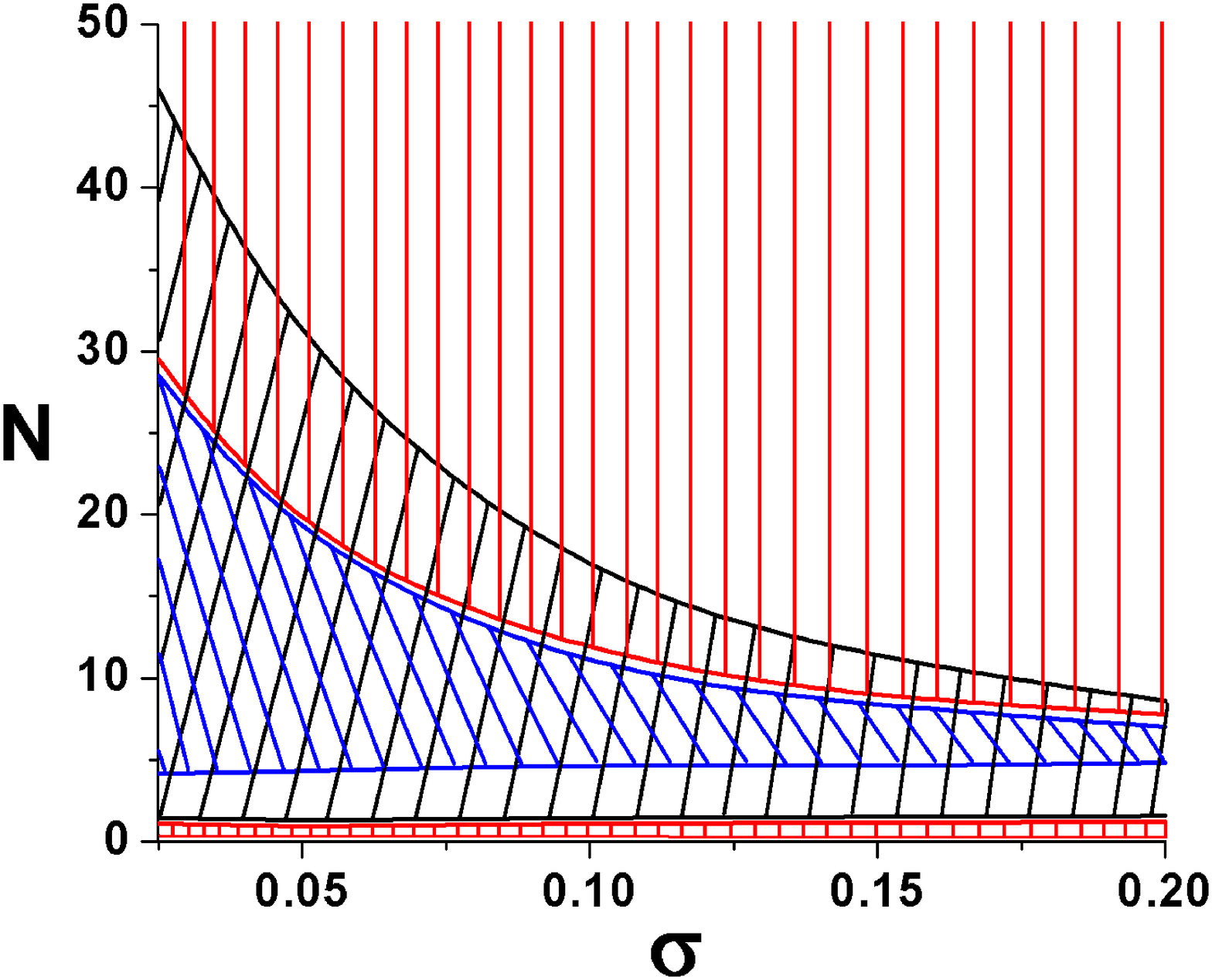}
\includegraphics[width=8.5cm]{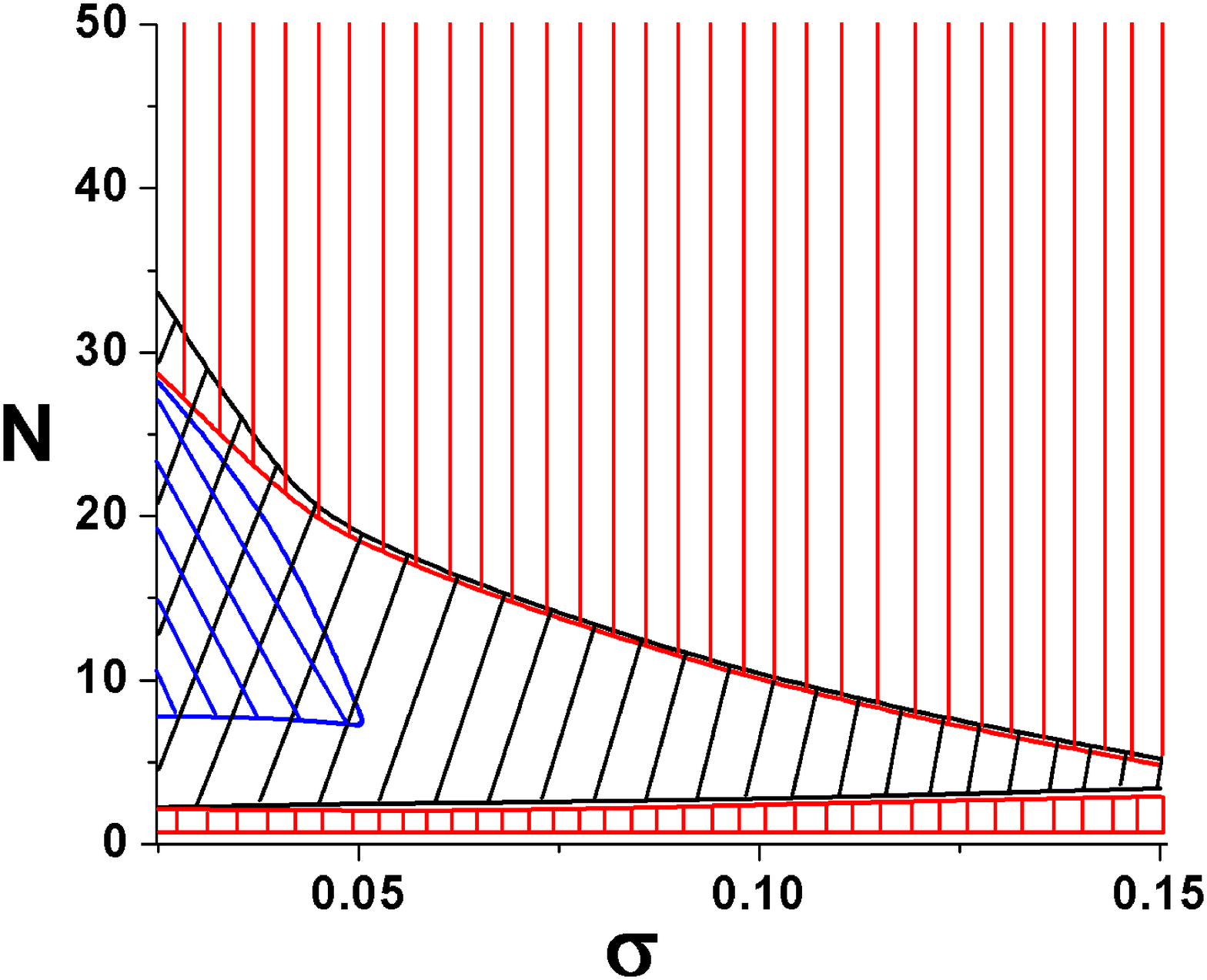}
\caption{(Color online) Stability areas for symmetric, antisymmetric, and
asymmetric solitons (red (vertical), blue (leftward), and black (rightward), respectively) in the plane of $%
\left( \protect\sigma ,N\right) $ in the 1D double-well system based on Eq. (%
\protect\ref{1D2}). The half-distance between the wells is, severally, $%
x_{0}=5$ and $1.25$ in the left and right panels.} \label{dia2d}
\end{figure}

The results are finally summarized in Fig. \ref{dia2d}, which displays
stability domains for the symmetric, antisymmetric, and asymmetric solitons,
in the plane of the quintic SDF coefficient, $\sigma $, and the total norm, $%
N$. The figure demonstrates that the stability areas for the antisymmetric
and asymmetric model shrink with the increase of the SDF coefficient, $%
\sigma $, as well as with the decrease of the distance ($2x_{0}$) between
the two wells. The latter trend is quite natural, as at $x_{0}<1/\sqrt{2}$
[see Eq. (\ref{min})] the double-well structure turns into the single-well
one, which cannot support states different from the symmetric ones. The
former feature is natural too, as the increase of the SDF strength makes the
modes broader, favoring the simple symmetric profiles. Note that the system
exhibits bistability, in the form of the overlap between the stability
regions of asymmetric solitons with those of the symmetric and antisymmetric
ones.

\section{Two-dimensional models}

\subsection{The formulation}

The 1D models considered in above can be naturally extended to 2D, similar
to how the 1D system with two nonlinear potential wells \cite{Dong} was
generalized into the 2D setting in Ref. \cite{Thawatchai}. Here we model our
2D systems by equations
\begin{equation}
iU_{z}=-\frac{1}{2}\left( U_{xx}+U_{yy}\right) -\tilde{\delta}_{\mathrm{2D}%
}(x,y)|U|^{2}U+\sigma |U|^{4}U,  \label{2Da}
\end{equation}%
in the case of single potential well and,
\begin{equation}
iU_{z}=-\frac{1}{2}\left( U_{xx}+U_{yy}\right) -\left[ \tilde{\delta}_{%
\mathrm{2D}}(x-x_{0},y)+\tilde{\delta}_{\mathrm{2D}}(x+x_{0},y)\right]
|U|^{2}U+\sigma |U|^{4}U,  \label{2D}
\end{equation}%
for the double-well case. The regularized 2D counterpart of the $\delta $%
-function used in these equations is defined as%
\begin{equation}
\tilde{\delta}_{\mathrm{2D}}(x,y)=\left\{
\begin{array}{c}
\left( \pi a^{2}\right) ^{-1},~\mathrm{at}~~x^{2}+y^{2}<a^{2}. \\
0,~\mathrm{at}~~x^{2}+y^{2}>a^{2}.%
\end{array}%
\right.  \label{tilde}
\end{equation}%
which satisfies the normalization condition, $\int \int \tilde{\delta}_{%
\mathrm{2D}}(x,y)dxdy=1$. By means of the rescaling, we set the
regularization scale $a=1/2$ in Eq. (\ref{tilde}). We consider cases
of the separated circles in Eq. (\ref{2D}), with $x_{0}>1/2$, and
partly overlapping ones, with $x_{0}<1/2$. In addition to $x_{0}$,
Eq. (\ref{2Da}) is governed by two remaining independent parameters:
$\sigma $ and the two-dimensional
total power (norm),%
\begin{equation}
N=\int \int \left\vert U\left( x,y\right) \right\vert ^{2}dxdy
\end{equation}%
(or the respective propagation constant, $\mu $), while Eq.
(\ref{2D}) features three parameters: $\sigma $, $N$ (or $\mu $) and
the distance between the wells, $2x_{0}$. The asymmetry of 2D modes
with respect to the two identical circles in Eq. (\ref{2D}) is
defined as a counterpart of the 1D definition (\ref{niu}):
\begin{equation}
\nu =N^{-1}\int_{-\infty }^{+\infty }dy\left[ \int_{0}^{+\infty
}\left\vert U(x,y)\right\vert ^{2}dx-\int_{-\infty }^{0}\left\vert
U(x,y)\right\vert ^{2}dx\right] .  \label{nu-2D}
\end{equation}%

The 2D models were investigated only in the numerical form, as an
analytical approach would be too difficult in this case. Stationary
solutions were
constructed by means of the Newton conjugate gradient method \cite%
{Yang1,Yang2}. The imaginary-time-integration method does not converge in
this case, but we used intermediate states generated by it as an input for
the Newton's method. The stability of the 2D states was tested by means of
the split-step Fourier method in direct simulations. Results of the direct
simulations were confirmed by computation of eigenvalues for small
perturbations, which was carried out with the help of the Newton conjugate
gradient method and the Fourier collocation method.

\subsection{The single-well model}

\begin{figure}[tbp]
\includegraphics[width=9cm]{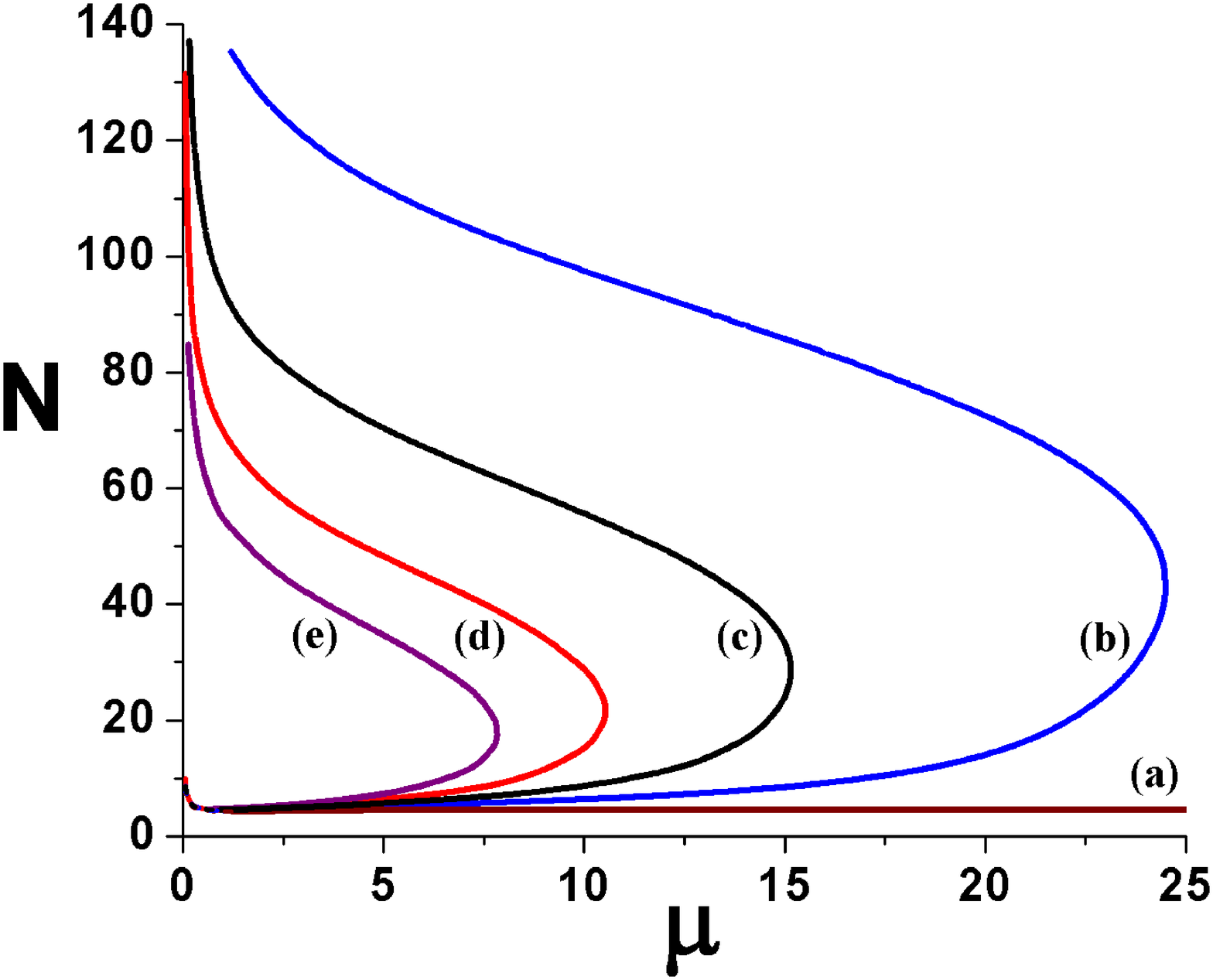}
\includegraphics[width=7.5cm]{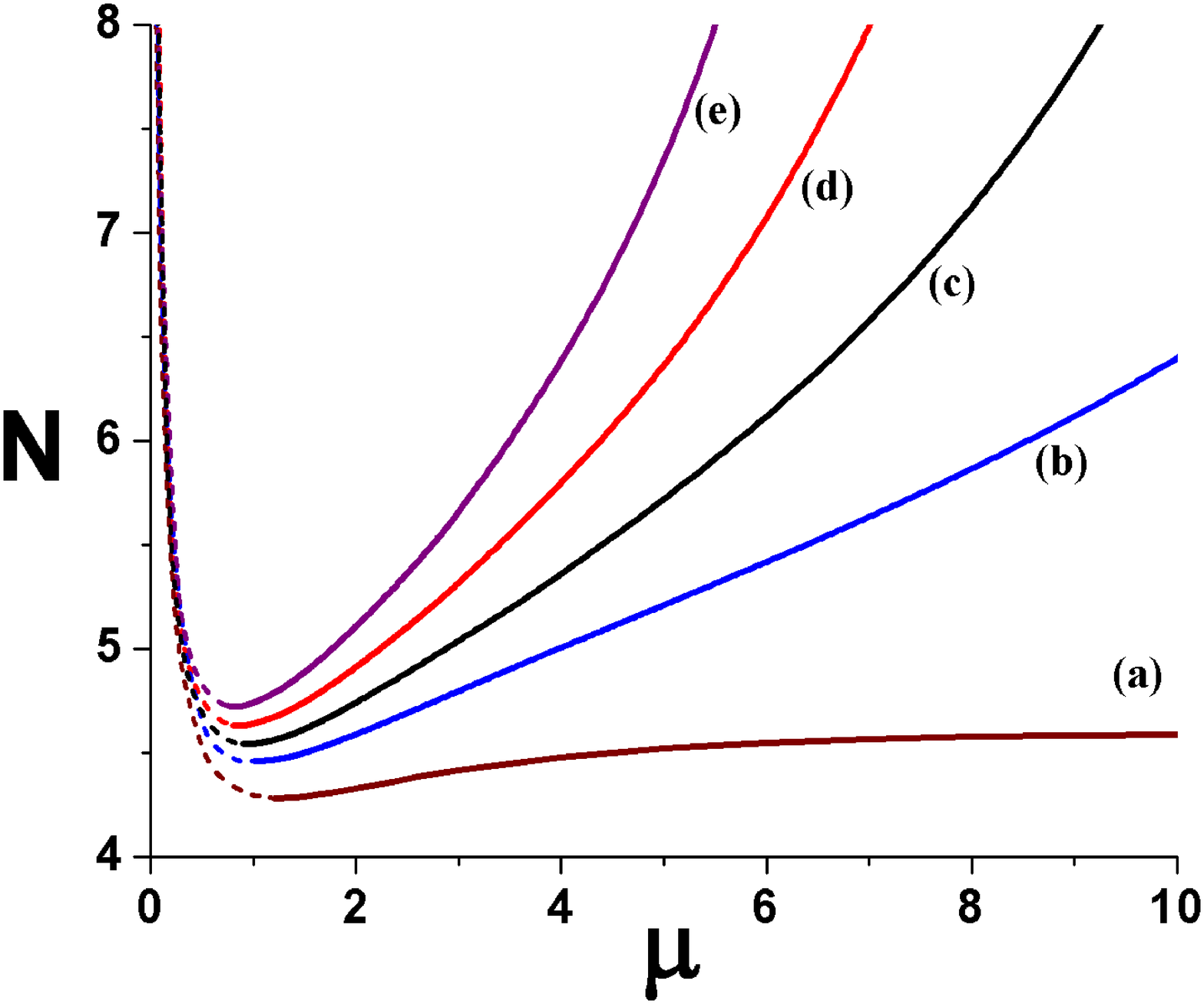} %
\includegraphics[width=7.5cm]{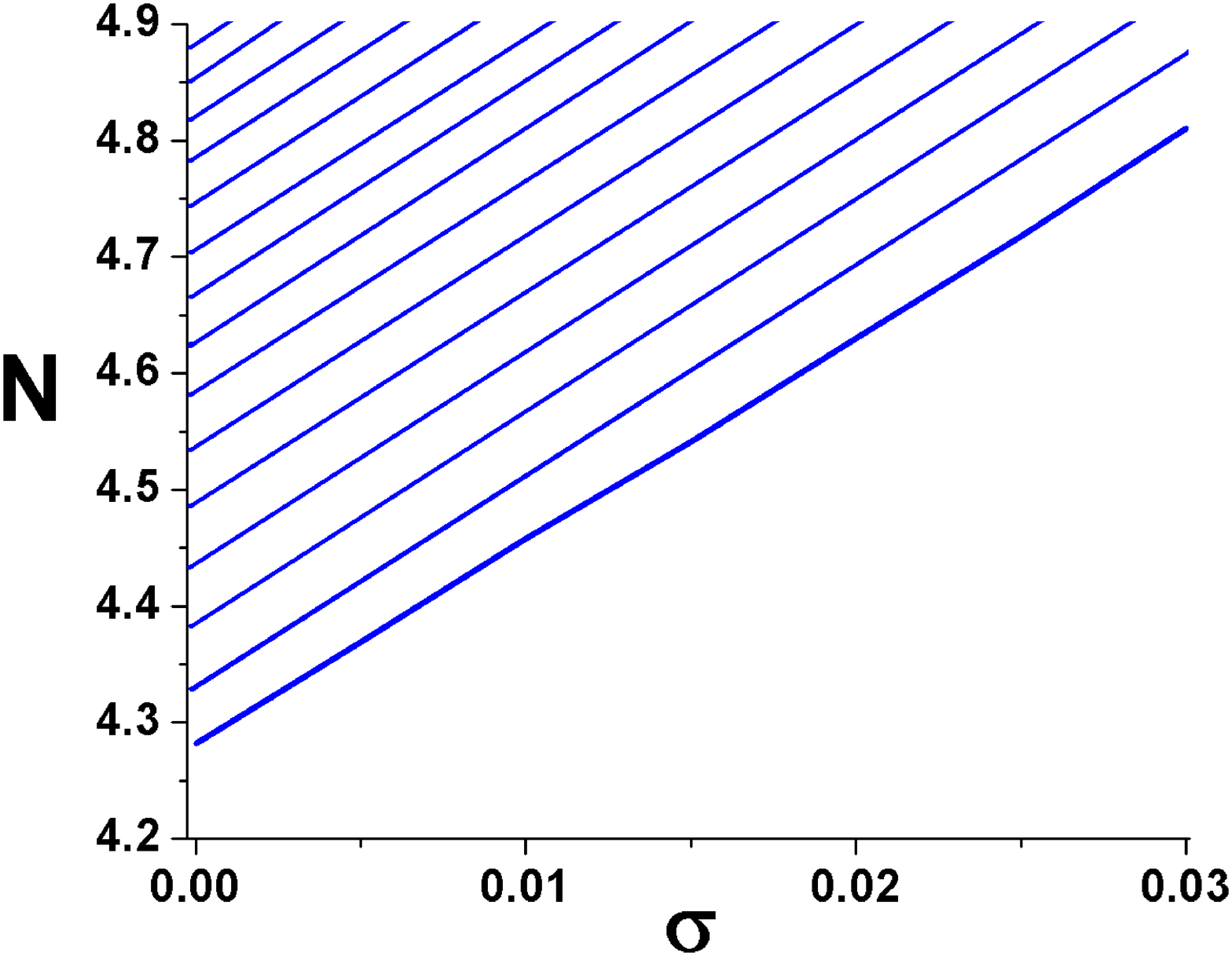}
\caption{(Color online) Numerical results for the 2D single-well model
represented by Eq. (\protect\ref{2Da}). The upper panel shows a set of $N(%
\protect\mu )$ curves for the following values of the SDF quintic
coefficient: $\protect\sigma =0$ (a), $\protect\sigma =0.01$ (b), $\protect%
\sigma =0.015$ (c), $\protect\sigma =0.02$ (d) and $\protect\sigma =0.025$
(e). The left bottom panel shows a closeup of the region of small $\protect%
\mu $, where the transition from stable (solid) branches to unstable
(dashed) ones occurs. In the right bottom panel, the domain where
the system supports stable solitons is shaded by oblique lines.}
\label{dia7a}
\end{figure}

Figure \ref{dia7a} shows curves $N(\mu )$ for solitons produced by Eq. (\ref%
{2Da}) for selected values of the quintic coefficient. In the bottom left
panel, we show a close-up of the picture at small values of propagation
constant $\mu $, where one can observe the transition from stable to
unstable solutions. In fact, curve (a), pertaining to $\sigma =0$,
reproduces the results reported in Refs. \cite{Thawatchai} and \cite{Saka-2D}%
, with the norm of the stable solitons bounded from above by the collapse
\cite{Berge}. In this case, the stability of the solitons is secured by the
VK criterion. In the presence of the quintic term [curves (b) - (e) in Fig. %
\ref{dia7a}], additional (upper) stable soliton branches appear, which obey
the anti-VK criterion, making the situation similar to that in 1D, cf. Fig. %
\ref{dia1}.
\begin{figure}[tbp]
\includegraphics[width=5cm]{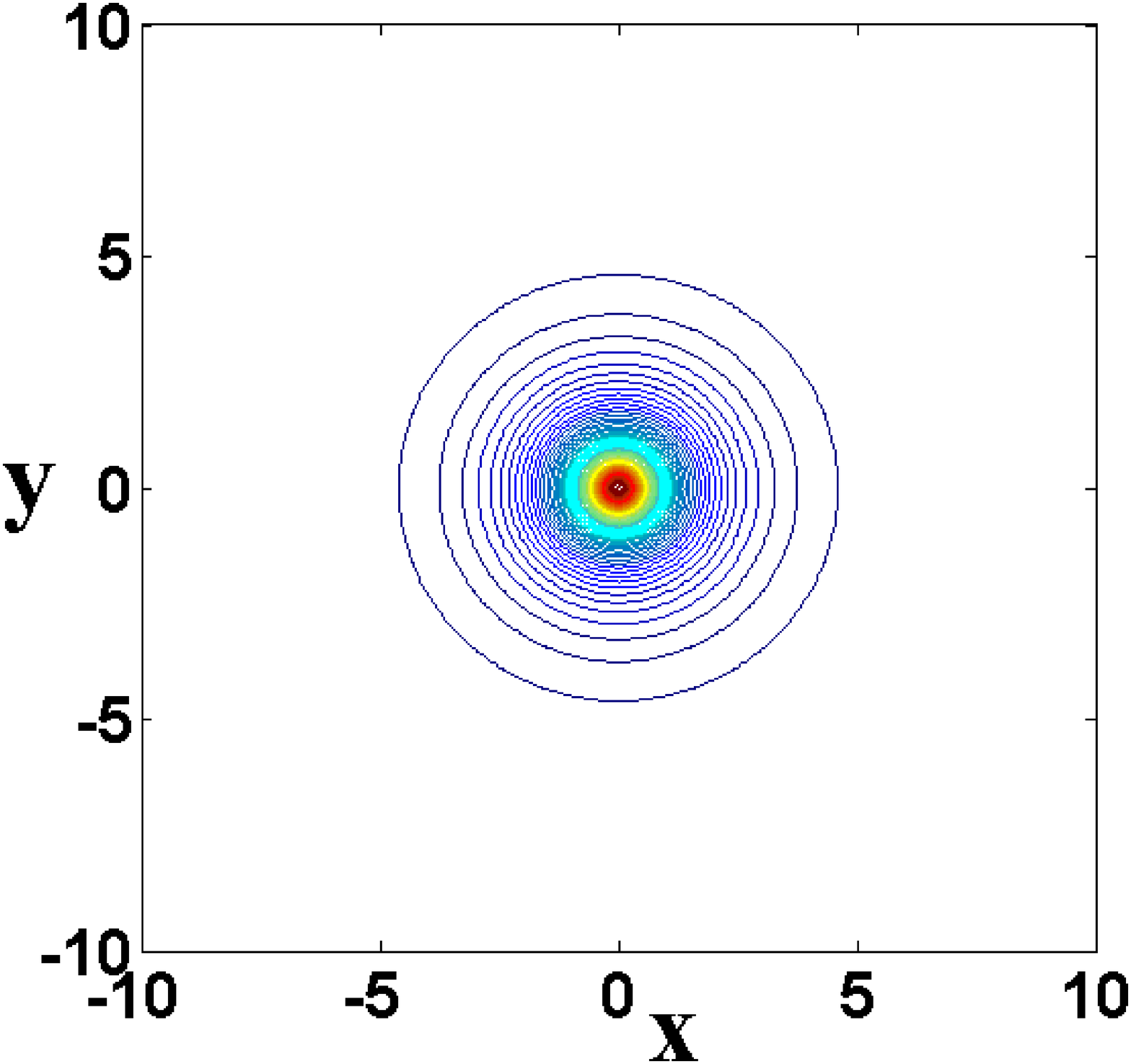} %
\includegraphics[width=5cm]{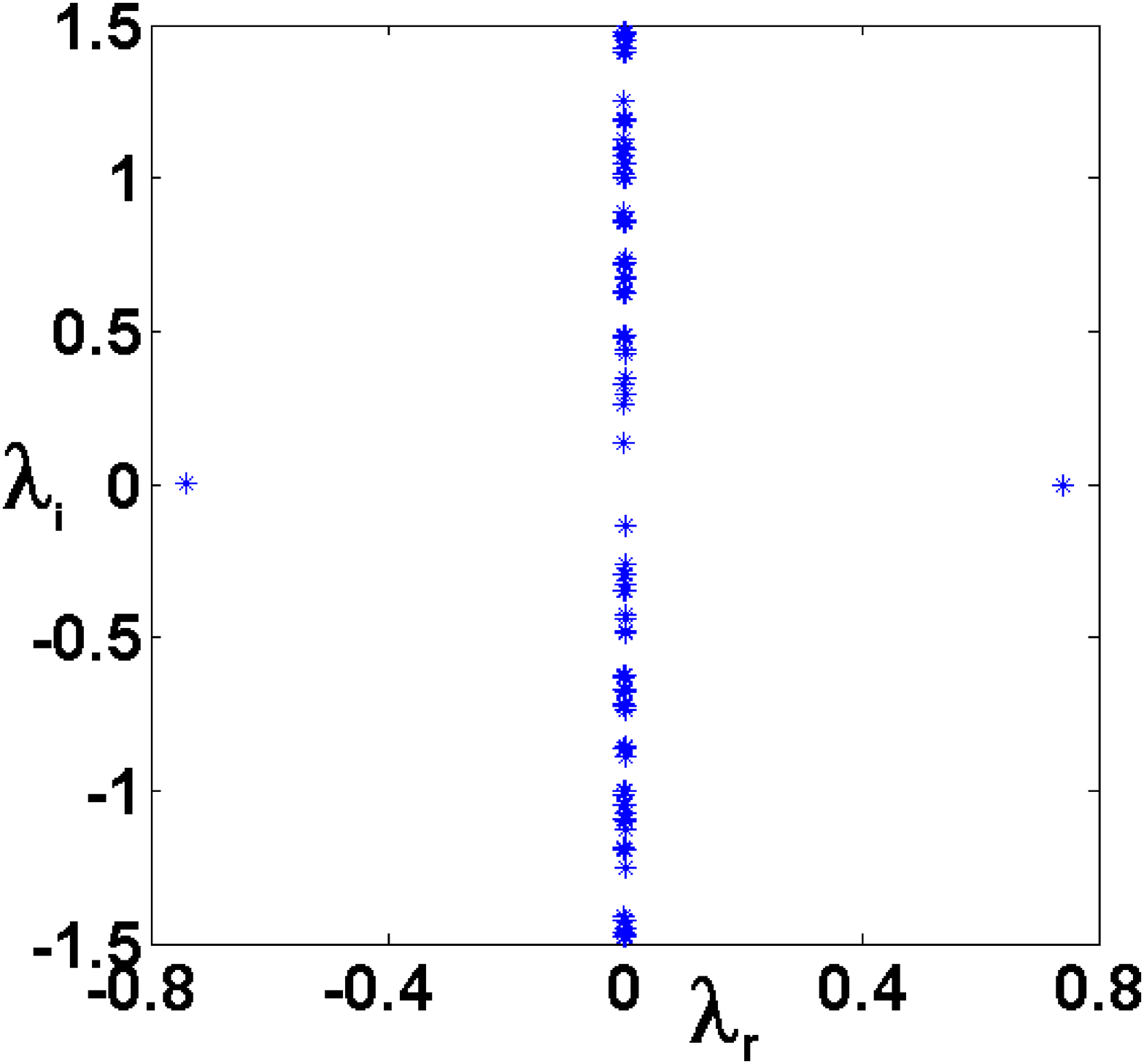}
\includegraphics[width=5cm]{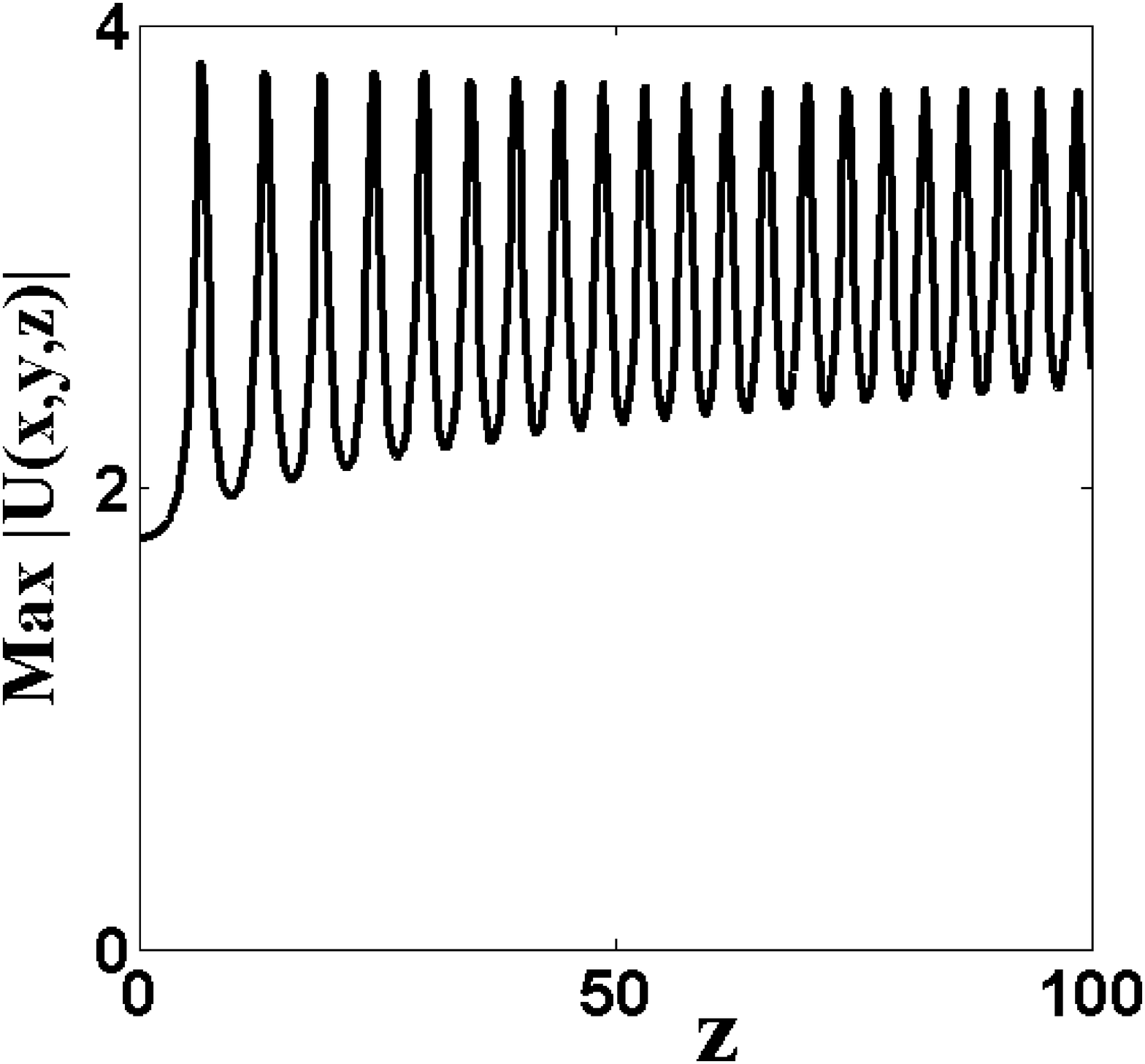} %
\caption{(Color online) An example of an unstable 2D soliton in the
single-well model, based on Eq. (\protect\ref{2Da}) with $\protect\sigma %
=0.02$ (the first panel). The propagation constant and total power of the soliton are $\protect%
\mu =0.25$ and $N=5.2517$. The second panel shows the linear
spectrum of small perturbations around the soliton. The third panel
illustrates the transformation of the unstable soliton into a robust
breather, which features regular oscillations of the amplitude.}
\label{dia7b}
\end{figure}
Additional unstable branches, specific to the 2D setting, which do
not satisfy the VK criterion, exist at small values of $\mu $, as
shown in detail in the left bottom panel of Fig. \ref{dia7a}. In
direct simulations, displayed in Fig. \ref{dia7b}, solitons
belonging to an unstable branch at first lose a part of the norm
through the emission of radiation, and later transform themselves
into robust breathers. As concerns the shape of the boundary of the
region where stable 2D solitons exist in the right panel of Fig.
\ref{dia7a}, the necessary value of the quintic coefficient, $\sigma
$, grows (roughly, linearly) with $N$, as the sufficiently strong
SDF quintic term is necessary to stabilize the solitons against the
collapse.

\subsection{The double-well model}

Extending the analysis to the 2D model with two regularized $\delta $%
-functions, described by Eq. (\ref{2D}), we have found that the model can
support stable symmetric, antisymmetric and asymmetric states, where, like
in the 1D setting, the asymmetry is defined with respect to the two
identical regularized $\delta $-functions, i.e., the circles in Eq. (\ref{2D}%
). The respective asymmetry measure, $\nu $, is defined by Eq. (\ref{nu-2D}).

\begin{figure}[tbp]
\includegraphics[width=5cm]{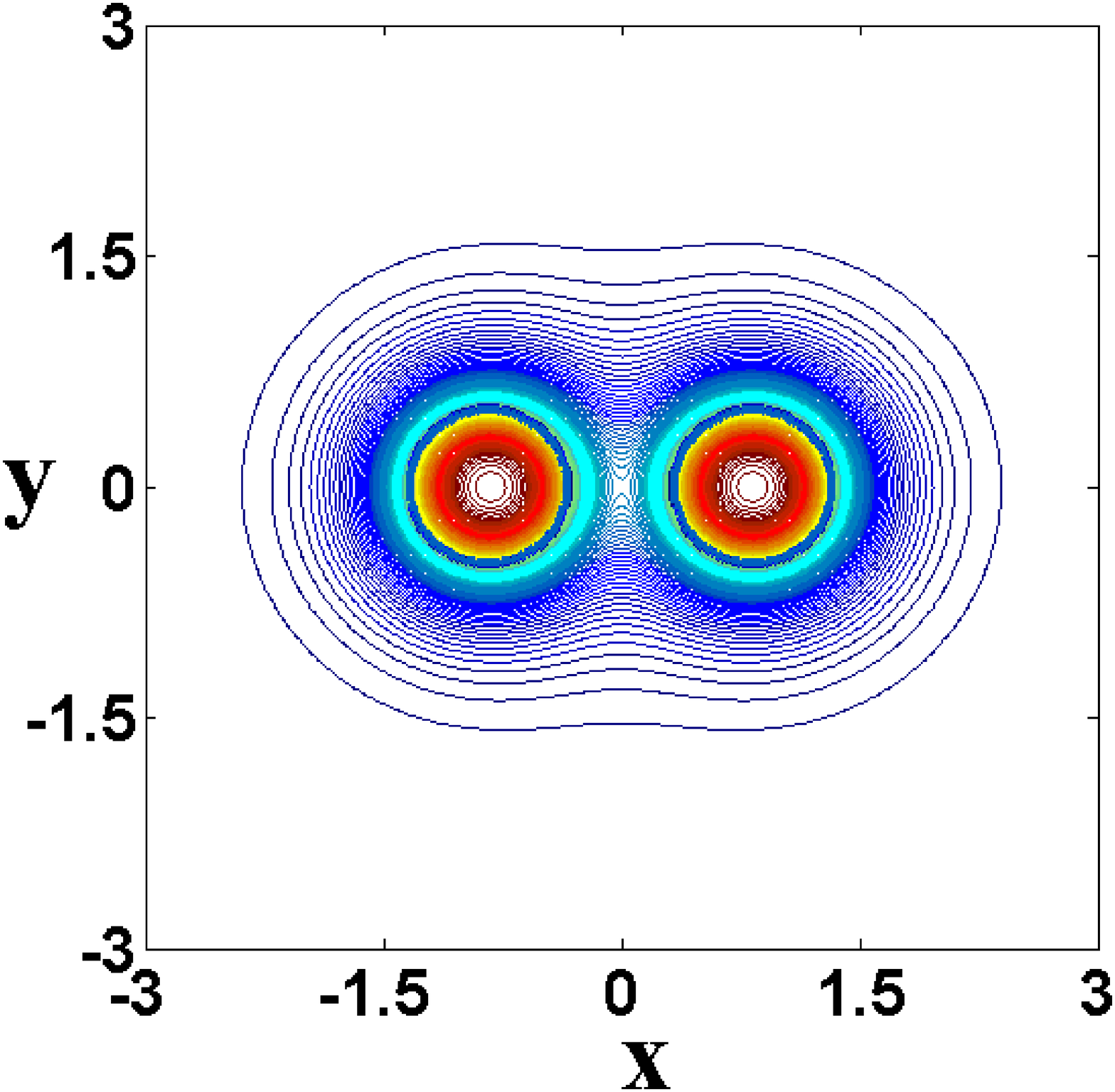} %
\includegraphics[width=5.1cm]{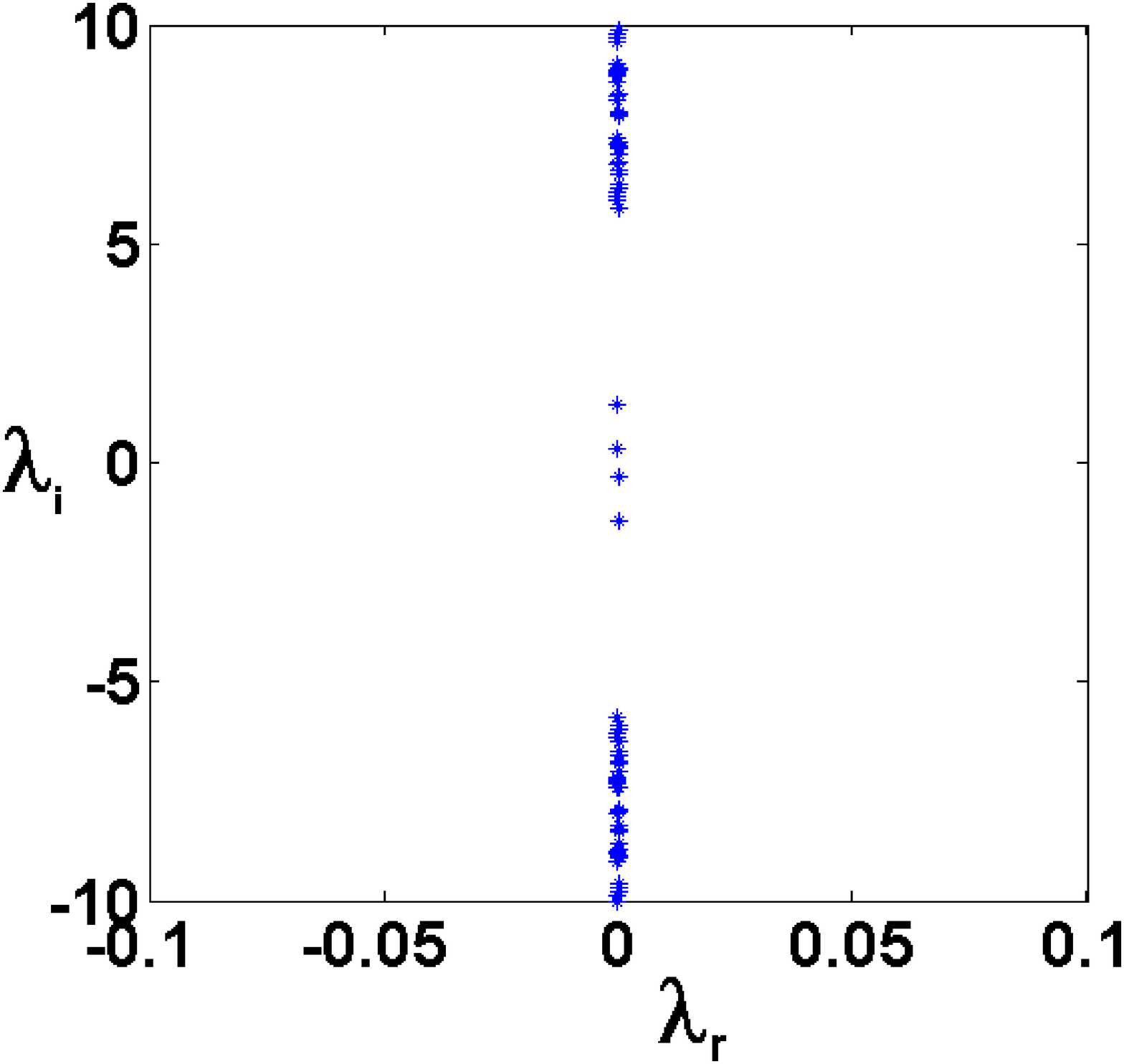}
\caption{(Color online) A stable symmetric double-peak soliton of equation (%
\protect\ref{2D}) with $x_{0}=0.83$, $\protect\sigma =0.0288$ and $\protect\mu %
=5.76$, $N=49.02$. Bold blue circles in the first panel and in
similar
figures below depict the shape of the regularization of the $\protect\delta $%
-functions per Eq. (\protect\ref{tilde}). Here and in similar
figures below, the second panel shows the spectrum of stability
eigenvalues for small perturbations around the soliton.}
\label{contour2da1}
\end{figure}

\begin{figure}[tbp]
\includegraphics[width=5cm]{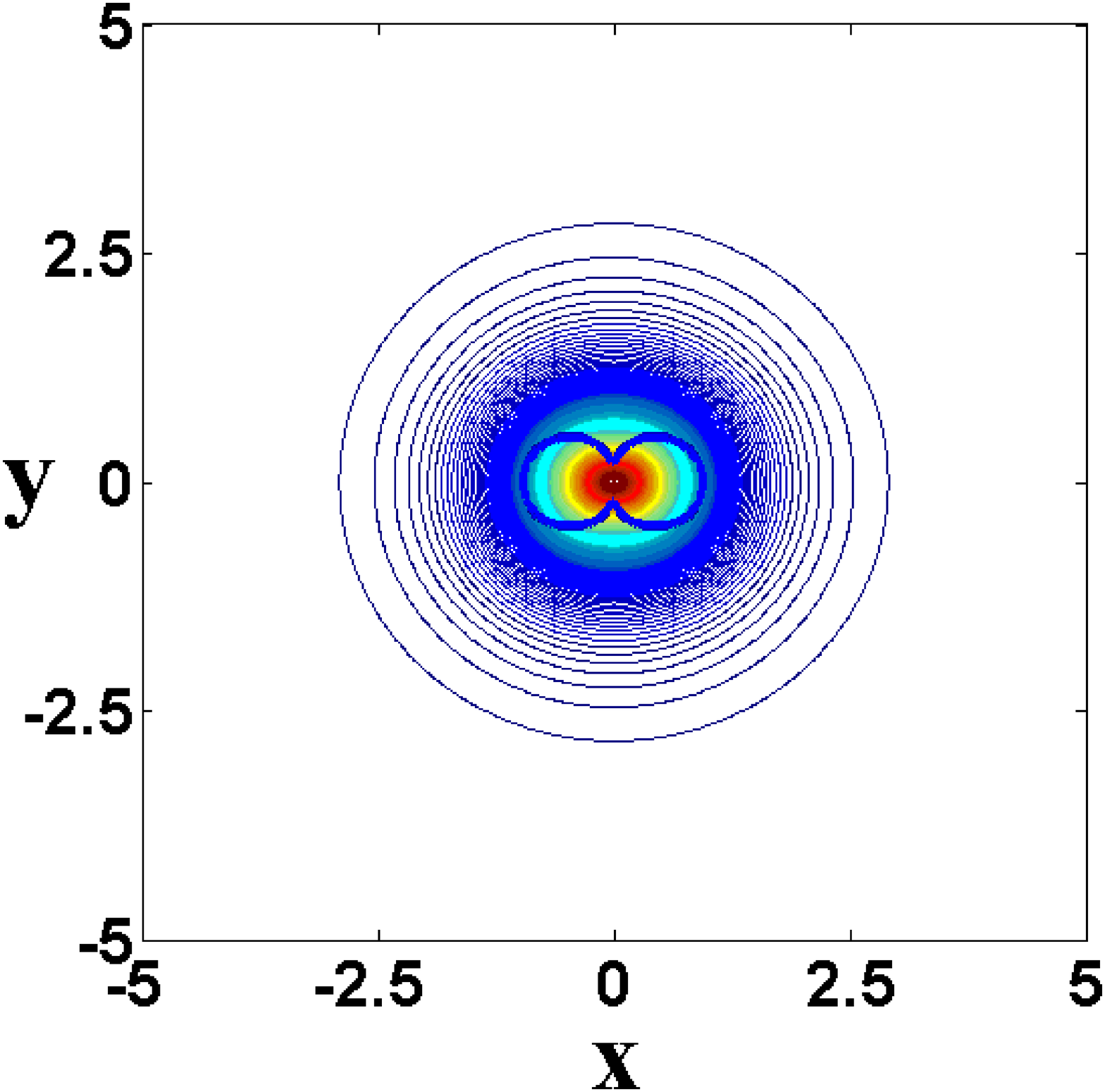} %
\includegraphics[width=5cm]{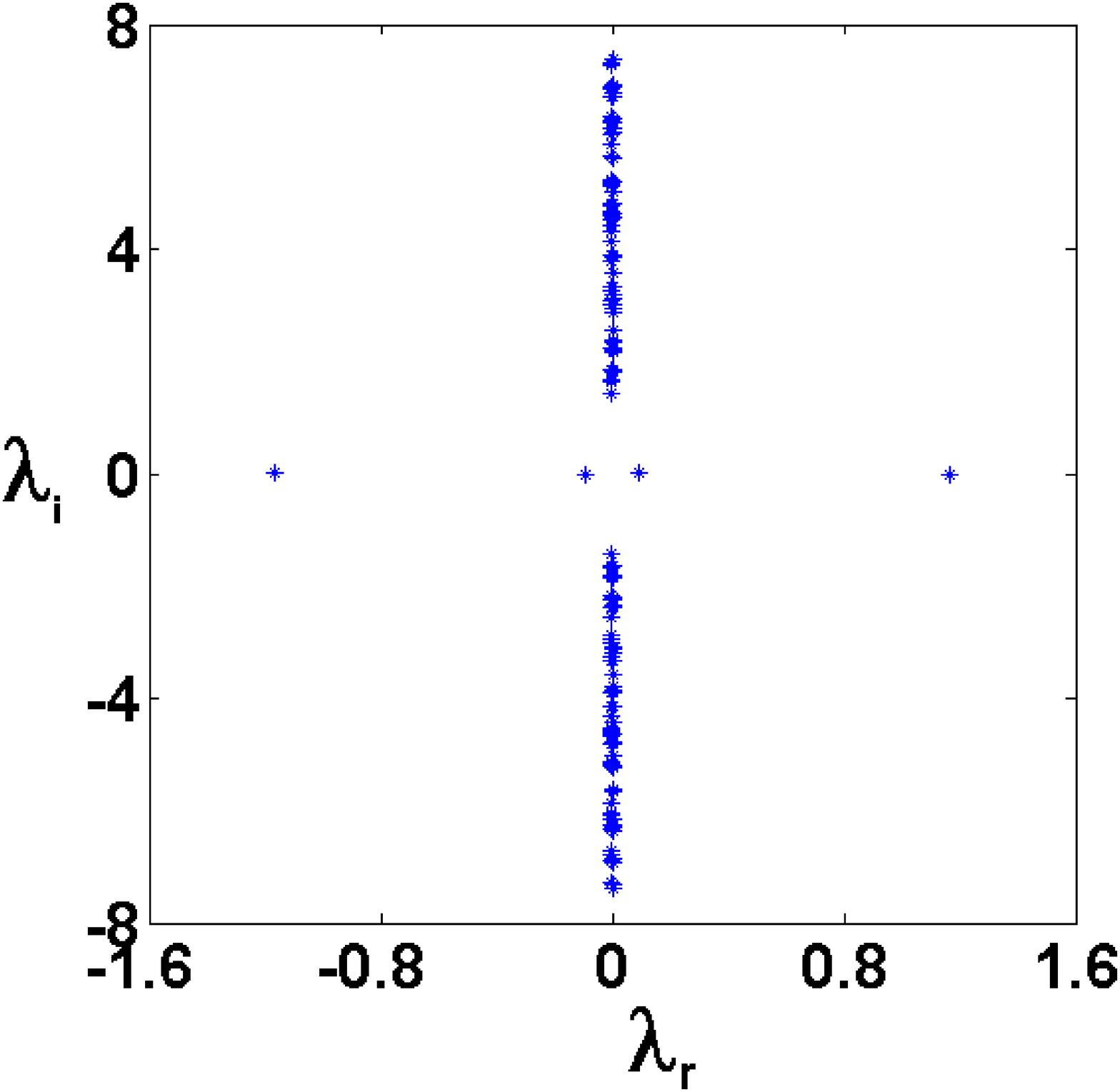}
\includegraphics[width=6cm]{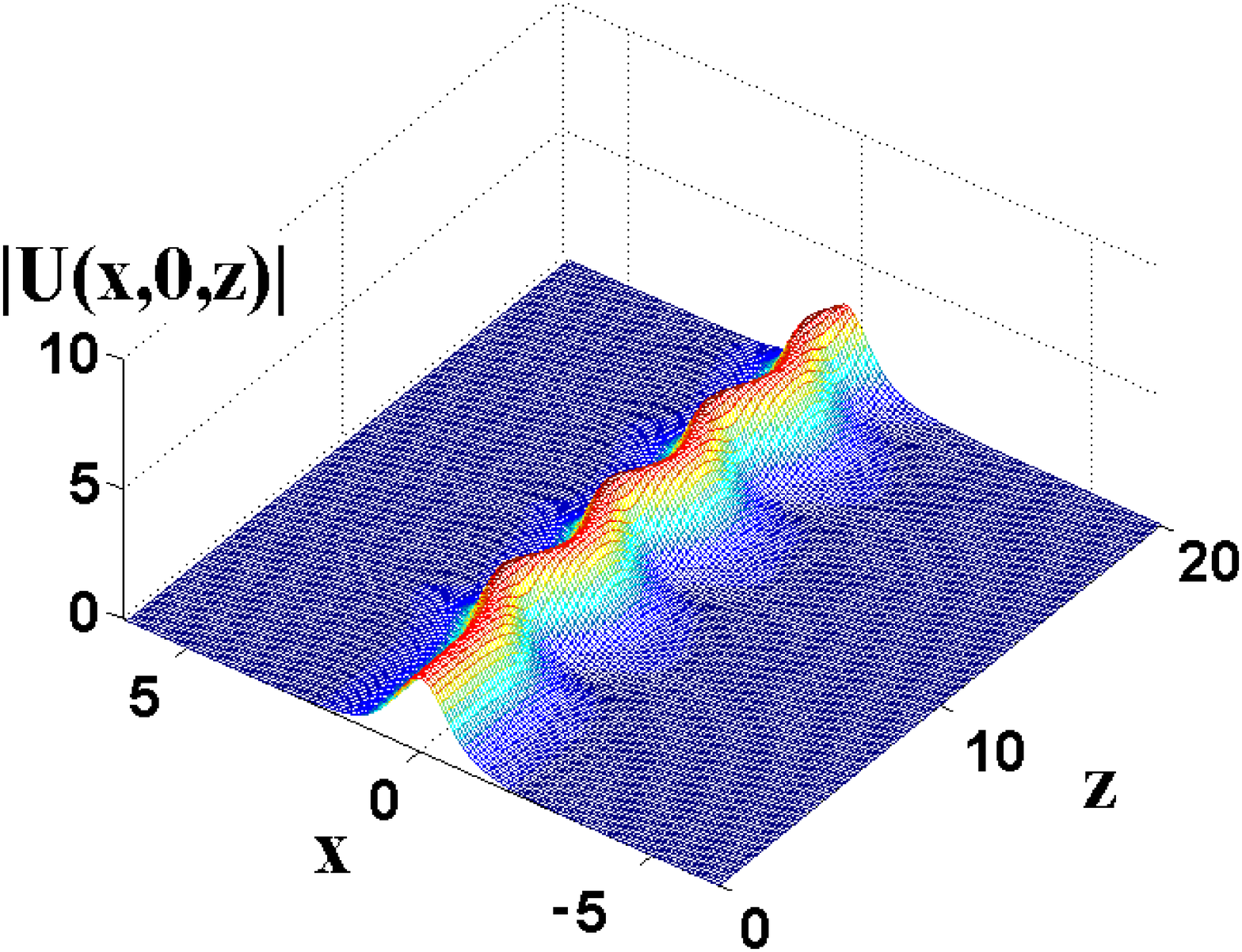} %
\caption{(Color online) An unstable symmetric single-peak soliton of
equation (\protect\ref{2D}) with $x_{0}=0.45$, $\protect\sigma =0.0484$, and $%
\protect\mu =1.452$, $N=6.1614$. Here and in similar figures below,
the third panel shows the cross section $y=0$ of the wave field in
the course of its evolution. } \label{contour2da}
\end{figure}

\begin{figure}[tbp]
\includegraphics[width=5cm]{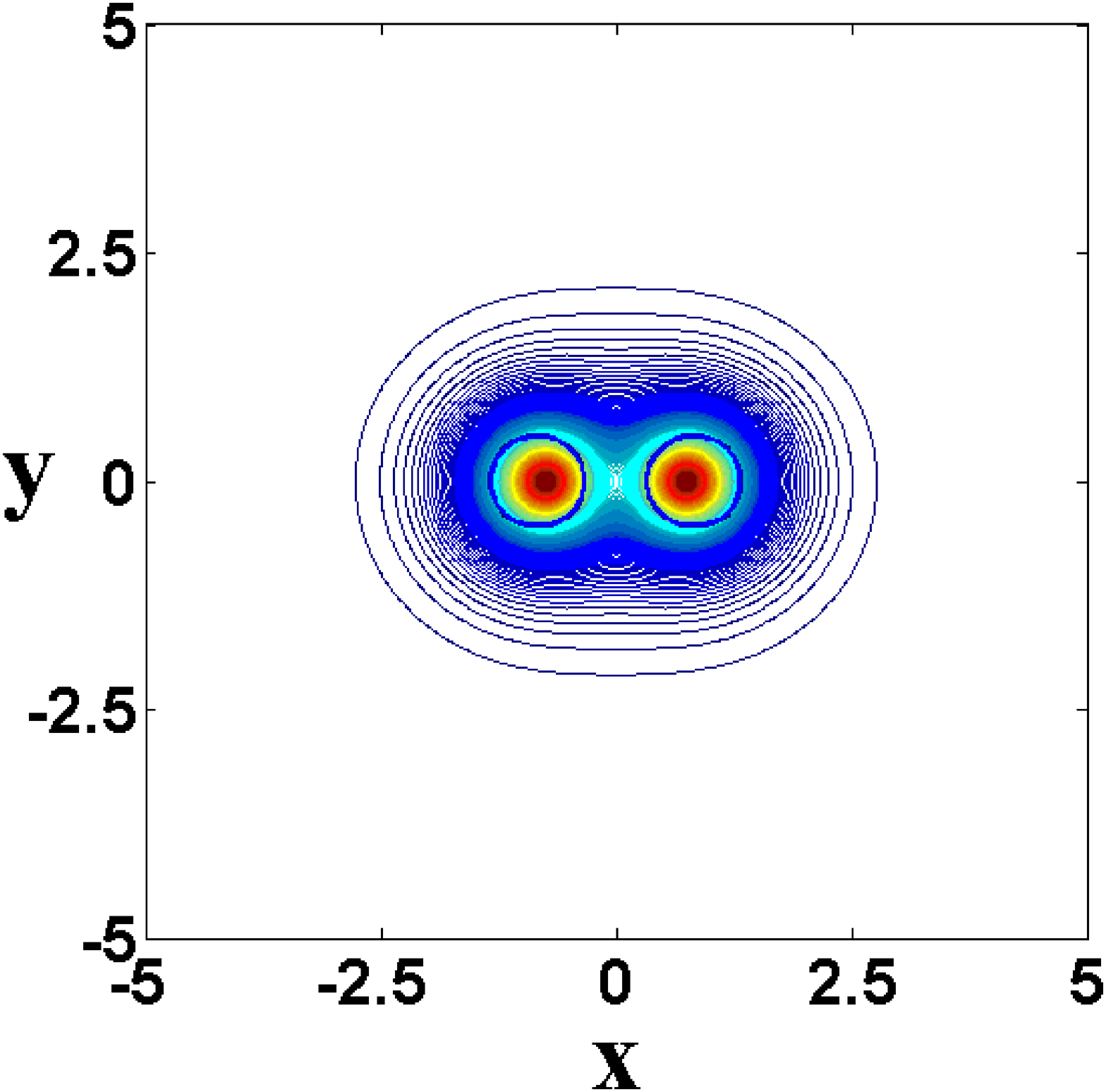} %
\includegraphics[width=5cm]{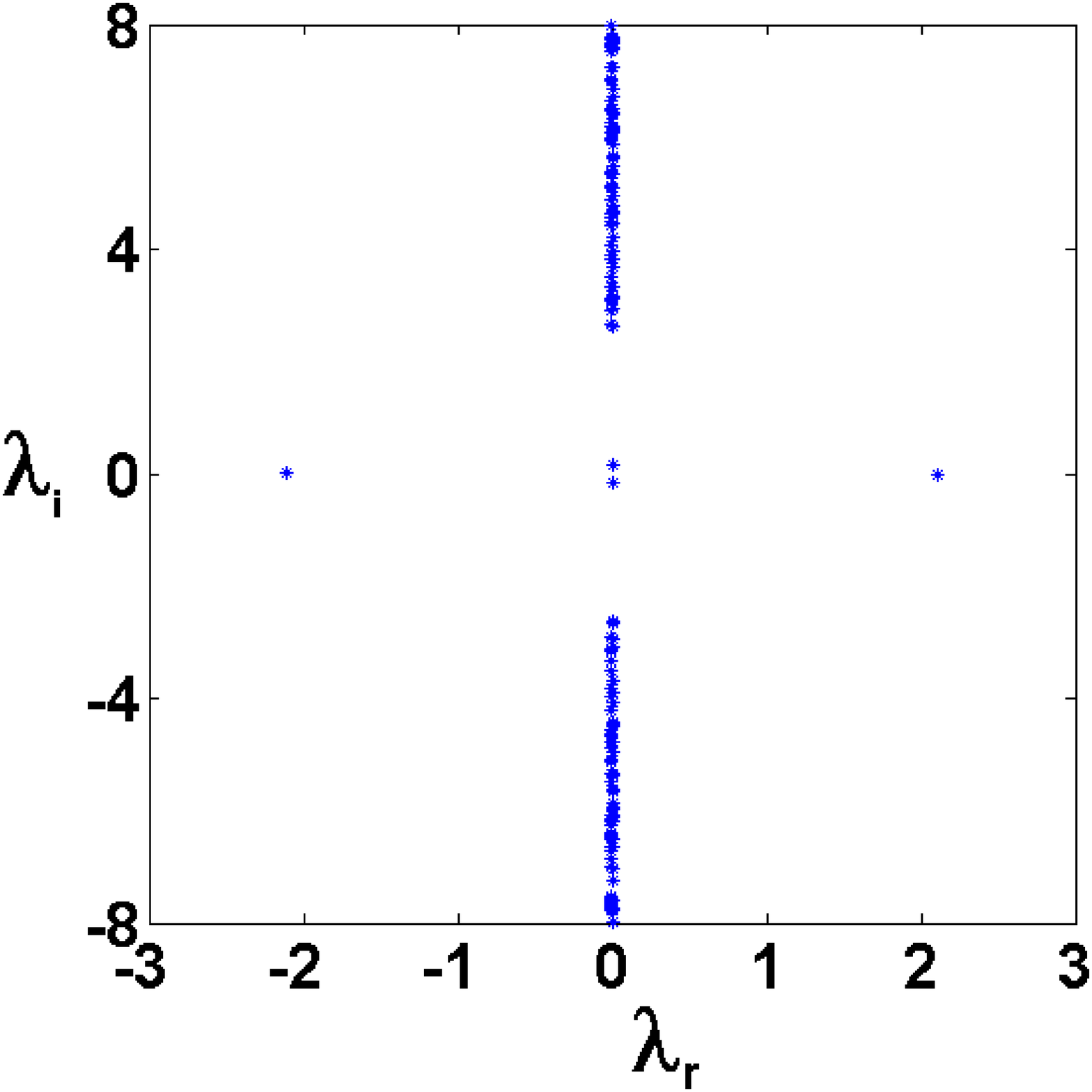}
\includegraphics[width=5.6cm]{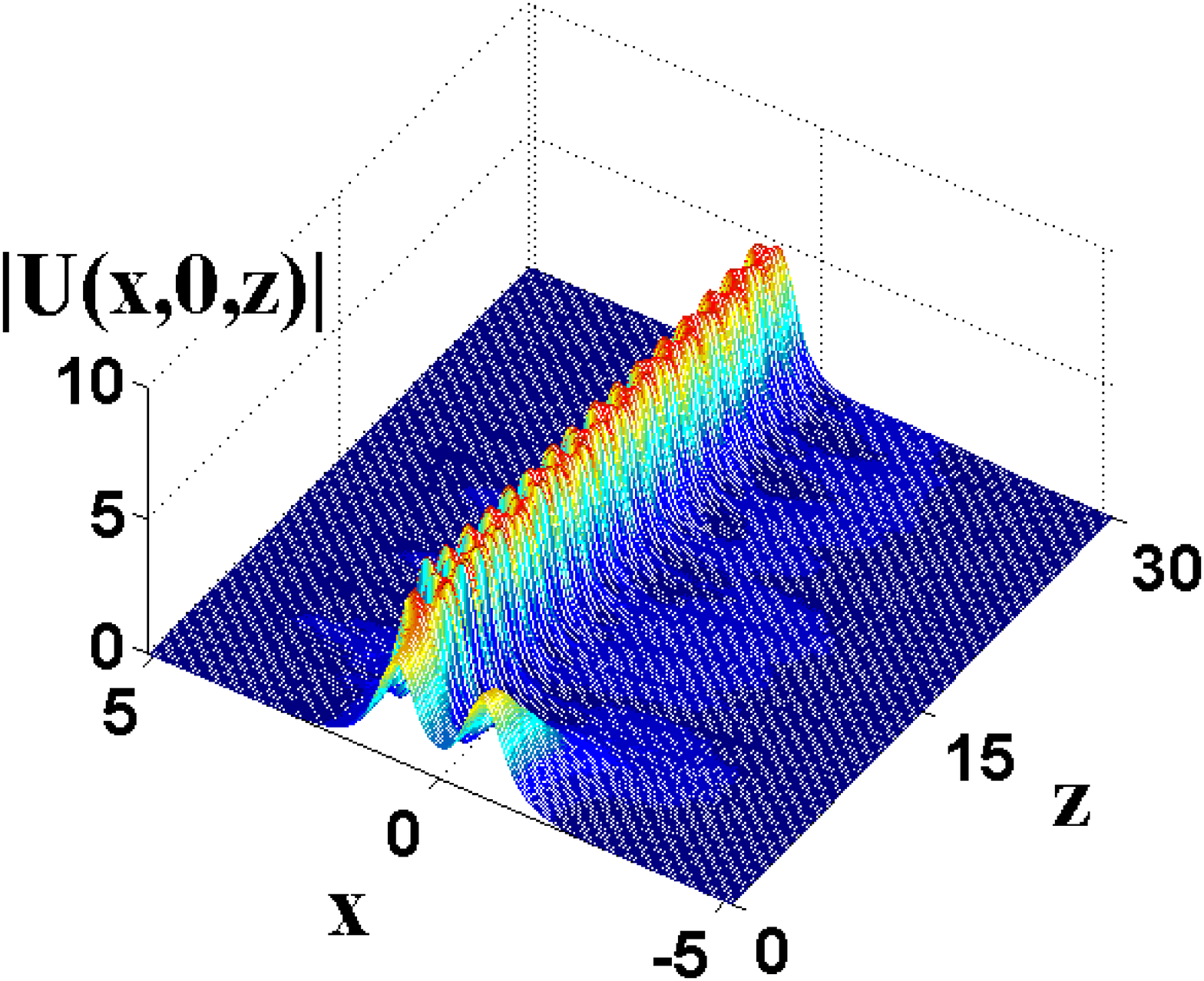} %
\caption{(Color online) An unstable symmetric double-peak soliton of Eq. (%
\protect\ref{2D}) with $x_{0}=0.83$, $\protect\sigma =0.0288$, and $\protect%
\mu =2.88$, $N=11.95$.} \label{contour2d2}
\end{figure}

\begin{figure}[tbp]
\includegraphics[width=5cm]{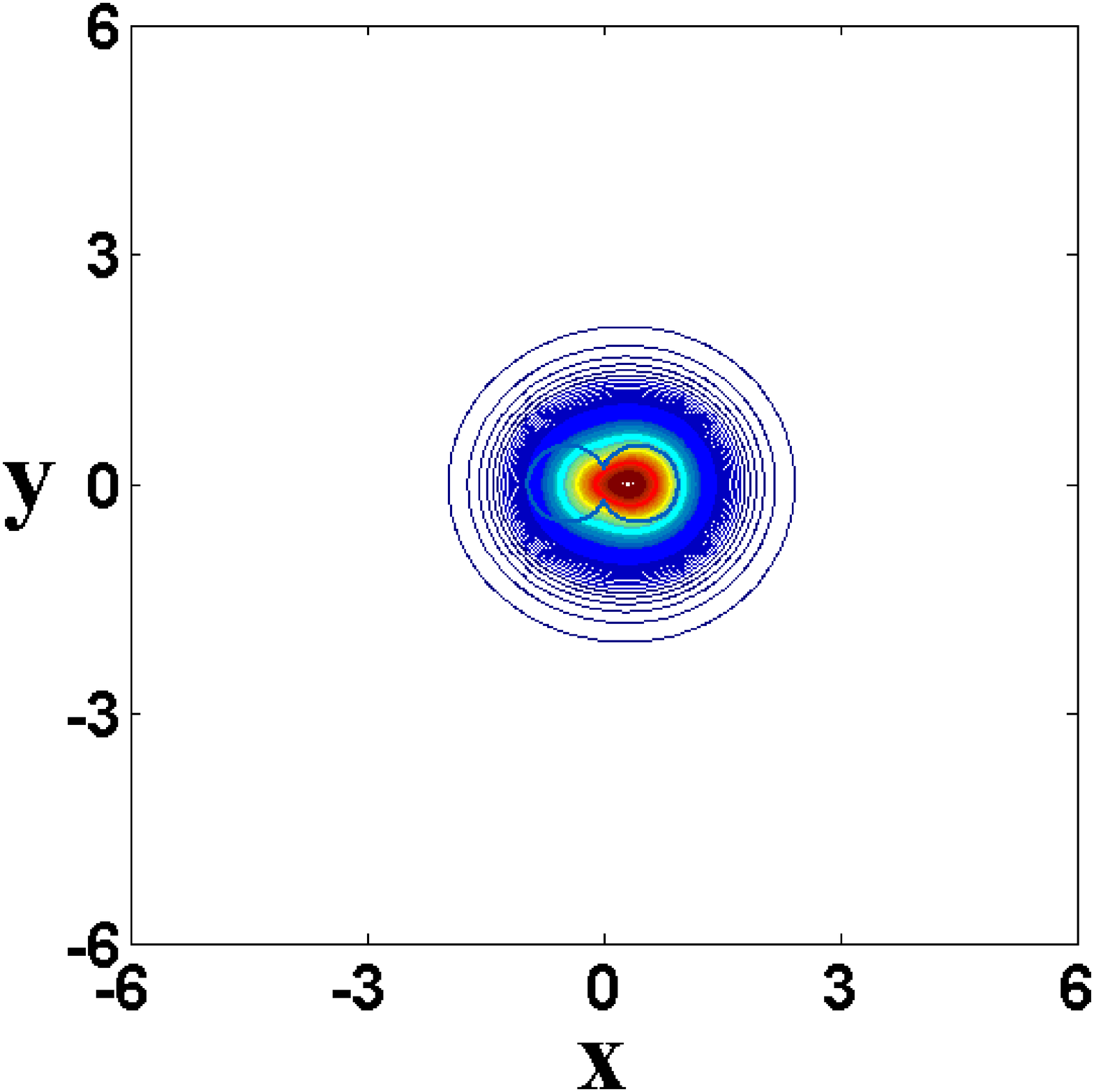} %
\includegraphics[width=5.1cm]{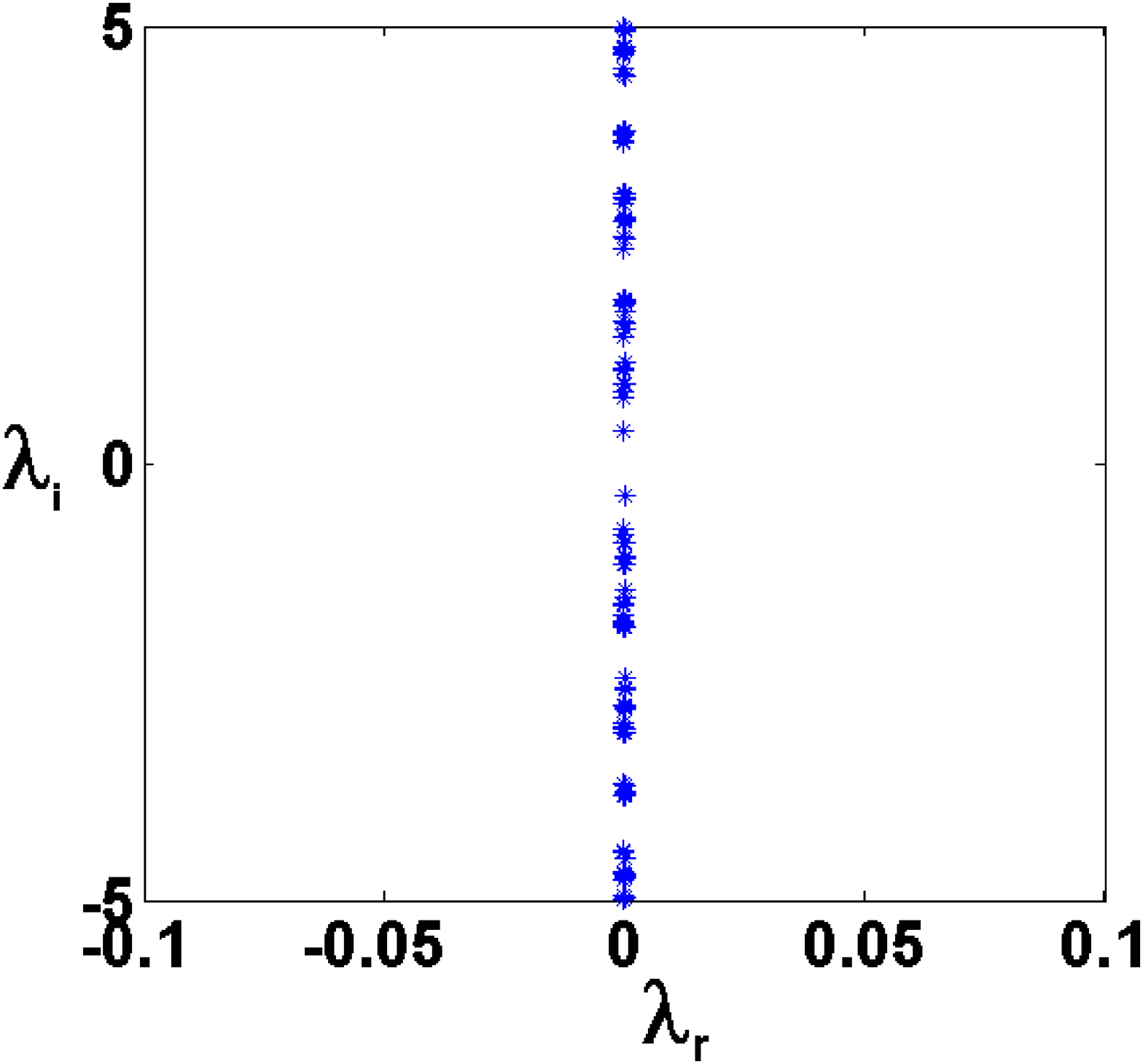}
\caption{(Color online) A stable asymmetric soliton of Eq. (\protect\ref{2D}%
) with $x_{0}=0.45$, $\protect\sigma =0.0484$, and $\protect\mu =3.267$, $%
N=17.246$, the asymmetry parameter being $\protect\nu =0.474$.}
\label{contour2db1}
\end{figure}

\begin{figure}[tbp]
\includegraphics[width=5cm]{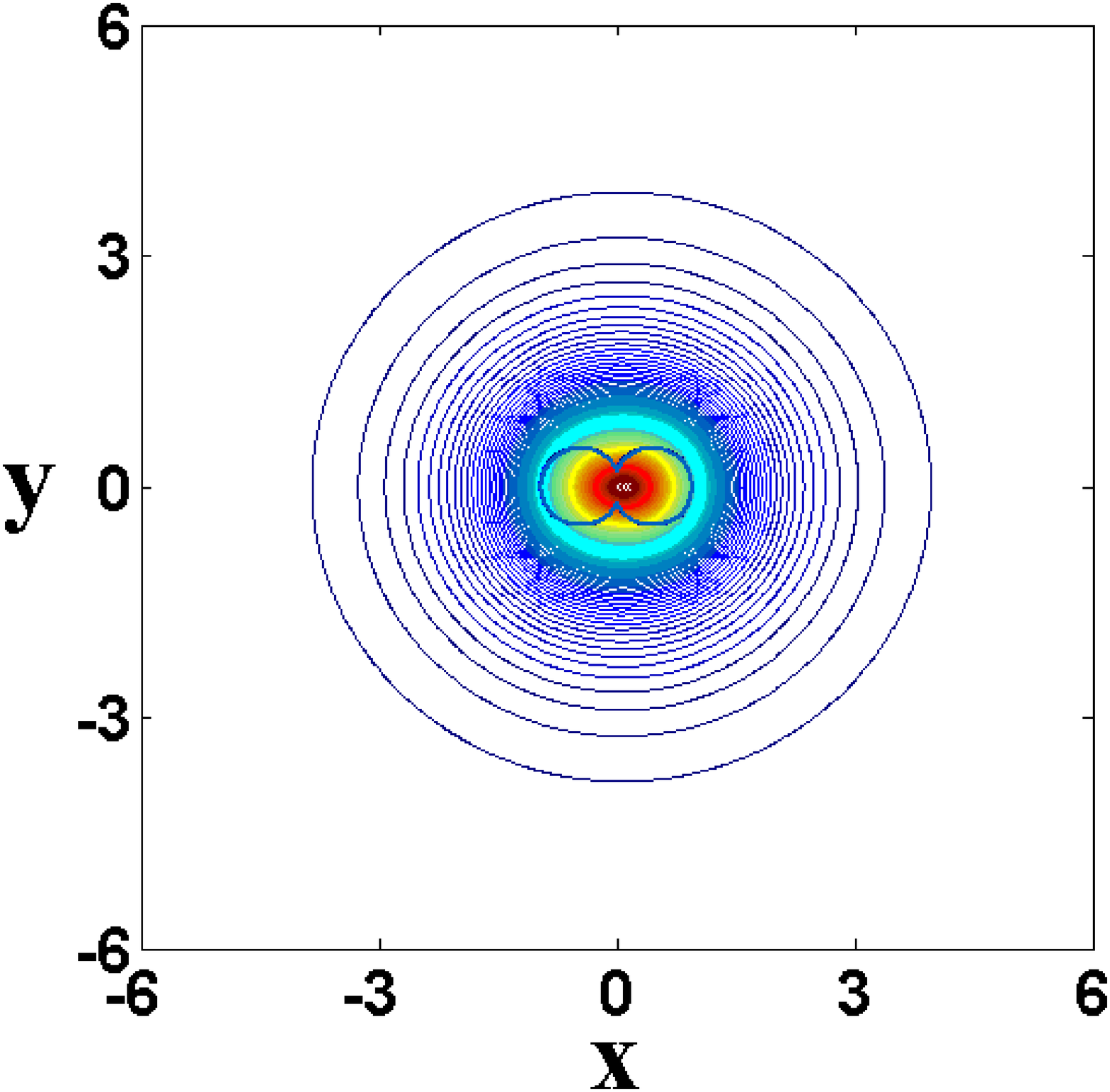} %
\includegraphics[width=5cm]{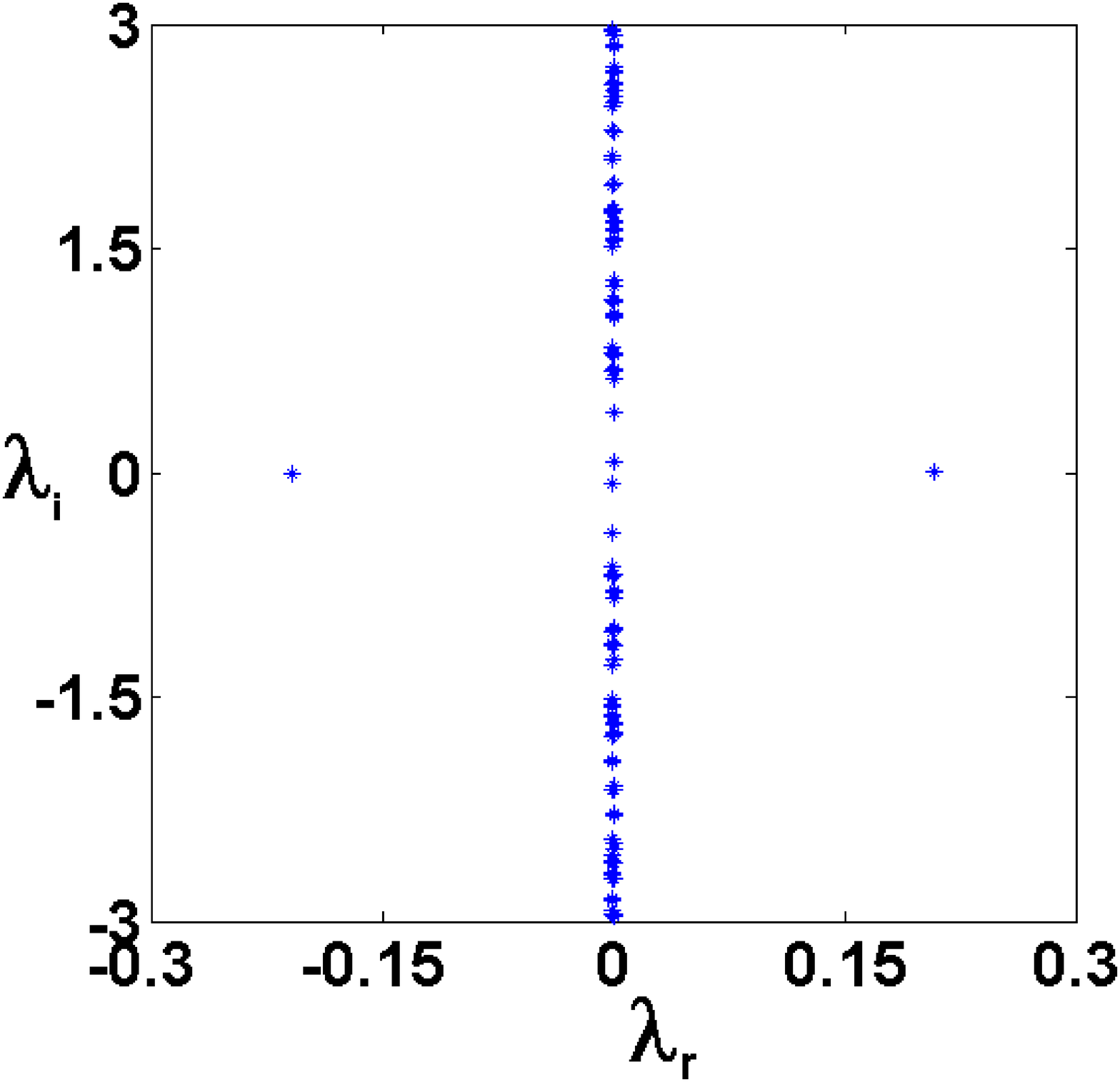}
\includegraphics[width=6cm]{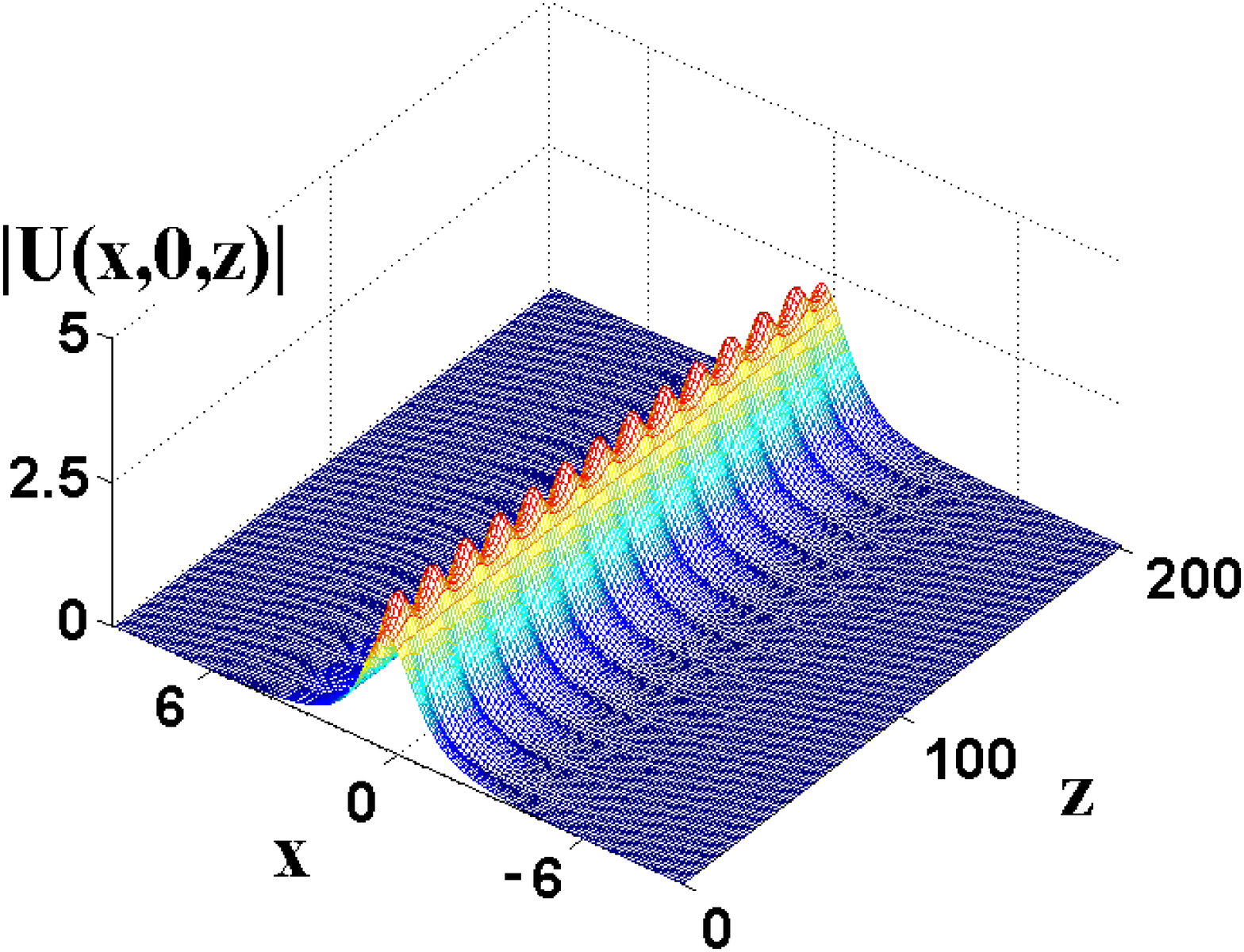} %
\caption{(Color online) An unstable asymmetric soliton of Eq. (\protect\ref%
{2D}) with $x_{0}=0.45$, $\protect\sigma =0.0484$ and $\protect\mu =0.557$, $%
N=5.324$, the asymmetry parameter being $\protect\nu =0.131$.}
\label{contour2db}
\end{figure}

\begin{figure}[tbp]
\includegraphics[width=4.5cm]{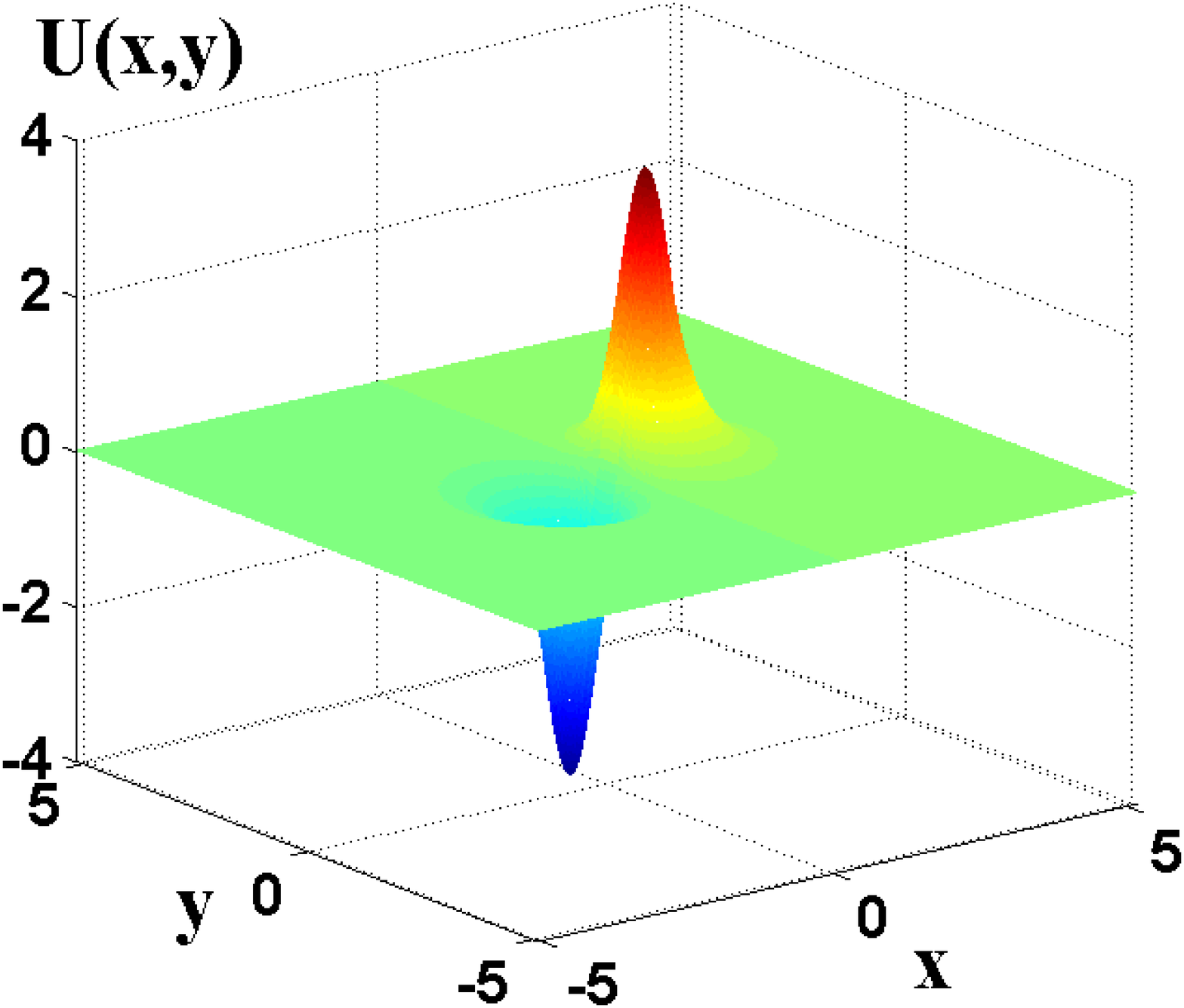} %
\includegraphics[width=4.5cm]{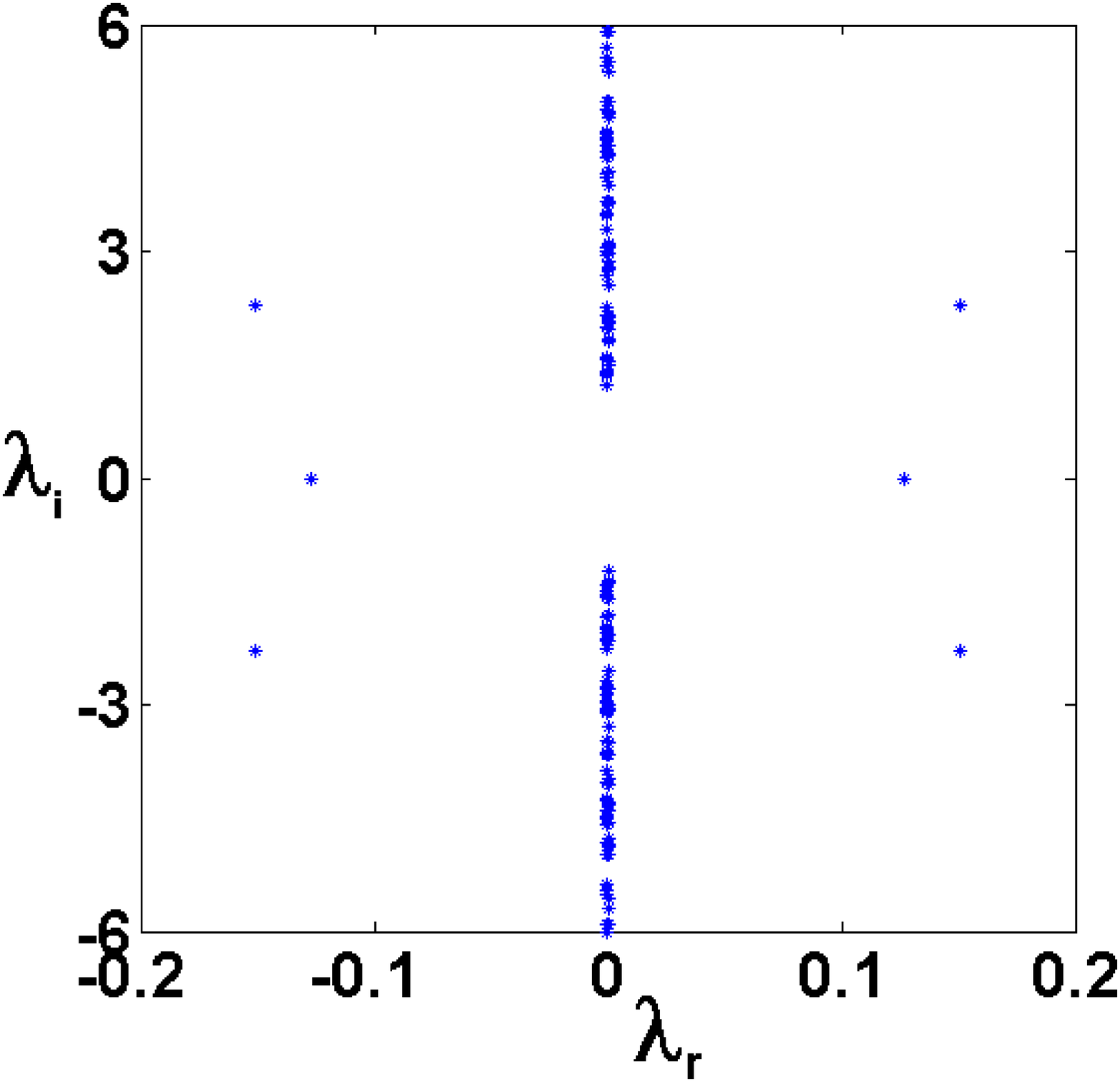}
\includegraphics[width=5.5cm]{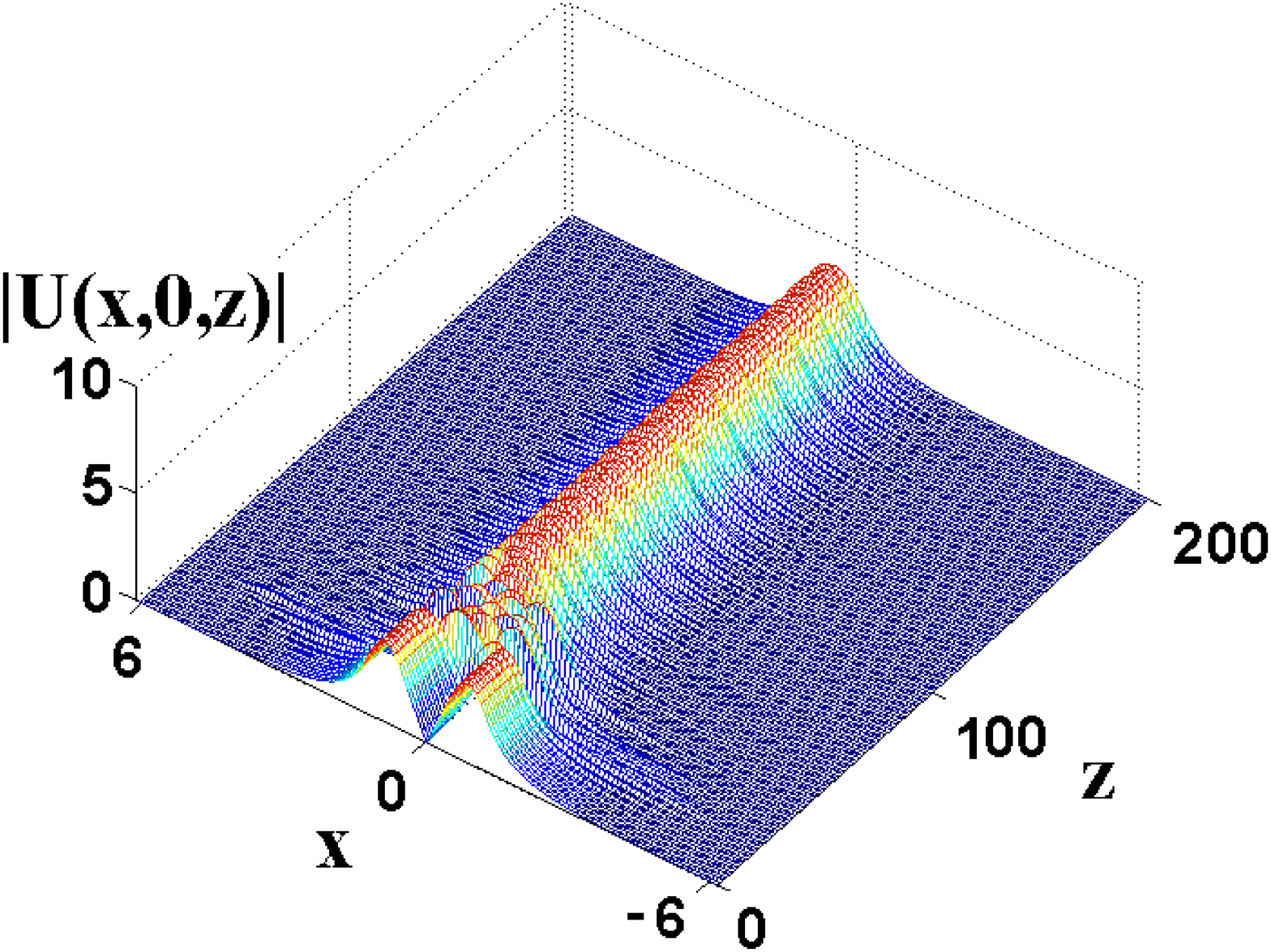} %
\caption{(Color online) An unstable antisymmetric soliton of equation (%
\protect\ref{2D}) with $x_{0}=0.45$, $\protect\sigma =0.0484$ and $\protect\mu %
=1.21$, $N=13.7$.} \label{contour2db2}
\end{figure}

\begin{figure}[tbp]
\includegraphics[width=5cm]{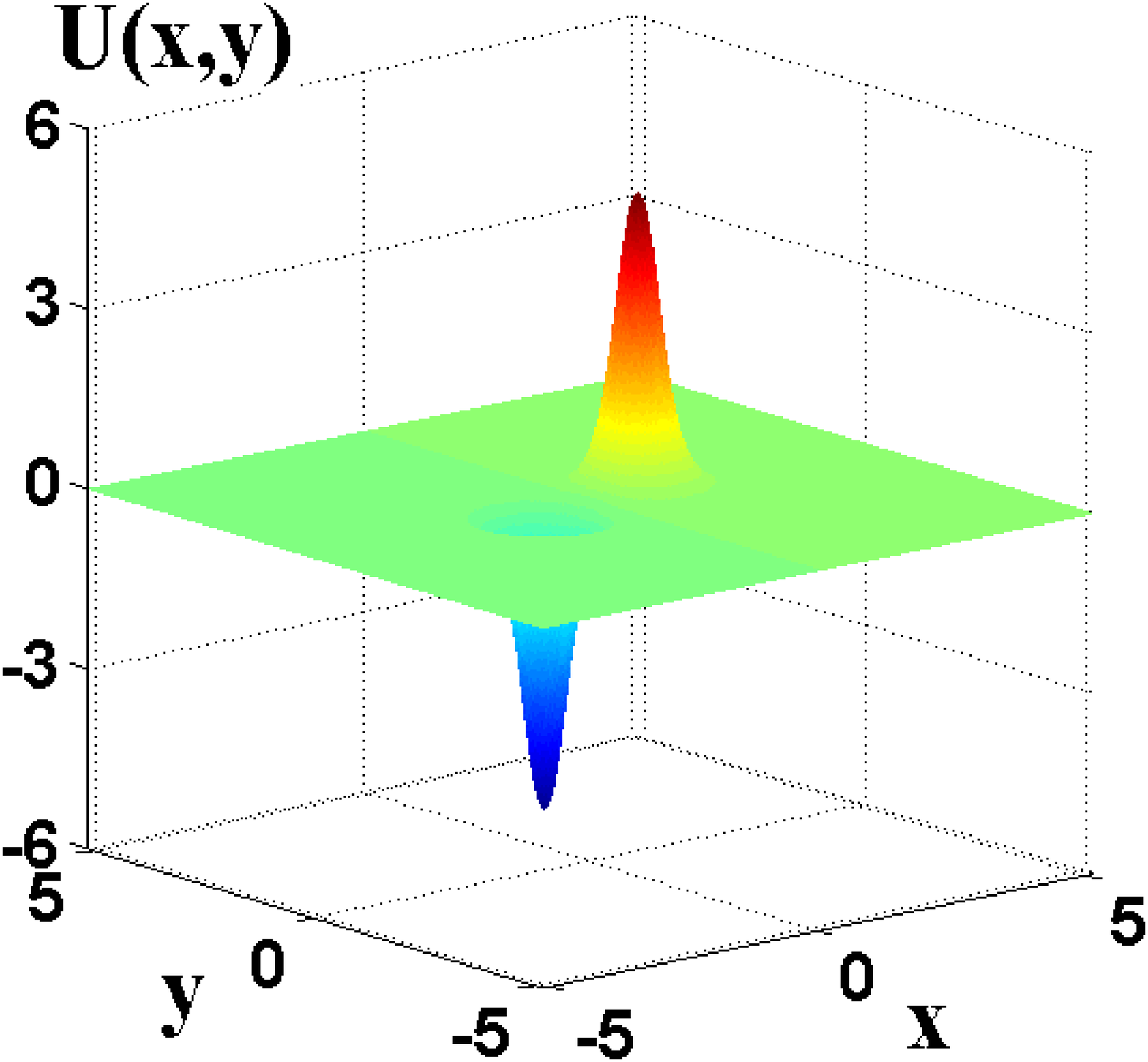} %
\includegraphics[width=5cm]{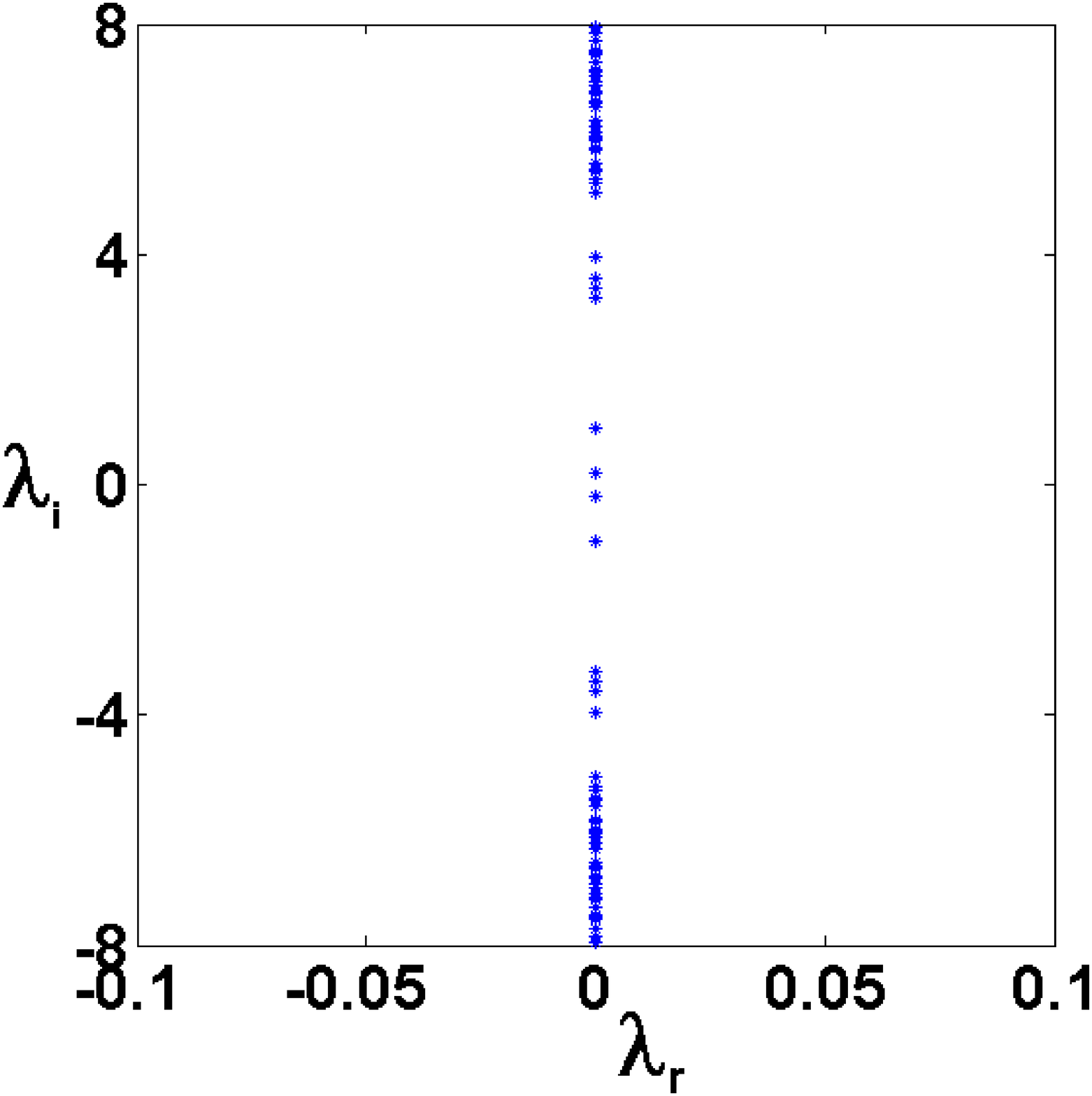}
\includegraphics[width=6.1cm]{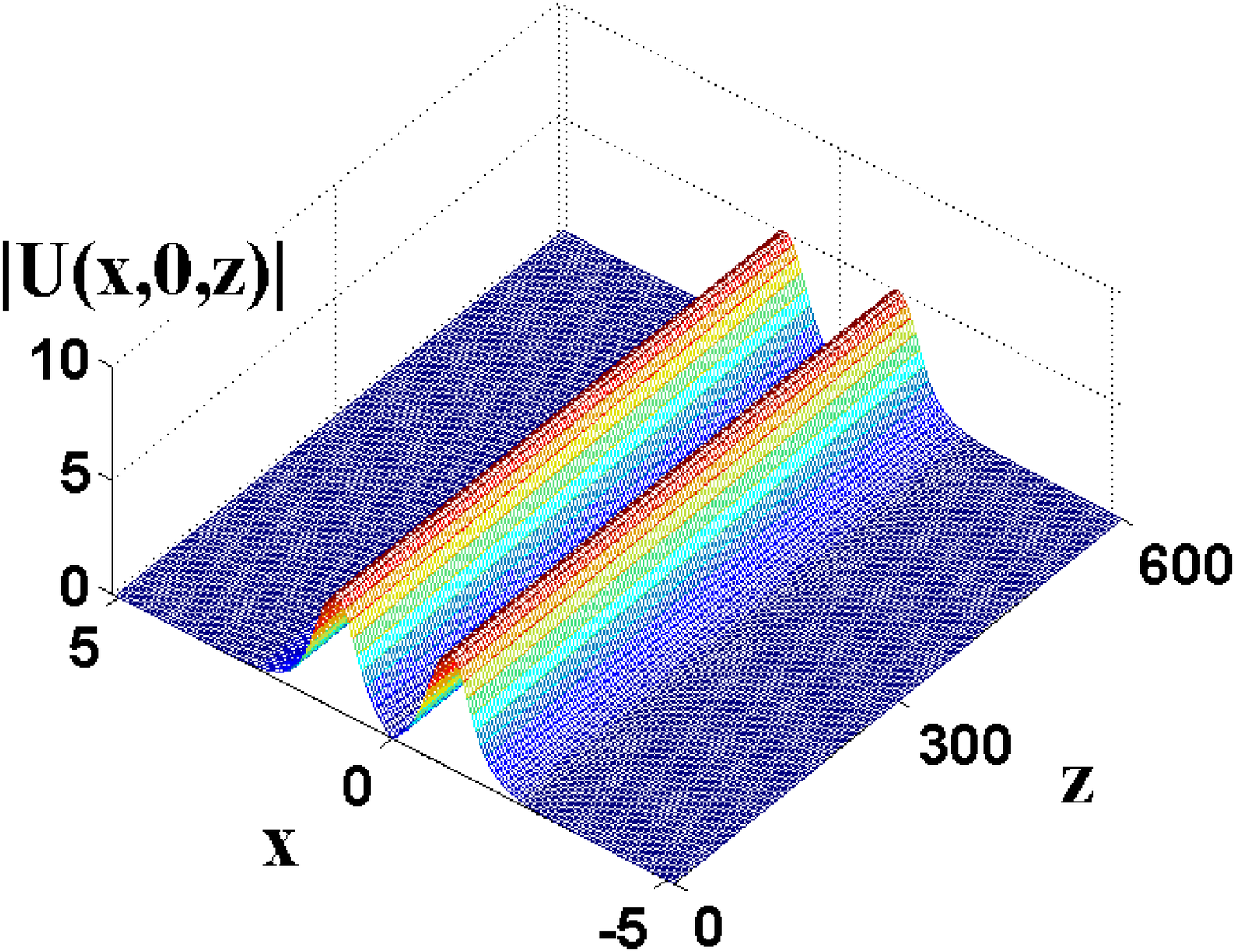} %
\caption{(Color online) A stable antisymmetric soliton of Eq. (\protect\ref%
{2D}) with $x_{0}=0.83$, $\protect\sigma =0.0288$ and $\protect\mu =5.04$, $%
N=16.96$.} \label{contour2db3}
\end{figure}

Typical examples of symmetric, asymmetric and antisymmetric solitons are
displayed in Figs. \ref{contour2da1} -- \ref{contour2db3}. Note that
symmetric solitons may feature both single-peak and double-peak shapes. In
particular, unstable single-peak symmetric solitons spontaneously turn into
excited asymmetric solitons, while unstable double-peak symmetric solitons
radiate away a part of their norm before turning into strongly excited
asymmetric solitons, see Figs. \ref{contour2da} and \ref{contour2d2}). A
typical example of the evolution of an unstable asymmetric soliton is
presented in Fig. \ref{contour2db}, where we observe its spontaneous
transformation into a robust asymmetric breather, following shedding off a
part of its norm with transient radiation.

\begin{figure}[tbp]
\includegraphics[width=7.5cm]{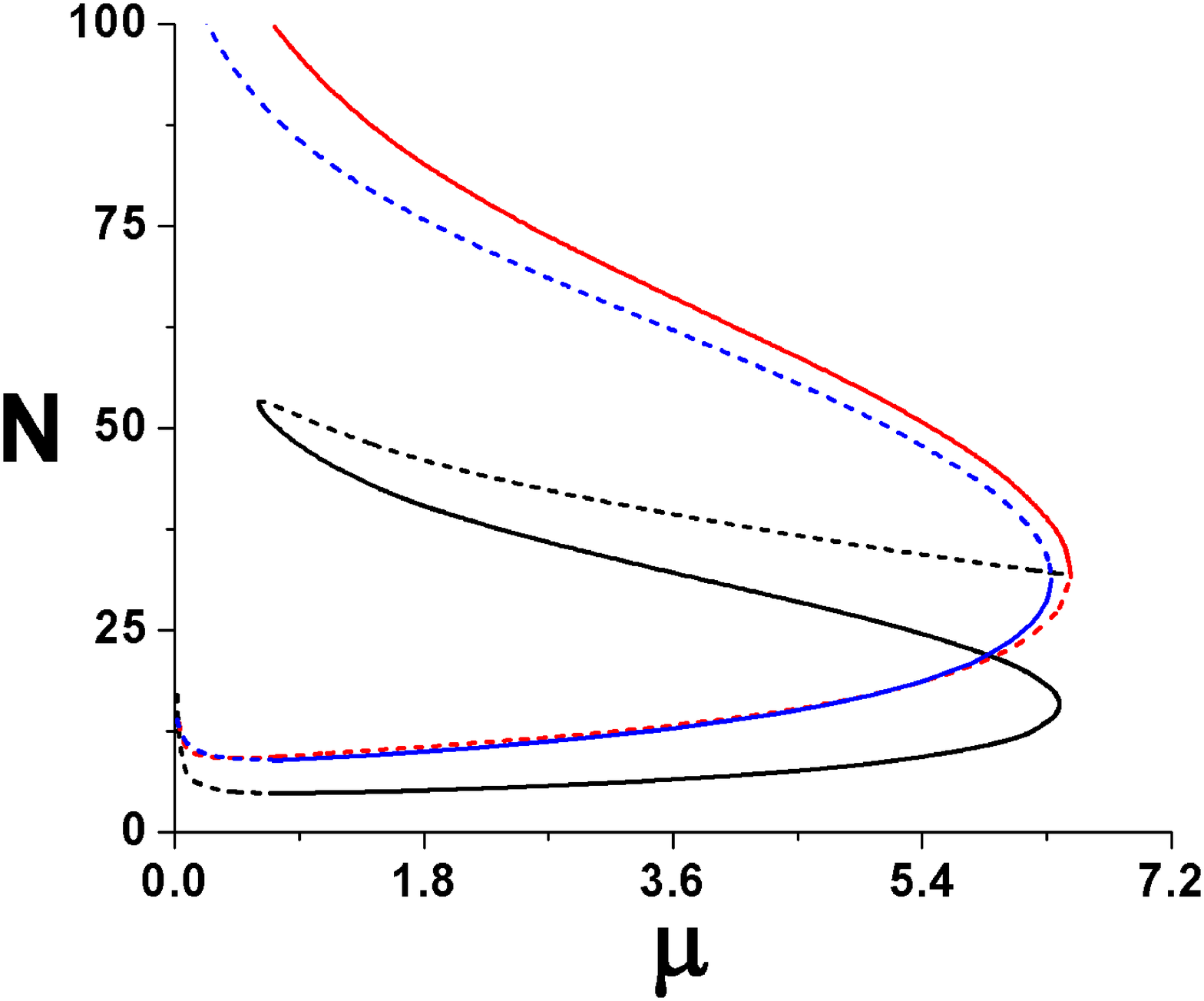} %
\includegraphics[width=7.5cm]{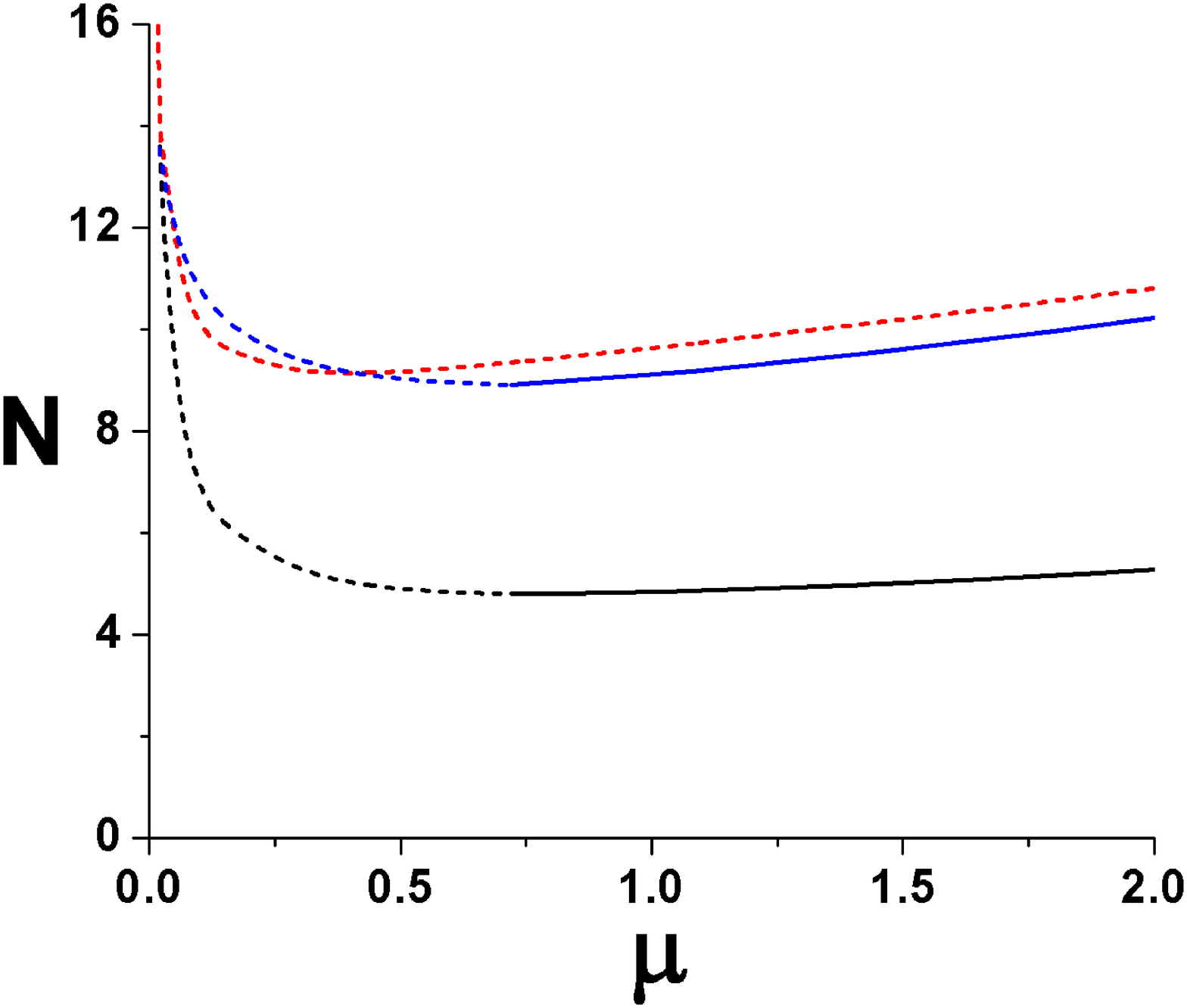}
\caption{(Color online) The first panel: the total power vs. the
propagation
constant, $N(\protect\mu )$, for solitons in the 2D double-well system with $%
\protect\sigma =0.0288$ and $x_{0}=0.83$ [i.e., with non-overlapping
circles in Eq. (\protect\ref{2D})]. The second panel is a closeup of
a region at small $\protect\mu $ of where the asymmetric branch
splits off from the symmetric one. Here and below, the solid and
dashed curves (or chains of symbols) represent, respectively, stable
and unstable families of solitons. The red, blue, and black colors
designate, severally, symmetric, antisymmetric, and asymmetric
modes.} \label{dia5a}
\end{figure}

\begin{figure}[tbp]
\includegraphics[width=8cm]{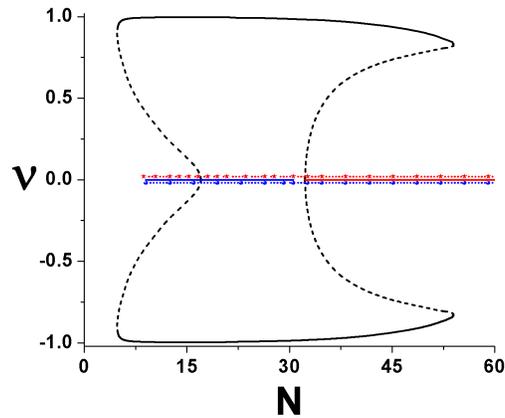}
\caption{(Color online) The symmetry-breaking diagram, $\protect\nu
(N)$, in
the 2D double-well model (\protect\ref{2D}) with $\protect\sigma =0.0288$ and $%
x_{0}=0.83$, i.e., non-overlapping circles.} \label{dia5}
\end{figure}

\begin{figure}[tbp]
\includegraphics[width=6.5cm]{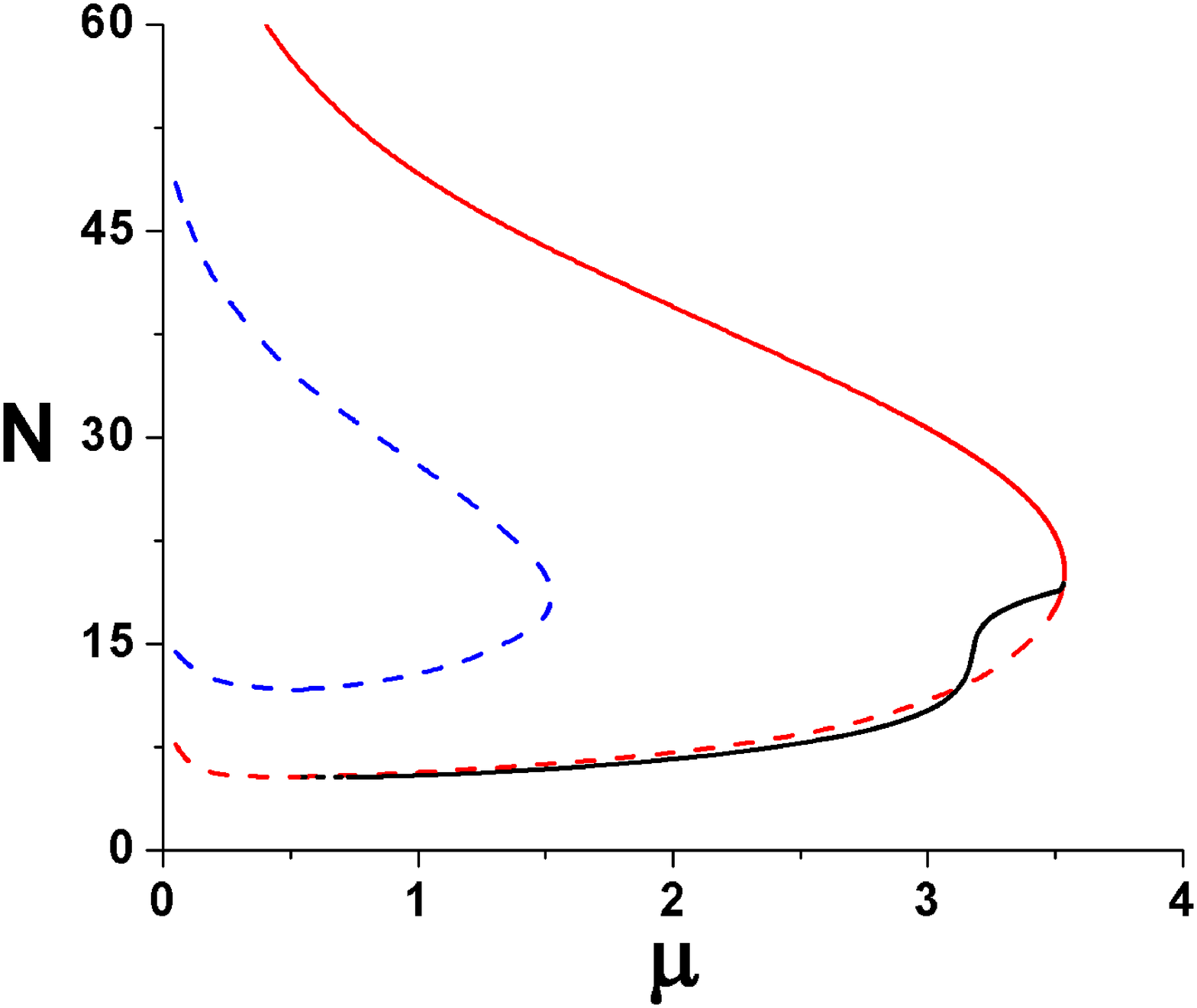} %
\includegraphics[width=6.5cm]{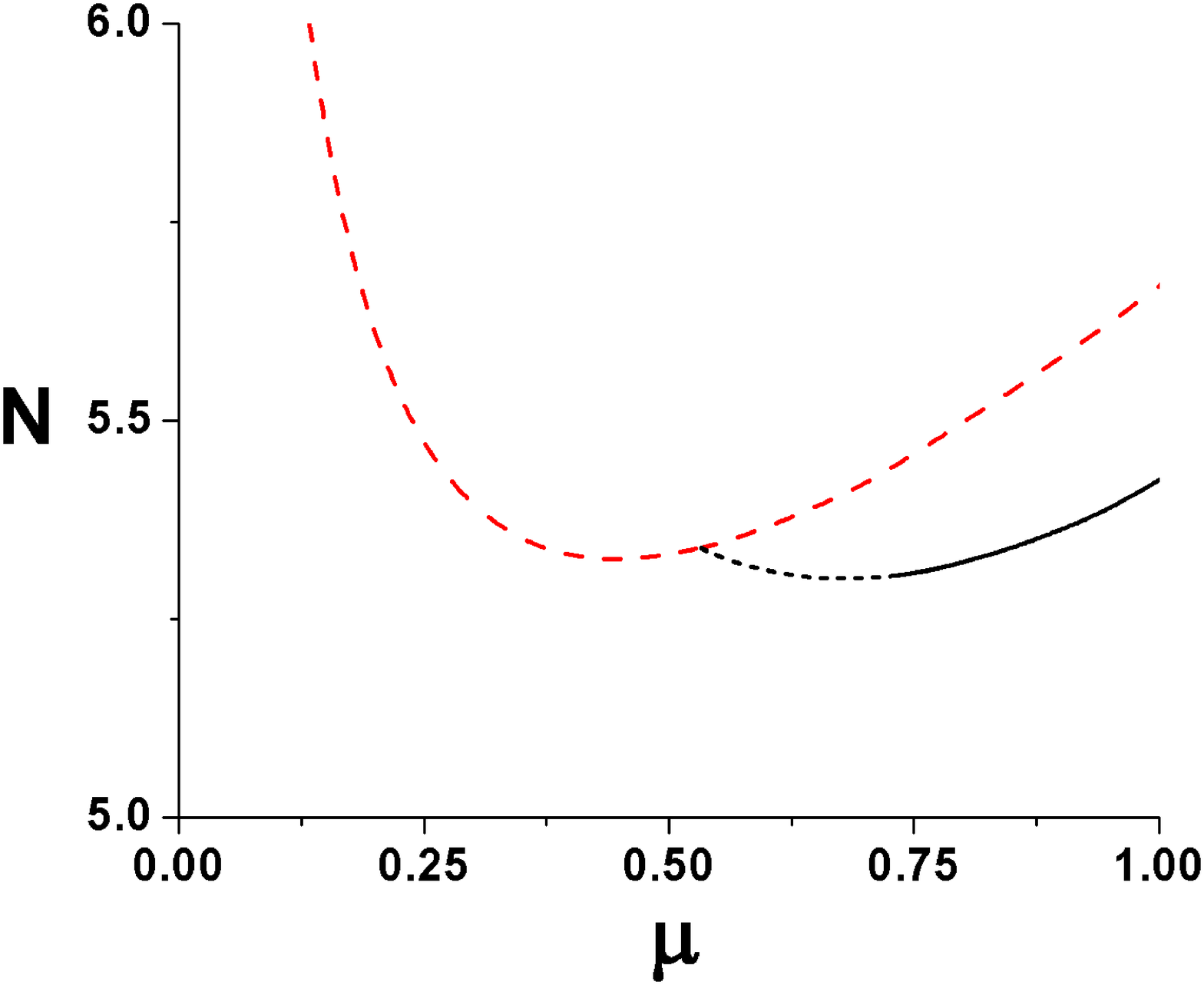}
\caption{(Color online) The first panel: the curves $N(\protect\mu
)$ for the 2D double-well system described by Eq. (\protect\ref{2D})
with partly ovelapping circles, $x_{0}=0.45$ and $\protect\sigma
=0.0484$. The second panel is a closeup of the region of small
$\protect\mu $, where the symmetry-breaking transition from
symmetric to asymmetric solitons occurs.} \label{dia4b}
\end{figure}

\begin{figure}[tbp]
\includegraphics[width=8cm]{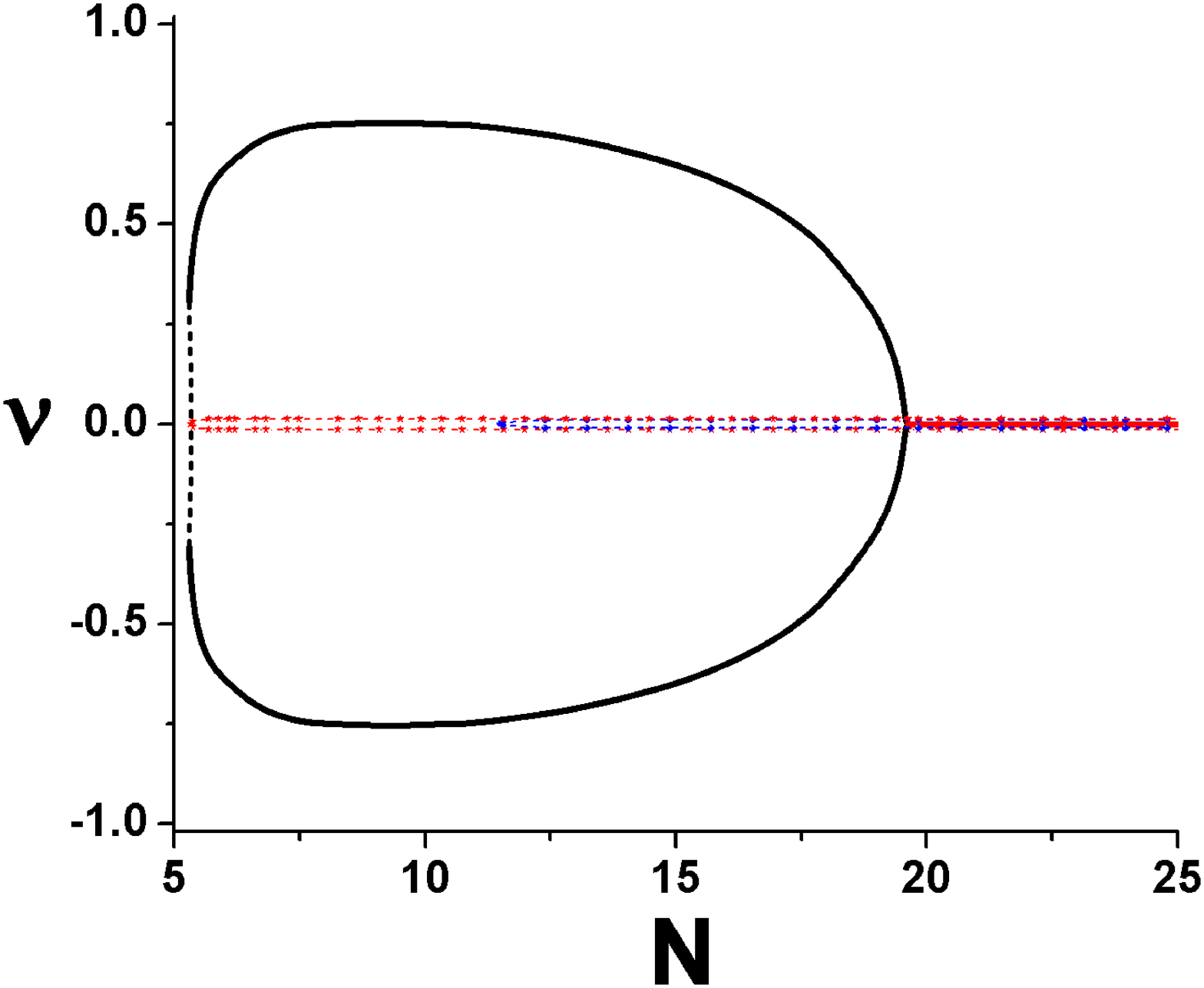}\\
\includegraphics[width=6cm]{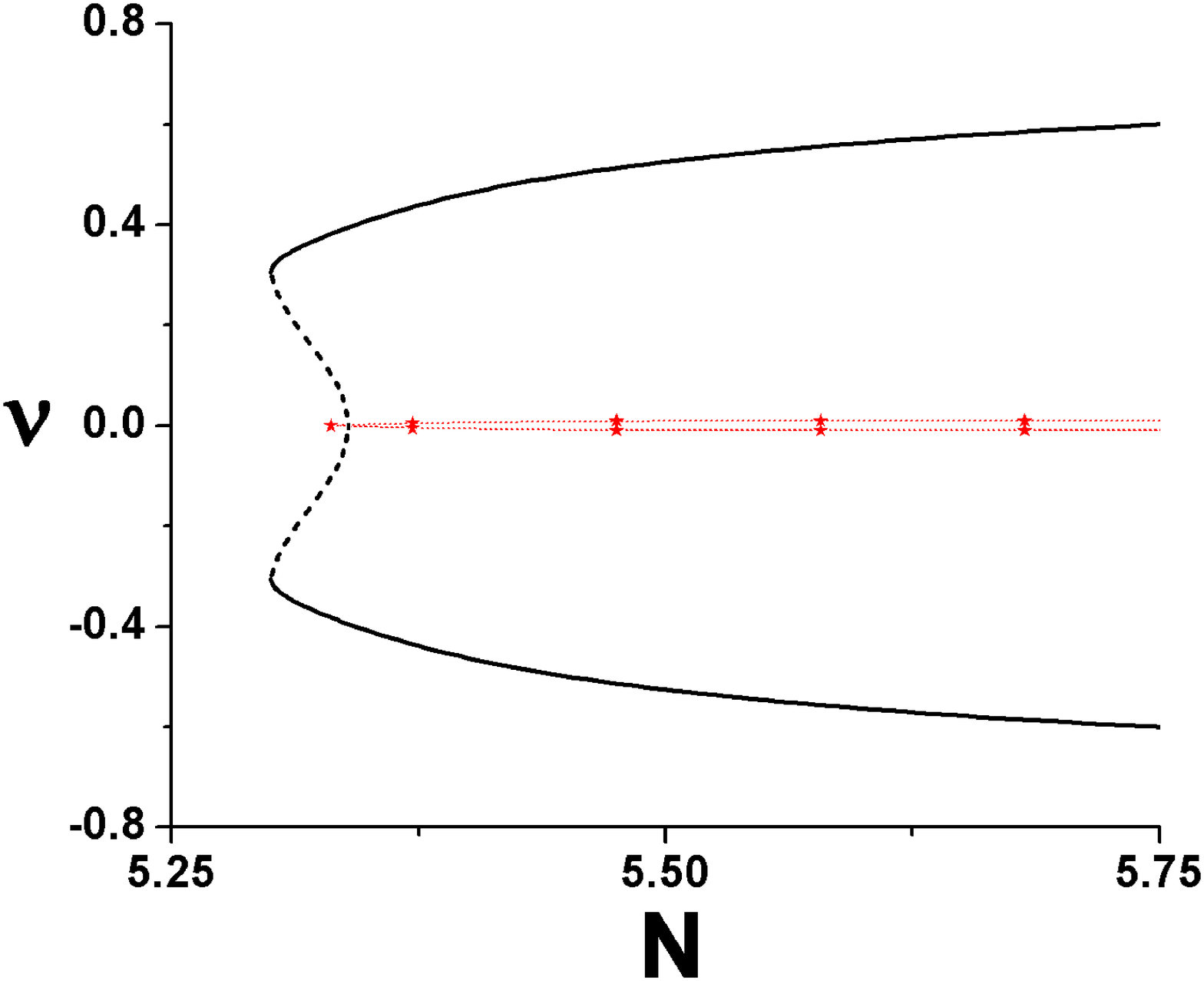} %
\includegraphics[width=6cm]{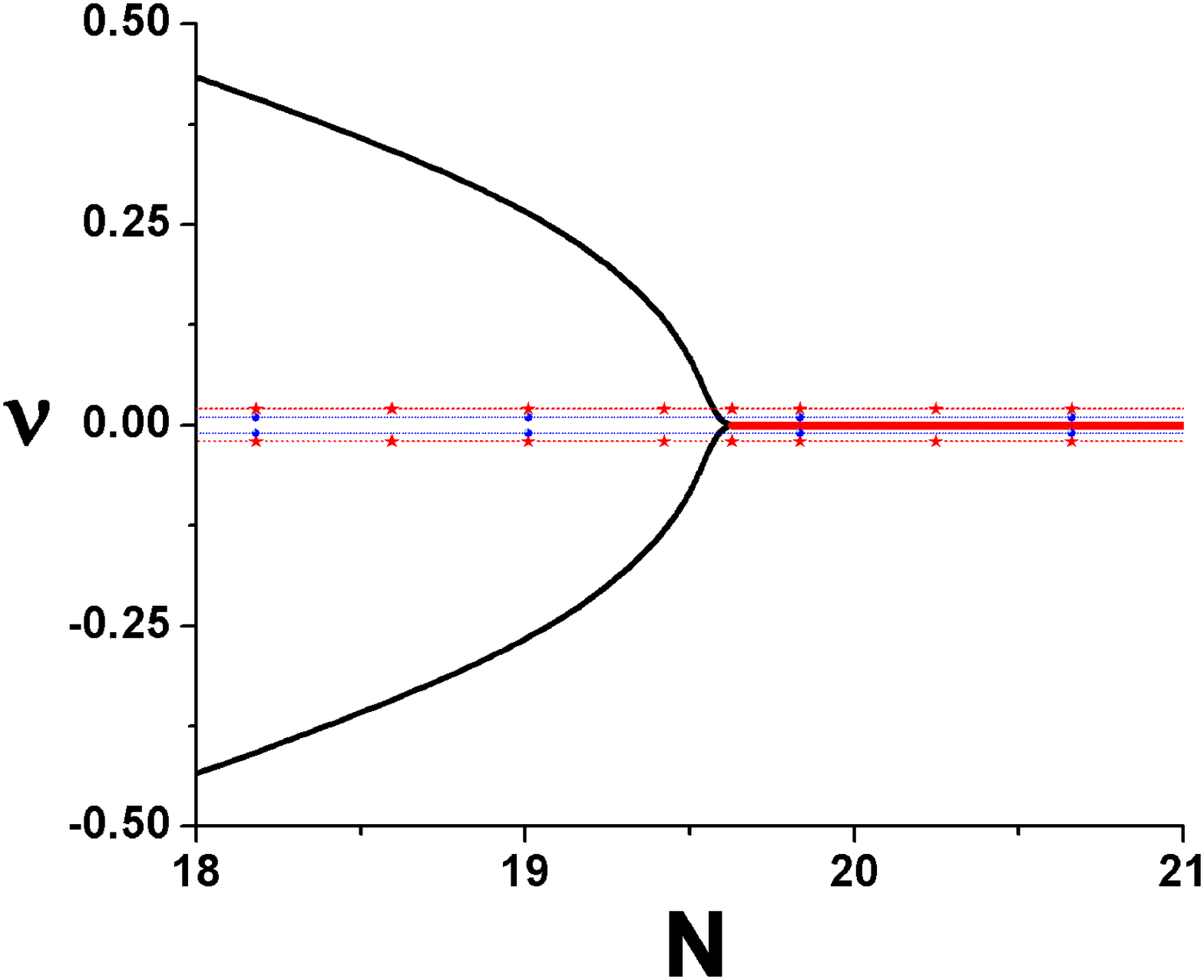}
\caption{(Color online) The symmetry-breaking diagram, $\protect\nu
(N)$
[the asymmetry measure, $\protect\nu $, is defined as per Eq. (\protect\ref%
{nu-2D})] for the 2D double-well system, based on Eq.
(\protect\ref{2D})
with the partly overlapping circles, $x_{0}=0.45$, and $\protect\sigma =0.0484$%
. The two bottom panels are closeups of regions where the
symmetry-breaking and restoring bifurcations (opening up and closing
down of the bifurcation loop) take place.} \label{dia4a}
\end{figure}
The results of the numerical analysis of the 2D double-well system
are summarized in Figs. \ref{dia5a} - \ref{dia4a}, which address the
existence and stability, represented by curves $N(\mu )$, and the
symmetry breaking between the wells, represented by the $\nu (N)$
curves, for two generic situations, which correspond to the
separated circles ($x_{0}>1/2$) or partly overlapping ones
($x_{0}<1/2$). In the latter case, the $N(\mu )$ curves are similar
to their counterpart in the 1D model, cf. Fig. \ref{dia1}. In
particular, different branches are stable according to the VK or
anti-VK criterion, depending on whether the SF or SDF term is the
dominant one. Collapse of 2D modes never occurs in the presence of
$\sigma >0$.

The bifurcation loop accounting for the breaking and restoration of
the symmetry is concave for the weak coupling between the circles ($x_{0}=0.83$%
), in Fig. \ref{dia5}, and convex for the strongly coupled (partly
overlapping) circles, with $x_{0}=0.45$ in Fig. \ref{dia4a}. This
observation agrees with what was found before in the above-mentioned
dual-core systems carrying the competing SF-SDF cubic-quintic
nonlinearity \cite{Albuch}. Note that the concave and convex
bifurcation loops correspond to strongly
differing sets of the $N(\mu )$ curves, as seen from the comparison of Figs. %
\ref{dia5a} and \ref{dia4b}.

\section{Conclusions}

The first objective of this work was to address the competition of SF and
SDF (self-focusing and self-defocusing) nonlinearities in the system where
both nonlinear terms have the same dimension (scaling behavior). The 1D
system offers a physically relevant setting for the realization of such a
system, in the form of the combination of the SF cubic term localized with
the delta-functional coefficient, and the spatially uniform SDF quintic
term. We have found, in the exact form, the most general family of the 1D
Townes' solitons. It remains degenerate and unstable irrespective of the
relative sign between the cubic and quintic terms. However, a weak
regularization of the $\delta $-function immediately stabilizes the solitons
in the case of the competing (opposite) signs of the SF cubic and SDF
quintic terms. These results were obtained, and mutually verified, in the
numerical and analytical forms. Then, the analysis was extended to the 1D
system with the pair of regularized $\delta $-functions, and, eventually, to
the 2D single- and double-delta-function systems. In all the cases, stable
families of solitons have been identified.

In the 2D geometry, a remaining problem is to construct vortex
solutions. A challenging possibility is to extend the analysis to 3D
settings, which may be as physically relevant as the BEC model.

\section{Acknowledgements}

This research is funded by Vietnam National Foundation for Science
and Technology Development (NAFOSTED) under Grant number
"103.01-2013.48" (N.V.H.).

\end{document}